%% file: Notes.tex
\documentclass[%
 reprint,
superscriptaddress,
nobibnotes,
 amsmath,amssymb,
 aps,
prl,
floatfix,
]{revtex4-1}

\usepackage{dsfont}
\usepackage{graphicx}% Include figure files
\usepackage{dcolumn}% Align table columns on decimal point
\usepackage{bm}% bold math
\usepackage{bbm}
\usepackage{soul}
\usepackage{hyperref}
\usepackage{cleveref}
\usepackage{breqn}
\usepackage{subfigure}
\usepackage{float}
\usepackage{flushend}
\usepackage{balance}

\newcommand{\be}{\begin{equation}}
\newcommand{\ee}{\end{equation} }

\newcommand{\us}{\underline{s} }

\newcommand{\bsl }{\backslash}

\begin{document}

%\preprint{APS/123-QED}
 \title{A fast and accurate algorithm for inferring sparse Ising models via parameters activation to maximize the pseudo-likelihood}% Force line breaks with \\

\makeatletter
\let\cat@comma@active\@empty
\makeatother

\author{Silvio Franz}%
\affiliation{LPTMS, Universit\`{e} Paris-Sud 11, UMR 8626 CNRS, B\^{a}t. 100, 91405 Orsay Cedex, France}
\affiliation{Dipartimento di Fisica Universit\`{a}, La Sapienza, Piazzale Aldo Moro 5, I-00185 Roma, Italy}

\author{Federico Ricci-Tersenghi}
\affiliation{Dipartimento di Fisica Universit\`{a}, La Sapienza, Piazzale Aldo Moro 5, I-00185 Roma, Italy}
\affiliation{INFN-Sezione di Roma 1, and CNR-Nanotec, Rome unit, P.le A. Moro 5, I-00185, Roma, Italy}

\author{Jacopo Rocchi}
\affiliation{LPTMS, Universit\`{e} Paris-Sud 11, UMR 8626 CNRS, B\^{a}t. 100, 91405 Orsay Cedex, France}


\date{\today}% It is always \today, today,

\begin{abstract}

We propose a new algorithm to learn the network of the interactions of pairwise Ising models. The algorithm is based on the pseudo-likelihood method (PLM), that has already been proven to efficiently solve the problem in a large variety of cases. Our present implementation is particularly suitable to address the case of sparse underlying topologies and it is based on a careful search of the most important parameters in their high dimensional space. We call this algorithm Parameters Activation to Maximize Pseudo-Likelihood (PAMPL). Numerical tests have been performed on a wide class of models such as random graphs and finite dimensional lattices with different type of couplings, both ferromagnetic and spin glasses. These tests show that PAMPL improves the performances of the fastest existing algorithms. 

%\begin{description}
%You may use the \texttt{description} environment to structure your abstract; use the optional argument of the \verb+\item+ command to give the category of each item.
%\end{description}
\end{abstract}

\pacs{Valid PACS appear here}% PACS, the Physics and Astronomy
\maketitle

The Ising model is a graphical model whose parameters $\{ J_{ij},h_i \}$ can be tuned in order to describe stationary distributions of binary variables, $s_i$, according to the weight $P(\underline{s}) \sim \exp \left( \sum_{i<j} J_{ij} s_i s_j + \sum_i h_i s_i \right)$. In many practical problems in different domains - e.g. physics, biology, neuroscience, finance, sociology - the topology of the graph and the values of the couplings are unknown and they need to be reconstructed from the data. The inverse Ising problem aims to find the parameters of the model that best fit the data. 

From the original attempt to solve this problem \cite{ackley1985learning}, many techniques of statistical mechanics and machine learning have been developed \cite{kappen1998efficient,tanaka2000information,sohl2011new,cocco2011adaptive,aurell2012inverse,ricci2012bethe,nguyen2012mean,cocco2012adaptive,raymond2013mean,
decelle2014pseudolikelihood,lokhov2018optimal} to study different cases. The need to develop approximate algorithms can be understood from the observation that the likelihood depends on the partition function, which is generally intractable. 
Among these methods, the pseudo-likelihood \cite{ravikumar2010high} has been proven to be particularly efficient, leading to polynomial algorithms which give the exact solution in the limit of infinite sampling. Methods based on the pseudo-likelihood need to be complemented with a threshold procedure, implemented a posteriori or through a regularization function. 
An improvement of this method based on a decimation scheme was presented in \cite{decelle2014pseudolikelihood}. The decimation based algorithm has been shown to outperform existing algorithms based on the pseudo-likelihood method in terms of the quality of the reconstructed graph and it has been commonly used in a variety of contexts \cite{marruzzo2017inverse,marruzzo2018improved,ancora2019learning}. Our aim is to improve this algorithm in the case of sparse graphs. In fact, in this case, the underlying structure is closer to an empty graph than to a fully connected graph and we would like to avoid to explore the full set of parameters in the inference process, while maintaining the same quality in the inferred graph. While the decimation step is $O(N^2)$, our elementary operation is $O(N)$. We begin formulating the Inverse Ising problem. We discuss the pseudo-likelihood method and the present implementation. Finally we describe the results of our algorithm in a wide class of Ising models with a comparison with the fast Minimum Probability Flow (MPF) \cite{sohl2011new}, showing that the two methods have similar execution times and that ours outperform the other in terms of the quality of the reconstructed graphs.

An Ising model in the absence of local fields is defined by the Hamiltonian $H(\us)=-\sum_{i<j}J_{ij} s_i s_j$. After the observation of $M$ independent equilibrium configurations, the problem of inferring the couplings can be formulated in terms of the Bayes theorem $P( J | \{ \us^{(\mu)}\}_{\mu=1,\ldots,M} ) \propto P( \{ \us^{(\mu)}\}_{\mu=1,\ldots,M} | J) P(J)$, where the two functions on the r.h.s. are named likelihood and prior, respectively. If we assume to be in a Bayes optimal case setting where we do not need to introduce local fields in the model, the log-likelihood function is defined by $\mathcal{L}(J) = M^{-1} \log P( \us^{(\mu)}\}_{\mu=1,\ldots,M} | J)$ and reads
\be   
\mathcal{L}(J) = \frac{\beta}{M} \sum_{\mu=1}^{M} \sum_{i<j}J_{ij} s^{(\mu)}_i s^{(\mu)}_j - \log Z(J),
\ee
where $Z(J)=\sum_{\us} e^{\beta \sum_{i<j}J_{ij} s_i s_j}$ is the partition function of the problem, and $\beta$ an inverse temperature. Optimizing $\mathcal{L}(J)$ over the parameters of the model leads to
\be
\frac{\partial \mathcal{L}(J)}{\partial J_{ij}} =  \beta \left( \frac{}{} \left<s_i s_j\right>_{\text{Data}} -\left<s_i s_j\right>_{\text{Model}} \right) \:,\:
\ee
where we defined $M \left< s_i s_j \right>_{\text{Data}} =  \sum_{\mu=1}^M s^{(\mu)}_i s^{(\mu)}_j$ and $\beta \left< s_i s_j \right>_{\text{Model}} =  \partial_{J_{ij}} \log Z $. The $J$ that maximizes the log-likelihood is such that
\be
\left< s_i s_j \right>_{\text{Data}} = \left< s_i s_j \right>_{\text{Model}} \:.
\label{eq:Lcond}
\ee
This formulation involves the computation of the partition function, which is a complicated object. The log-pseudo-likelihood \cite{besag1977efficiency} is introduced to deal with this difficulty. It is defined by
\be
\mathcal{S}(J) = \frac{1}{M} \sum_{r=1}^N \sum_{\mu=1}^M \log p(s_r^{(\mu)} | \us^{(\mu)}_{\bsl r})
\label{eq:Sdef}
\ee
where $p(s_r | \us_{\bsl r}) = \left[1 + e^{-2 \beta s_r \sum_{j\neq r} J_{rj} s_j }\right]^{-1}$, and as discussed in the Appendix A, maximizing $\mathcal{S}$ leads to the correct solution in the infinite sampling limit.

\begin{figure}[h]
\centerline{\resizebox{0.5\textwidth}{!}{\input{Activation-36-2D-FB.tex}}}
\caption{2-D ferromagnetic lattice with $N=36$ and free boundary conditions, $M=5000$, $\beta=0.5$. Main Figure: (a) $-\text{BIC}/M$ and $10 \epsilon$ as a function of $x$, fraction of couplings, for the activation-based method. $x$ is $0$ at the beginning and it increases as the iteration proceeds. Only the first part of the iteration is presented. The Arrow specifies the directions in which the iteration moves. Inset: ROC curve, with the TNR on the x-axis and the TPR on the y-axis.}
\label{fig:figA}
\end{figure}
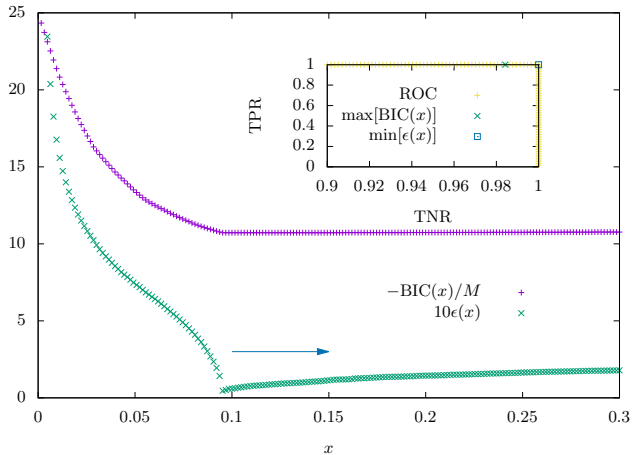

The use of the pseudo-likelihood has been already shown to be very useful in the context of the Inverse Ising problem \cite{ravikumar2010high,aurell2012inverse}.
The standard implementation of this method consists in
maximizing each of the $N$ local likelihood functions
separately, thus getting two different estimates for each
coupling. When complemented with a post-optimization parameter thresholding procedure, this method can be shown to reconstruct arbitrary Ising models \cite{lokhov2018optimal}. Anyway, this scheme relies on the delicate choice of the threshold and leads to extimated couplings that are systematically smaller than the true values. This problem was first addressed in \cite{decelle2014pseudolikelihood} to eliminate the bias in the coupling estimation, where an iterative decimation based approach was developed. A maximization over the pseudo-likelihood is alternated with a decimation step where the smallest estimated couplings are set to zero. More details and examples are provided in Appendix B.

Here we propose an improvement of this algorithm especially suitable for sparse graphs. In fact, in this case, starting from the complete graph and decimating couplings requires a long time before reaching the correct stopping point. On the contrary, it would be wiser to have an iterative algorithm that starts from the empty graph, and add links sequentially. We call this algorithm Parameters Activation to Maximize Pseudo-Likelihood (PAMPL). In order to add the correct links, we search for the directions, in the parameter space, that give the largest gain in the log-pseudo-likelihood. The change in $\mathcal{S}$ due to a change in the coupling $J_{ij}$ is estimated using a second order approximation, 
\be
\mathcal{S}(J+\Delta J_{ij})=\mathcal{S}(J)+\mathcal{S}'(J)\Delta J_{ij}+\frac{ \mathcal{S}''(J)}{2}\Delta J_{ij}^2+ \ldots\:,
\label{eq:GainS}
\ee
where prime denotes differentiation with respect to $J_{ij}$. Couplings are updated with one step of the Newton method, 
\be
0 = \mathcal{S}'(J_{ij}) + \mathcal{S}''(J_{ij}) \Delta J_{ij}\:,
\label{eq:Newton}
\ee 
and ranked in an ascending order according to the values of the quantities
\be
\Delta \mathcal{S}_{ij} = -\frac{1}{2} \frac{ {\mathcal{S}'}^2(J_{ij}) }{ \mathcal{S}''(J_{ij}) },
\ee
obtained by plugging eq. (\ref{eq:Newton}) in eq. (\ref{eq:GainS}). Finally, the first $K$ are included in the set of non-zero couplings $\mathbb{J}$ and optimized as explained below. This procedure is iterated adding more and more couplings in $\mathbb{J}$ at each iteration. In order to keep this elementary step $O(N)$, updating and sorting need to be done carefully. In particular, there is no need to update all of the $\Delta \mathcal{S}_{ij}$ at each step, since only $O(N)$ of them are affected by the presence of a new coupling in $\mathbb{J}$. Moreover, since most of them are small, we don't need to order $O(N^2)$ elements, but only $O(N)$. We use the Bayesian Information Criterion (BIC), introduced in \cite{schwarz1978estimating}, to locate the stopping point of the iteration. For our purposes, it is defined by
\be
\text{BIC}= 2 M \mathcal{S}^{*} - k \log M\:,
\label{eq:BIC}
\ee
where $k = || \mathbb{J} || $ is the cardinality of the set  $\mathbb{J}$, corresponding to the number of links used to describe the observations, and $\mathcal{S}^{*}$ is the maximum of $\mathcal{S}$ found optimizing the couplings in $\mathbb{J}$. In order to perform the optimization step on the couplings of $\mathbb{J}$, we used the LBFSG \cite{liu1989limited} algorithm and a simple gradient ascent. Results obtained in the two cases are the same within numerical errors and, since the second one is faster, it is particularly appropriate for large systems. In the following we update one coupling per iteration time. More details about the algorithm are provided in Appendix C. 

We generate independent equilibrium configurations from given graphs using a Monte Carlo sampling algorithm. Then, during the inference process, we compare the inferred graph with the original one using the measure
\be
\epsilon=\sqrt{ \frac{ \sum_{i<j}(J_{ij}-J^*_{ij})^2 }{\sum_{i<j} J^2_{ij}} } \:,
\label{eq:eddefepserr}
\ee
were $J^*$ is the original set of couplings. 
\begin{figure}
    \centering
    \subfigure[]
    {
        \centerline{\resizebox{0.5\textwidth}{!}{\input{differentTM-av-50-RR-05.tex}}}
    }
    \\
    \subfigure[]
    {
        \centerline{\resizebox{0.5\textwidth}{!}{\input{diff_M_50-RR-05.tex}}}
    }
    \caption{
Random regular spin glass with $N=50$ spins, $c=4$, $100$ couplings, $\beta_c=0.524$. (a): Main Figure: average of $\text{BIC}/M$ as a function of the iteration time over $10$ runs of the activation algorithm, for different values of $M$ at $\beta=0.8$. Inset: average of $\text{BIC}/M$ as a function the iteration time over $10$ runs of the activation algorithm, for different values of the inverse temperature $\beta$, at $M = 8000$. $\text{BIC}$ corresponding to different $\beta$ has been shifted in order to fit in the same figure.
(b): Average of $\Delta \text{BIC}(t)$  over $10$ runs of the activation algorithm, for different values of the inverse temperature $\beta$.
}
\label{fig:setteT}
\end{figure}
\begin{figure}
    \centering
    \subfigure[]
    {
        \centerline{\resizebox{0.5\textwidth}{!}{\input{differentTM-av-44-Dia.tex}}}
    }
    \\
    \subfigure[]
    {
        \centerline{\resizebox{0.5\textwidth}{!}{\input{diff_M_44-Dia.tex}}}
    }
    \caption{
2D ferromagnetic diamond lattice with $N=44$ spins, $64$ couplings, $\beta_c=0.609$. (a): Main Figure: average of $\text{BIC}/M$ as a function of the iteration time over $10$ runs of the activation algorithm, for different values of $M$ at $\beta=0.5$. Inset: average of $\text{BIC}/M$ as a function the iteration time over $10$ runs of the activation algorithm, for different values of the inverse temperature $\beta$ at $M = 8000$. $\text{BIC}$ corresponding to different $\beta$ has been shifted in order to fit in the same figure.
(b): Average of $\Delta \text{BIC}(t)$  over $10$ runs of the activation algorithm, for different values of the inverse temperature $\beta$.
}
\label{fig:ottoT}
\end{figure}
We study graphs with ferromagnetic and spin glass interactions. We denote with spin glass graph systems with couplings equal to $\pm 1$ with probability $0.5$. We stress that no extra time is required to infer the structure of a spin glass topology compared to the ferromagnetic case.
In Fig. \ref{fig:figA} we show the behavior of our algorithm on a 2-D lattice with $N=36$ and free boundary conditions. Inference is made after observing $M=5000$ samples extracted at equilibrium at $\beta=0.5$. In the inset we plot the ROC curves that give information on the fraction of true couplings retrieved (true positive rate, TPR), and the fraction of non-existent couplings not created by the algorithm (true negative rate, TNR). Each point of the curve corresponds to the graph inferred at a particular stage of the iterative process. Ideally, for a perfect reconstruction, the inferred graph corresponds to the point $(1,1)$. We observe that the maximum of the BIC does not coincide with the point where $\epsilon$ is minimum. This is due to the fact that after the activation of all the correct couplings of the graph, the BIC keeps growing for some other time steps before the penalty terms start to be effective. Despite this issue, we notice that the correct stopping point is clearly recognizable by a visual inspection: this problem can be overcame easily, as will be shown below.

In Fig. \ref{fig:setteT}-\ref{fig:ottoT} we study the performances of the algorithm with $M$ and $\beta$ for different topologies and sizes. As expected, inference becomes hard in the low temperature phase and when the dataset is too small. In fact, for small $M$ the singular behavior of the $\text{BIC}$ becomes smoother, and the detection of the stopping point is impossible. On the other hand, at larger values of $\beta$, more and more samples are required for a correct inference because most of the samples are very close to the ground state(s) and we lose information from fluctuations. We consider a Random Regular (RR) spin glass graph with $N=50$ and $c=4$ and a ferromagnetic diamond lattice of $N=44$ spins. Diamond lattices \cite{berker1979renormalisation}, \cite{kaufman1981exactly} are graphs constructed recursively from a single link corresponding to the generation $n=0$. The generation $n=1$ consists of 2 branches in parallel, each one made by 2 links in series. The generation $n=2$ is obtained by applying the same transformation to the each link. The present case corresponds to the case $n=3$. In this graphs there is a clear hierarchy between couplings and we show that the learning algorithm is clearly sensitive to it. In both cases we study the quantity $\Delta \text{BIC} (t) = [\text{BIC}(t) - \text{BIC}(t-1)]/M$, averaged over $10$ inference iterations, as a function of the iteration time. We observe that it becomes very small as soon as the correct graph structure is recovered, and that this threshold behavior is more evident in the vicinity of the phase transition. We find that a good stopping point can be defined when $\Delta \text{BIC}<0.01$.
\begin{figure}[h]
\centerline{\resizebox{0.5\textwidth}{!}{\input{ROC_M_DiaRR.tex}}}
\caption{ROC values for the (a) random graph case considered in Fig. (\ref{fig:setteT}), (b) diamond lattice case considered in Fig. (\ref{fig:ottoT}). In both cases, each point corresponds to a different value of $M$, for $M=100,200,300,400,500,1000,2000,4000,8000$, and they approach $(1,1)$ as $M$ increases.}
\label{fig:nove}
\end{figure}
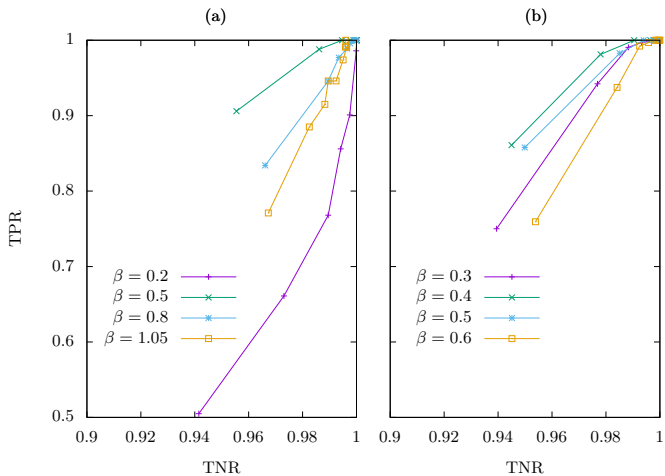
The quality of the reconstructed graphs using this criterion is studied in detail in Fig. \ref{fig:nove}, where we plot the ROC parameters TNR and TPR for different temperatures, for the cases discussed above. 
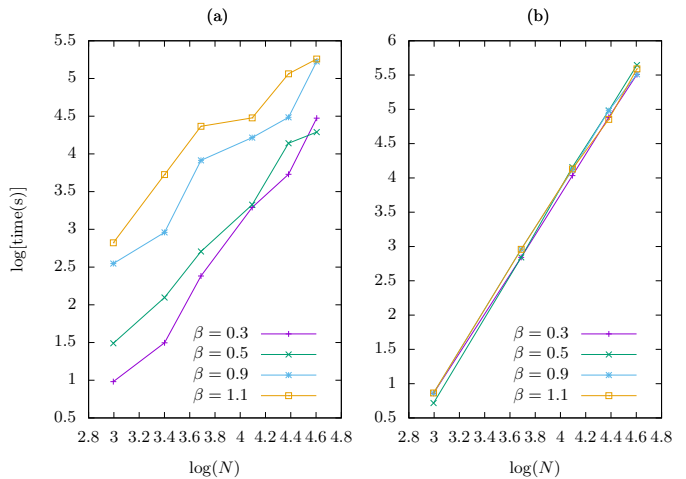
\begin{figure}[h]
\centerline{\resizebox{0.5\textwidth}{!}{\input{time-RR.tex}}}
\caption{Comparison between the $\log$ of the execution time (seconds) needed to find a solution as function of $N$ and $\beta$ for a RR spin glass with $c=3$ using (a) PAMPL and (b) MPF. Each point corresponds to an average over $10$ runs of the algorithm with $M=4000$. $\beta_c=0.615$.}
\label{fig:time-RR}
\end{figure}

\begin{figure}[h]
\centerline{\resizebox{0.5\textwidth}{!}{\input{eps-RR.tex}}}
\caption{Error $\epsilon$, defined in eq. (\ref{eq:eddefepserr}), as a function of $N$ and $\beta$ for a RR spin glass with $c=3$ using (a) PAMPL and (b) MPF. Each point corresponds to an average over $10$ runs of the algorithm at $M=4000$.}
\label{fig:eps-RR}
\end{figure}
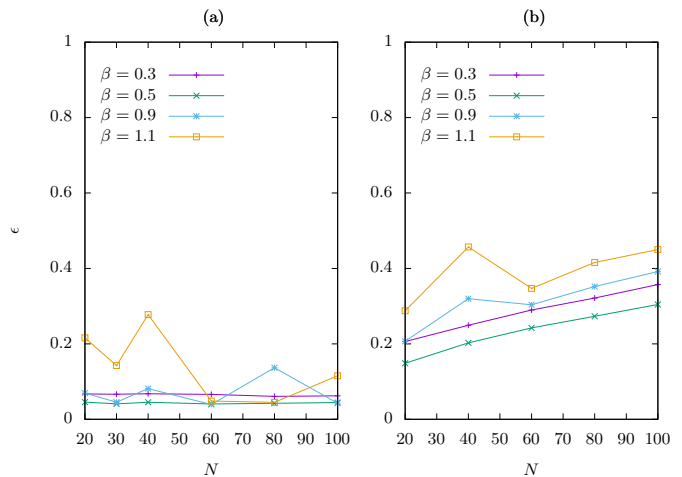

We compare the performances of PAMPL with those of another fast inference method, namely the MPF \cite{sohl2011new}. MPF is as fast as a single maximization of the pseudo-likelihood and needs to be complemented with a threshold procedure. In Fig. \ref{fig:time-RR}-\ref{fig:eps-RR} we analyze a RR graph with $c=3$ with these two methods. MPF is expected to be $O(M N^2)$, as ours. While the tests show a more pronounced temperature dependence for PAMPL, we observe that the execution times are of the same order, both being very fast. Moreover, as the decimation algorithm improved the performances of methods based on the maximization of the pseudo-likelihood, similarly the errors in the reconstructed graph made by PAMPL are 2-3 time smaller than those made by MPF. We also note that while the errors made by PAMPL does not depend on $N$, the error made by MPF do. More details on MPF and the case of a 2-D ferromagnetic lattice is discussed in details in Appendix D. 

In summary we presented a new method, to reconstruct the hidden structure of Ising models based on the pseudo-likelihood and an activation procedure that includes recursively new parameters into a set whose elements are then optimized over. The method is exact in the limit of very large number of samples $M$ and does not require setting ad-hoc extra parameters, apart from the choice of $K$ which is mostly irrelevant. Performances of PAMPL are as good as or better than existing algorithms both based on PSL and other approaches, and the method is especially suitable to study inference problems with underlying sparse graphs. 

This project has received funding from the European Research Council (ERC) under the European Union’s Horizon 2020 research and innovation program (grant agreement No [694925]). S. Franz and J. Rocchi acknowledge the support of a grant from the Simons Foundation (No. 454941, Silvio Franz). 

\bibliography{apssamp}% Produces the bibliography via BibTeX.
\bibliographystyle{unsrt}

\onecolumngrid

\clearpage
\newpage

\twocolumngrid

\section{Supplemental Material}

\subsection{Appendix A: Pseudo-likelihood}

We show that in the infinite sampling limit $\mathcal{S}$ and $\mathcal{L}$ are maximized by the same $J$. From the definition of $\mathcal{S}$ in eq. (\ref{eq:Sdef}), we obtain the property
\begin{dmath}
  \frac{\partial \mathcal{S}}{\partial J_{ij}} = 2 \beta \langle s_i s_j \rangle_{\text{Data}}  - \beta \left \langle s_i \tanh \beta \sum_{m\neq j} J_{jm} s_m \right \rangle_{\text{Data}} - \beta \left \langle s_j \tanh \beta \sum_{m\neq i} J_{im} s_m \right \rangle_{\text{Data}}
  \label{eq:diff1uno}
\end{dmath}
and thus $\mathcal{S}$ is maximum on the parameters $J$ such that
\be
 \langle s_i s_j \rangle_{\text{Data}} = \left \langle s_i \tanh \beta \sum_{m\neq j} J_{jm} s_m \right \rangle_{\text{Data}} \:.
\ee
On the other hand, using the identity
\be
\left< s_i s_j \right>_{\text{Model}} = \left< s_i \tanh \beta \sum_{m \neq j} J_{jm} s_m\right>_{\text{Model}}\:
\ee
in eq. (\ref{eq:Lcond}), we notice that $\mathcal{L}$ is maximum when
\be
\left< s_i s_j \right>_{\text{Data}} = \left< s_i \tanh \beta \sum_{m \neq j} J_{jm} s_m\right>_{\text{Model}}\:,
\ee
and thus, since in the infinite sampling limit $\lim_{M\rightarrow \infty} \left< f (\us,J)\right>_{\text{Data}} = \left< f (\us,J)\right>_{\text{Model}}\: $, we observe that $\mathcal{S}$ and $\mathcal{L}$ are maximized by the same $J$. 

\subsection{Appendix B: Decimation algorithm}

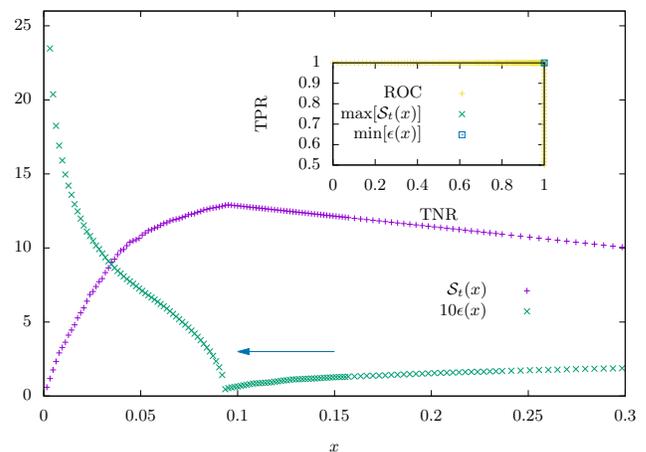
\begin{figure}[h]
\centerline{\resizebox{0.5\textwidth}{!}{\input{Decimation-36-2D-FB.tex}}}
\caption{2-D ferromagnetic lattice with $N=36$ and free boundary conditions, $M=5000$, $\beta=0.5$. Main Figure: $\mathcal{S}_t$ and $10 \epsilon$ as a function of $x$ for the decimation based method. $x$ is $1$ at the beginning and decreases as the iteration proceeds. Only the last part of the iteration is presented. The Arrow specifies the directions in which the iteration moves. Inset: ROC curve, with the TNR on the x-axis and the TPR on the y-axis.}
\label{fig:figD}
\end{figure}

The idea of the decimation algorithm \cite{decelle2014pseudolikelihood} is that starting from the complete graph, the full $\mathcal{S}$ is maximized (maximization step) and the $K$ couplings with the smallest values are set to zero (decimation step). The two steps are iterated until when no more couplings are present in the graph. In order to locate the stopping point, a new function is defined. Be $\mathcal{S}^c_{max}$ the maximum of the pseudo-likelihood on the complete graph. When no couplings remain, the pseudo-likelihood is $-N \log 2$. The new function is given by $\mathcal{S}_t(x)=\mathcal{S}-[x \mathcal{S}^c_{max} - (1-x) N \log 2]$, where $x \in [0,1]$ is the fraction of couplings, being $1$ on the complete graph and $0$ at the end of the decimation process. This function is $0$ at the beginning and at the end of the process, by construction, and it is positive in the intermediate steps. The stopping point $x^*$ is chosen by looking at the maximum of $\mathcal{S}_t$. 
The solutions found with this method are much better than those found with other methods based on PSL. In Fig. \ref{fig:figD} we show the behavior of this algorithm in the study of a 2-D ferromagnetic lattice with free boundary conditions with $N=36$. In the inset we show the TPR and the TNR evolving with the iterations. The point where $\mathcal{S}_t$ is maximum coincides with the point where $\epsilon$ is minimum. In Fig. \ref{fig:figA} we analyze the same dataset with PAMPL and find a solution much faster because we starts from the empty graph, rather than from the fully connected graph. 
Each step of the decimation algorithm is $O(M N^2)$, since it needs to optimize the PSL over the number of couplings that haven't been  decimated yet. If the true graph is sparse we need to run the iterations for $O( N^2)$ times. In comparison, ours takes $O(M N^2)$ operations to provide a solution in sparse graphs, as explained in the following section and it is thus much faster.

\subsection{Appendix C: Details of the implementation}

\paragraph{Bayesian Information Criterion:}
In PAMPL we cannot use the tilted pseudo-likelihood to locate the stopping point because this would require the maximization of the pseudo-likelihood on the complete graph. 
As stated in the main text, the quantity that we observe during learning is thus the Bayesian Information Criterion, defined in eq. (\ref{eq:BIC}). Let us consider $P(\{\us^{(\mu)}\}_{\mu=1,\ldots,M}) = \int dJ P(\{\us^{(\mu)}\}_{\mu=1,\ldots,M}|J) P(J)$. Under the assumption of a flat prior $P(J)$, we expand $\mathcal{L}(J)=M^{-1} \log P(\{\us^{(\mu)}\}_{\mu=1,\ldots,M}|J)$ to the second order around the parameters $J^*$ for which the likelihood $P(\{\us^{(\mu)}\}_{\mu=1,\ldots,M}|J)$ is maximum. A Gaussian integration leads to
\be
P(\{\us^{(\mu)}\}_{\mu=1,\ldots,M}) = e^{M \mathcal{L}(J^*)} \left( \frac{2\pi}{I(J) M} \right)^{\frac{k}{2}}
\ee 
where $I(J) =  \mathcal{L}''(J^*) \sim O(1)$. Thus, we see that $P(\{\us^{(\mu)}\}_{\mu=1,\ldots,M}) \sim e^{M \mathcal{L}^* - k/2 \log M }$. In our analysis, we may replace $\mathcal{L}^*$ with $\mathcal{S}^*$ and obtain $P(\{\us^{(\mu)}\}_{\mu=1,\ldots,M})=e^{\text{BIC}/2}$, using eq. (\ref{eq:BIC}): the largest the $\text{BIC}$, the better the parameters $J$ describe the observations.

\paragraph{Complexity:}
Updating and sorting the elements $\Delta \mathcal{S}_{ij}$, defined in eq. (\ref{eq:Sdef}), requires a careful discussion. In fact, if we want to keep the iterative step $O(M N)$, we cannot afford $O(N^2)$ operations. The way we overcome this problem is explained below. At the initial time step we evaluate and sort all the gains in pseudo-likelihood $\Delta \mathcal{S}_{ij}$.
After the sorting, we create the vector $\underline{V}$ with the largest $O(N)$ elements. This vector is updated at each step and its size is kept to be $O(N)$. This makes the sorting less expansive. The cost of updating $O(N^2)$ element is alleviated by considering only $\Delta \mathcal{S}_{ij}$ whose nodes are involved in the activation of the couplings of the previous time step. More precisely, if coupling $J_{ij}$ has been updated at time $t$, at $t+1$ we update $\Delta \mathcal{S}_{ik}$ and $\Delta \mathcal{S}_{jk}$ with $k= 1,\ldots,N$ and neglect the changes in the others. These operations cost $O(N)$. If some of these values happen to be larger than the mean value of the elements of $\underline{V}$, they are included in $\underline{V}$. Finally, all the elements $v_i < 0.01 \max\{\underline{v_i}\}$ of $\underline{V}$ are excluded from it. These wise precepts allows the size of $\underline{V}$ to remain $O(N)$ and, thus, the ensemble of iterative steps to be $O(M N^2)$. The optimization of the couplings in the set $\mathbb{J}$ is performed with a gradient ascent with a learning rate of $0.0001$ until $\Delta \text{PSL} / \text{PSL} < 0.00001$, where $\Delta \text{PSL}$ is the difference between the $\text{PSL}$ computed before and after the updating. 

\paragraph{Derivatives:}
The expression of the first and second derivatives of $\mathcal{S}$ with respect to $J_{ij}$ in terms of the average over samples read

\begin{dmath}
  \frac{\partial \mathcal{S}}{\partial J_{ij}} = \frac{2 \beta}{M}  \sum_{\mu=1}^Ms^{(\mu)}_i s^{(\mu)}_j \left[ \frac{1}{1+e^{2\beta s^{(\mu)}_i \sum_m J_{jm} s^{(\mu)}_m }} + \frac{1}{1+e^{2\beta s^{(\mu)}_j \sum_m J_{im} s^{(\mu)}_m }} \right]\:,
  \label{eq:diff1}
\end{dmath}
\begin{dmath}
  \frac{\partial^2 \mathcal{S}}{\partial^2 J_{ij}} =  - \frac{4 \beta^2}{M}  \sum_{\mu=1}^M \left[ \frac{1}{2+2 \cosh \left [2\beta s^{(\mu)}_i \sum_m J_{jm} s^{(\mu)}_m \right]} + \frac{1}{2+2 \cosh \left [2\beta s^{(\mu)}_j \sum_m J_{im} s^{(\mu)}_m \right]}    \right]\:.
    \label{eq:diff2}
\end{dmath}
It is easy to check that eq. (\ref{eq:diff1}) coincide with eq. (\ref{eq:diff1uno}). 

\begin{figure}
    \centering
    \subfigure[]
    {
        \centerline{\resizebox{0.5\textwidth}{!}{\input{Av-49-2D-FB_05.tex}}}
    }
    \\
    \subfigure[]
    {
        \centerline{\resizebox{0.5\textwidth}{!}{\input{Av-100-RR-05.tex}}}
    }
    \caption{
Average and standard deviation of the BIC over $10$ runs of the activation-based algorithm. Blue and red arrows indicate the stopping point and the point at which the BIC is maximum.
Main Figures: (a) 2-D spin glass lattice with $N=49$ spins and $84$ couplings with free boundary conditions. Each run of the algorithm is made on a subset of $M=3000$ samples extracted from a dataset of $20000$ samples at equilibrium at $\beta=0.6$. (b) Random regular spin glass with $N=100$ spins, $c=4$, corresponding to $200$ couplings. Each run of the algorithm is made on a subset of $M=8000$ samples extracted from a dataset of $20000$ samples at equilibrium at $\beta=0.8$.
Insets: Average and standard deviation of the error over the same $L=10$ runs of the activation-based algorithm.
}
\label{fig:figAv}
\end{figure}
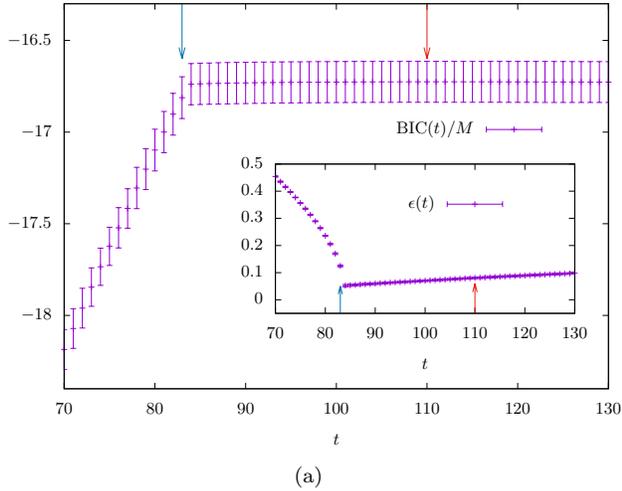
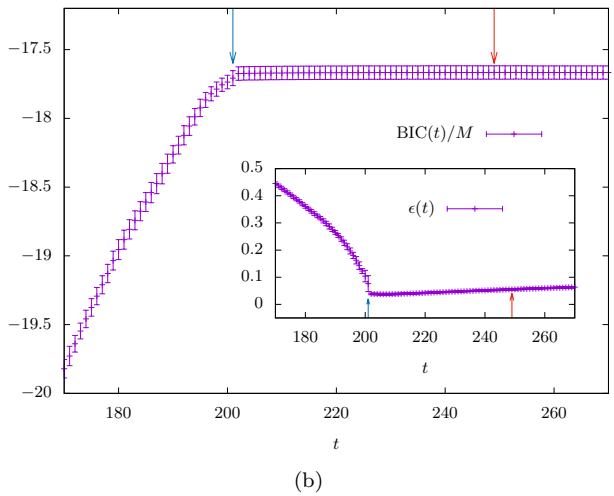

\paragraph{Stopping point:}
The increase of the BIC after the correct stopping point is due to the fluctuations induced by the finite size sampling. This can be understood considering different datasets, each one made by $M$ configurations. We use each dataset to extract the inferred graphs, and we observe the behavior of the BIC and the error. Then we compute the mean and the standard deviation of the two quantities and we observe that the minimum of the error is reached when the $\text{BIC}$ reaches for the first time the value $\text{BIC}_{max}-\sigma_{\text{BIC}}$, being $\sigma_{\text{BIC}}$ the standard deviation. In Fig. {\ref{fig:figAv}}a we show the results for a 2-D lattice with $N=49$ and free boundary conditions. In Fig. {\ref{fig:figAv}}b we show the results for a random regular graph with $N=100$ and mean connectivity equal to $4$. In both cases we observe that the increase of the $\text{BIC}$ after the correct stopping point (corresponding to the minimum of the error) is irrelevant.
In order to locate the stopping point we thus adopt the criterion explained in the main text.

\paragraph{ROC curves:}
We consider a 2-D ferromagnetic lattice with periodic boundary condition in Fig. \ref{fig:figROC}. In these figures we plot the TPR and the TNR for the inferred graph as a function of the (inverse of the) number of observed samples. We observe a weak dependence on the size of the system and a more severe one on the temperature. This is in line with the performances of specific algorithms for which it is possible to compute the scaling of the mininum number of samples for a perfect inference, where the dependence on $N$ is logarithmic and that in $\beta$ is exponential \cite{lokhov2018optimal}. We notice that a TNR smaller than $1$ means that the reconstructed graph contains couplings that are not present in the original graph, i.e. that our criterion does not detect correctly the stopping point and couplings keep being activated for some other step. 

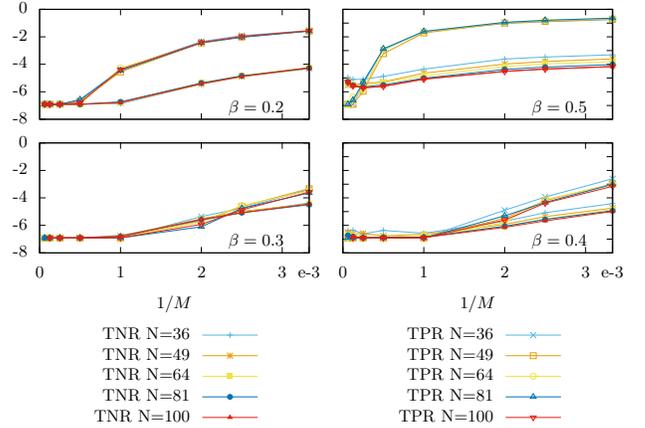
\begin{figure}[h]
\centerline{\resizebox{0.5\textwidth}{!}{\input{ROCvsM-2D-PB.tex}}}
\caption{$\log(1-\text{TNR}+0.001)$ and $\log(1-\text{TPR}+0.001)$ at different $N$ for a 2D lattice with periodic boundary conditions at different $\beta$ as a function of $1/M$. We notice that as $M$ increases, both the TPR and TNR approach $1$. Each point corresponds to the average of $\sim 50$ runs of the activation algorithm, stopped using the criterion defined in the text.}
\label{fig:figROC}
\end{figure}

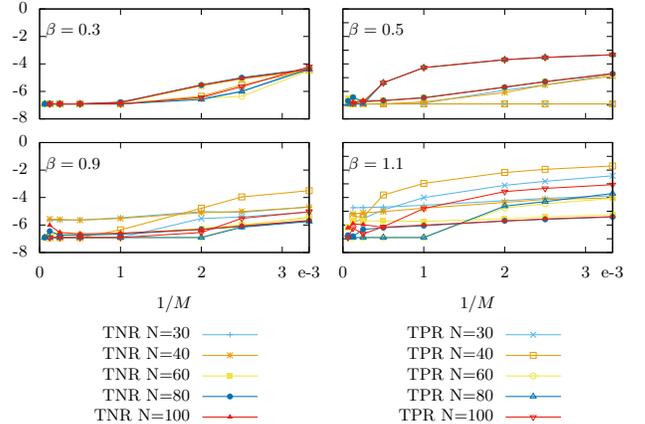
\begin{figure}[h]
\centerline{\resizebox{0.5\textwidth}{!}{\input{ROCvsM-RR.tex}}}
\caption{$\log(1-\text{TNR}+0.001)$ and $\log(1-\text{TPR}+0.001)$ at different $N$ for a RR graph with $c=3$ at different $\beta$ as a function of $1/M$. We notice that as $M$ increases, both the TPR and TNR approach $1$. Each point corresponds to the average of $\sim 50$ runs of the activation algorithm, stopped using the criterion defined in the text.}
\label{fig:figROC}
\end{figure}

\begin{figure}
    \centering
    \subfigure[]
    {
        \centerline{\resizebox{0.5\textwidth}{!}{\input{hist_RR_03.tex}}}
    }
    \\
    \subfigure[]
    {
        \centerline{\resizebox{0.5\textwidth}{!}{\input{hist_RR_11.tex}}}
    }
    \caption{
$\epsilon$ and $\delta K$ as a function of the iteration time $t$ for the MPF learning on a RR graph with $c=3$, $N=40$ and $M=4000$. (a) $\beta=0.3$ (b) $\beta=1.1$. The total execution time is roughly $30$ seconds on a normal laptop, learning is stopped after $700$ steps. Insets: histograms of the learned couplings $J$ at different stages of learning. After the last time, $\delta K$ is equal to $0.0011$ (a) and $0.0057$ (b). In both cases the majority of the true couplings have been found: setting a threshold at $|J|=0.5$, TPR and TNR are respectively 1-1 (a) and 0.95-1 (b).  
}
\label{fig:hists-RR}
\end{figure}

\begin{figure}
    \centering
    \subfigure[]
    {
        \centerline{\resizebox{0.5\textwidth}{!}{\input{hist_2D_PB_02.tex}}}
    }
    \\
    \subfigure[]
    {
        \centerline{\resizebox{0.5\textwidth}{!}{\input{hist_2D_PB_05.tex}}}
    }
    \caption{
$\epsilon$ and $\delta K$ as a function of the iteration time $t$ for the MPF learning on a 2D lattice with periodic boundary condition, $N=49$ and $M=4000$. (a) $\beta=0.2$ (b) $\beta=0.5$. The total execution time is roughly $55$ seconds on a normal laptop, learning is stopped after $700$ steps. Insets: histograms of the learned couplings $J$ at different stages of learning. After the last time, $\delta K $ is equal to $0.0065$ (a) and $0.0025$ (b). In both cases the majority of the true couplings have been found: setting a threshold at $|J|=0.5$, TPR and TNR are respectively 1-1 (a) and 0.983-1 (b).
}
\label{fig:hists-2D-PB}
\end{figure}

\subsection{Appendix D: Minimum probability flow}

\paragraph{Algorithm:} 
The Minimum Probability Flow (MPF) learning algorithm \cite{sohl2011new} is based on an hypothetical dynamics in the parameter space $\{J\}$, that we use to parametrize the probability distribution $P(\underline{s}|J) = \exp[-E_s(J)] / Z(J) = \exp \left( \sum_{i<j} J_{ij} s_i s_j  \right) / Z(J)$. This dynamics starts from the data distribution and ends up in the point $J$ that minimize the Kullback-Leibler $D_{KL}$ divergence between the distribution of data and $P(\underline{s}|J)$. Using detailed balance, it is possible to define a transition matrix that allows the dynamics to relax to the chosen probability distribution,
\begin{equation}
\Gamma_{s,s'}=g_{s,s'}\exp\left[\frac{1}{2}(E_{s'}-E_s)\right]
\end{equation} 
with $g_{s,s'}$ being a sparse matrix with $1$ between configurations differing by one-spin flip, and $0$ elsewhere. The dynamics considered is thus
\begin{equation}
\partial_t p^{(t)}_s = \sum_{\underline{s}'\neq \underline{s}} \Gamma_{s,s'} p^{(t)}_{s'} - \sum_{\underline{s}'\neq \underline{s}} \Gamma_{s',s} p^{(t)}_s\:,
\end{equation}
where $\Gamma_{s,s'}$ is the transition rate from configuration $\underline{s}'$ to $\underline{s}$. 
This dynamics may take several time steps to converge to the desired distribution and it is not practical. Rather than waiting such a long time, MPF considers a small time $t=\epsilon$. In fact, among the trajectories that leads to $J^*$, a special role is played by the one that points in the direction of $J^*$ already in the early steps. In this limit, it is possible to show that 
\begin{equation}
D_{KL}(p^{(0)},p^{(t)})= \frac{\epsilon}{M}\sum_{\underline{s}'\in \mathcal{D}}\sum_{\underline{s} \notin \mathcal{D}} \Gamma_{s,s'} \equiv K(J)\:,
\end{equation}
where $\mathcal{D}$ denotes the dataset. Parameter estimation is provided by $J^* = \arg \min_J K(J)$. If the system is big enough, and the configurations of the dataset sampled independently, it is likely that configuration space is not sampled extensively and thus, for each configuration $s'\in \mathcal{D}$ of the dataset, all those that differ from it for a spin flip are not part of $\mathcal{D}$. The second sum in the definition of $ K(J)$ is thus replaced by $\sum_{\underline{s} : g_{s,s'}= 1}$. This makes each step of the minimization process $O(M N)$. On the other hand, given that the algorithm optimizes over all the parameters $J$, we observe that the actual cost of each learning step is $O(M N^2)$. For the case under consideration, $E_s = - \sum_{i<j} J_{ij} s_i s_j$ and 
\begin{equation}
K(J) = \frac{\epsilon}{M}\sum_{\mu = 1}^M \sum_{t=1}^N e^{-\beta \sum_{q\neq t} J_{t q} s^{(\mu)}_t s^{(\mu)}_q}\:.
\end{equation}
$K(J)$ can be optimized either with a simple gradient descent and with a more sophisticated LBFSG \cite{liu1989limited} algorithm with similar results. Performances depend on the learning rate. In particular, when $\beta$ is large, generally a smaller learning rate needs to be used. Moreover, a small mini-batch allows to find solutions in a smaller amount of time. Mini-batch size should not be smaller than a few dozens in any case. A comparison between execution times in Figs  \ref{fig:time-RR}-\ref{fig:time-2D} shows that MPF is as fast as PAMPL. In the paper we present results with $M=4000$, $100$ mini-batch and $\epsilon=0.025$, that plays the role of a learning rate:
\begin{dmath}
\frac{\partial K}{\partial J_{ij}} = -\frac{\epsilon \beta}{M}  \sum_{\mu=1}^M s^{(\mu)}_i s^{(\mu)}_j 
\times \\ \times \left[\frac{}{} e^{-\beta \sum_{q\neq i} J_{i q} s^{(\mu)}_i s^{(\mu)}_q }+ e^{-\beta \sum_{q\neq j} J_{j q} s^{(\mu)}_j s^{(\mu)}_q}\right]\:.
\end{dmath}
Moreover, we observe that MPF provides an alternative method to infer couplings that is as fast as a single maximization of pseudo-likelihood on the complete graph. This is more clear by a comparison with eq. (\ref{eq:diff1}): the updating rules used to maximize the MPF and the pseudo-likelihood $ \mathcal{S}$ are very similar, with the first one consisting in neglecting the two denominators $2\cosh(\beta h_i)$ and $2\cosh(\beta h_j)$ appearing in the second one, where they act as normalizations factors. As other methods based on the pseudo-likelihood, MPF needs to be complemented with a threshold procedure.

\begin{figure}[h]
\centerline{\resizebox{0.5\textwidth}{!}{\input{time-2D-PB.tex}}}
\caption{Comparison between the $\log$ of the execution time (seconds) needed to find a solution as function of $N$ and $\beta$ for a 2D lattice with periodic boundary conditions using (a) PAMPL and (b) MPF. Each point corresponds to an average over $10$ runs of the algorithm with $M=4000$. }
\label{fig:time-2D}
\end{figure}
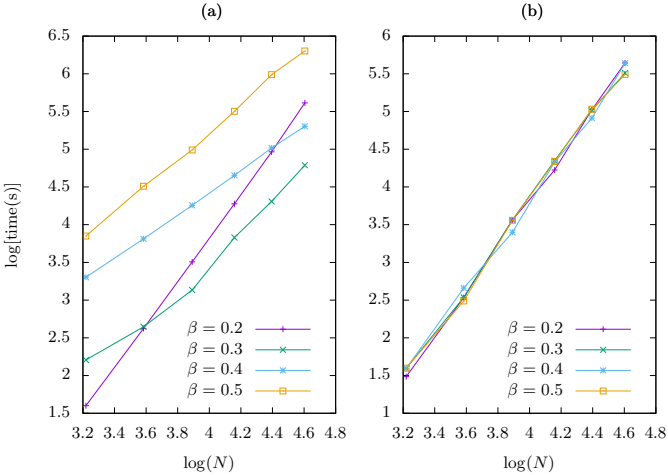
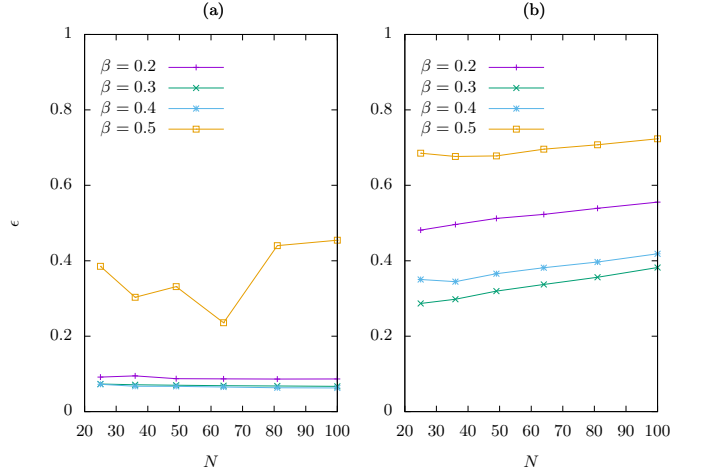
\begin{figure}[h]
\centerline{\resizebox{0.5\textwidth}{!}{\input{eps-2D-PB.tex}}}
\caption{Error $\epsilon$, defined in eq. (\ref{eq:eddefepserr}), as a function of $N$ and $\beta$ for a 2D ferromagnetic lattice with periodic boundaries using (a) PAMPL and (b) MPF. Each point corresponds to an average over $10$ runs of the algorithm at $M=4000$.}
\label{fig:eps-2D}
\end{figure}

%\vfill\null

\subsection{Stopping point}
Finding a stopping point for this algorithm is not easy. In particular, even if the error with respect the original graph decreases quickly, the values of the couplings are still far from the actual ones and are refined only in later time steps. In particular, for real application cases where the actual topology is unknown, one should rely on other measures of convergence, like for instance $\delta K = \Delta K / K$, where $\Delta K$ is the difference on $K$ computed every epoch running over the whole dataset. This quantity decreases during learning but our experiments do not provide a meaningful value where to stop the iteration. Unsurprisingly, parameters like mini-batch size, learning rate, stopping point are model, size and temperature dependent. In particular at large temperatures fewer iterations are needed to find satisfying results, as observed in the case of Fig. \ref{fig:hists-RR}-\ref{fig:hists-2D-PB}. As discussed above, MPF has to be complemented with a threshold procedure. In the main text we consider a RR spin glass with $c=3$ while here we consider a 2D lattice with ferromagnetic interactions and periodic boundary conditions. In both cases, after $\sim 700$ iterations, setting a threshold at $|J|=0.5$, both the TPR and the TNR are very close to 1. On the other hand, corresponding errors are still large as seen in the main text in Fig. \ref{fig:eps-RR} for the RR case and here in Fig. \ref{fig:eps-2D} for the 2D lattice. This is not surprising. In fact, although being very versatile and fast, it performs a single optimization.

\end{document}

%% file: Activation-36-2D-FB.tex
% GNUPLOT: LaTeX picture with Postscript
\begingroup
  \makeatletter
  \providecommand\color[2][]{%
    \GenericError{(gnuplot) \space\space\space\@spaces}{%
      Package color not loaded in conjunction with
      terminal option `colourtext'%
    }{See the gnuplot documentation for explanation.%
    }{Either use 'blacktext' in gnuplot or load the package
      color.sty in LaTeX.}%
    \renewcommand\color[2][]{}%
  }%
  \providecommand\includegraphics[2][]{%
    \GenericError{(gnuplot) \space\space\space\@spaces}{%
      Package graphicx or graphics not loaded%
    }{See the gnuplot documentation for explanation.%
    }{The gnuplot epslatex terminal needs graphicx.sty or graphics.sty.}%
    \renewcommand\includegraphics[2][]{}%
  }%
  \providecommand\rotatebox[2]{#2}%
  \@ifundefined{ifGPcolor}{%
    \newif\ifGPcolor
    \GPcolortrue
  }{}%
  \@ifundefined{ifGPblacktext}{%
    \newif\ifGPblacktext
    \GPblacktexttrue
  }{}%
  % define a \g@addto@macro without @ in the name:
  \let\gplgaddtomacro\g@addto@macro
  % define empty templates for all commands taking text:
  \gdef\gplbacktext{}%
  \gdef\gplfronttext{}%
  \makeatother
  \ifGPblacktext
    % no textcolor at all
    \def\colorrgb#1{}%
    \def\colorgray#1{}%
  \else
    % gray or color?
    \ifGPcolor
      \def\colorrgb#1{\color[rgb]{#1}}%
      \def\colorgray#1{\color[gray]{#1}}%
      \expandafter\def\csname LTw\endcsname{\color{white}}%
      \expandafter\def\csname LTb\endcsname{\color{black}}%
      \expandafter\def\csname LTa\endcsname{\color{black}}%
      \expandafter\def\csname LT0\endcsname{\color[rgb]{1,0,0}}%
      \expandafter\def\csname LT1\endcsname{\color[rgb]{0,1,0}}%
      \expandafter\def\csname LT2\endcsname{\color[rgb]{0,0,1}}%
      \expandafter\def\csname LT3\endcsname{\color[rgb]{1,0,1}}%
      \expandafter\def\csname LT4\endcsname{\color[rgb]{0,1,1}}%
      \expandafter\def\csname LT5\endcsname{\color[rgb]{1,1,0}}%
      \expandafter\def\csname LT6\endcsname{\color[rgb]{0,0,0}}%
      \expandafter\def\csname LT7\endcsname{\color[rgb]{1,0.3,0}}%
      \expandafter\def\csname LT8\endcsname{\color[rgb]{0.5,0.5,0.5}}%
    \else
      % gray
      \def\colorrgb#1{\color{black}}%
      \def\colorgray#1{\color[gray]{#1}}%
      \expandafter\def\csname LTw\endcsname{\color{white}}%
      \expandafter\def\csname LTb\endcsname{\color{black}}%
      \expandafter\def\csname LTa\endcsname{\color{black}}%
      \expandafter\def\csname LT0\endcsname{\color{black}}%
      \expandafter\def\csname LT1\endcsname{\color{black}}%
      \expandafter\def\csname LT2\endcsname{\color{black}}%
      \expandafter\def\csname LT3\endcsname{\color{black}}%
      \expandafter\def\csname LT4\endcsname{\color{black}}%
      \expandafter\def\csname LT5\endcsname{\color{black}}%
      \expandafter\def\csname LT6\endcsname{\color{black}}%
      \expandafter\def\csname LT7\endcsname{\color{black}}%
      \expandafter\def\csname LT8\endcsname{\color{black}}%
    \fi
  \fi
    \setlength{\unitlength}{0.0500bp}%
    \ifx\gptboxheight\undefined%
      \newlength{\gptboxheight}%
      \newlength{\gptboxwidth}%
      \newsavebox{\gptboxtext}%
    \fi%
    \setlength{\fboxrule}{0.5pt}%
    \setlength{\fboxsep}{1pt}%
\begin{picture}(7200.00,5040.00)%
    \gplgaddtomacro\gplbacktext{%
      \csname LTb\endcsname%%
      \put(462,704){\makebox(0,0)[r]{\strut{}$0$}}%
      \put(462,1527){\makebox(0,0)[r]{\strut{}$5$}}%
      \put(462,2350){\makebox(0,0)[r]{\strut{}$10$}}%
      \put(462,3173){\makebox(0,0)[r]{\strut{}$15$}}%
      \put(462,3996){\makebox(0,0)[r]{\strut{}$20$}}%
      \put(462,4819){\makebox(0,0)[r]{\strut{}$25$}}%
      \put(594,484){\makebox(0,0){\strut{}$0$}}%
      \put(1629,484){\makebox(0,0){\strut{}$0.05$}}%
      \put(2664,484){\makebox(0,0){\strut{}$0.1$}}%
      \put(3699,484){\makebox(0,0){\strut{}$0.15$}}%
      \put(4733,484){\makebox(0,0){\strut{}$0.2$}}%
      \put(5768,484){\makebox(0,0){\strut{}$0.25$}}%
      \put(6803,484){\makebox(0,0){\strut{}$0.3$}}%
    }%
    \gplgaddtomacro\gplfronttext{%
      \csname LTb\endcsname%%
      \put(3698,154){\makebox(0,0){\strut{}$x$}}%
      \csname LTb\endcsname%%
      \put(5327,1829){\makebox(0,0)[r]{\strut{}$- \text{BIC}(x) / M$}}%
      \csname LTb\endcsname%%
      \put(5327,1609){\makebox(0,0)[r]{\strut{}$10 \epsilon(x)$}}%
    }%
    \gplgaddtomacro\gplbacktext{%
      \csname LTb\endcsname%%
      \put(3550,3173){\makebox(0,0)[r]{\strut{}$0$}}%
      \put(3550,3391){\makebox(0,0)[r]{\strut{}$0.2$}}%
      \put(3550,3610){\makebox(0,0)[r]{\strut{}$0.4$}}%
      \put(3550,3828){\makebox(0,0)[r]{\strut{}$0.6$}}%
      \put(3550,4047){\makebox(0,0)[r]{\strut{}$0.8$}}%
      \put(3550,4265){\makebox(0,0)[r]{\strut{}$1$}}%
      \put(3682,2953){\makebox(0,0){\strut{}$0.9$}}%
      \put(4133,2953){\makebox(0,0){\strut{}$0.92$}}%
      \put(4585,2953){\makebox(0,0){\strut{}$0.94$}}%
      \put(5036,2953){\makebox(0,0){\strut{}$0.96$}}%
      \put(5488,2953){\makebox(0,0){\strut{}$0.98$}}%
      \put(5939,2953){\makebox(0,0){\strut{}$1$}}%
    }%
    \gplgaddtomacro\gplfronttext{%
      \csname LTb\endcsname%%
      \put(2934,3719){\rotatebox{-270}{\makebox(0,0){\strut{}$\text{TPR}$}}}%
      \put(4810,2623){\makebox(0,0){\strut{}$\text{TNR}$}}%
      \csname LTb\endcsname%%
      \put(4858,3937){\makebox(0,0)[r]{\strut{}$\text{ROC}$}}%
      \csname LTb\endcsname%%
      \put(4858,3717){\makebox(0,0)[r]{\strut{}$\max[\text{BIC}(x)]$}}%
      \csname LTb\endcsname%%
      \put(4858,3497){\makebox(0,0)[r]{\strut{}$\min[\epsilon(x)]$}}%
    }%
    \gplbacktext
    \put(0,0){\includegraphics{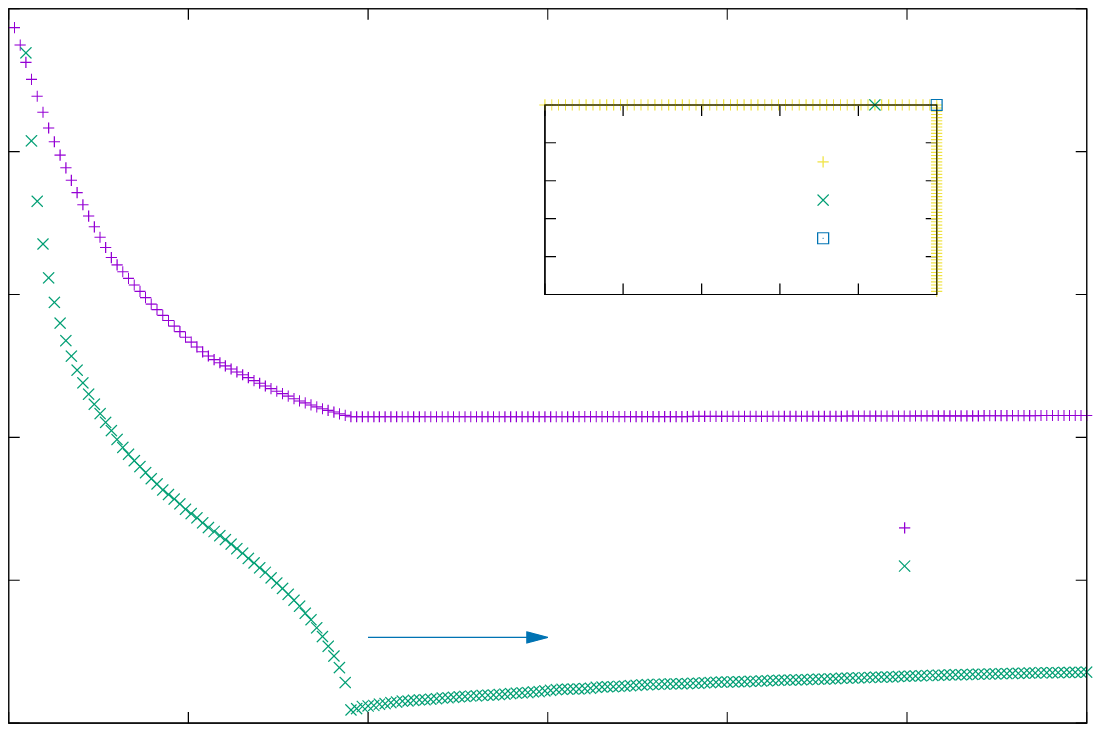}}%
    \gplfronttext
  \end{picture}%
\endgroup

%% file: differentTM-av-50-RR-05.tex
% GNUPLOT: LaTeX picture with Postscript
\begingroup
  \makeatletter
  \providecommand\color[2][]{%
    \GenericError{(gnuplot) \space\space\space\@spaces}{%
      Package color not loaded in conjunction with
      terminal option `colourtext'%
    }{See the gnuplot documentation for explanation.%
    }{Either use 'blacktext' in gnuplot or load the package
      color.sty in LaTeX.}%
    \renewcommand\color[2][]{}%
  }%
  \providecommand\includegraphics[2][]{%
    \GenericError{(gnuplot) \space\space\space\@spaces}{%
      Package graphicx or graphics not loaded%
    }{See the gnuplot documentation for explanation.%
    }{The gnuplot epslatex terminal needs graphicx.sty or graphics.sty.}%
    \renewcommand\includegraphics[2][]{}%
  }%
  \providecommand\rotatebox[2]{#2}%
  \@ifundefined{ifGPcolor}{%
    \newif\ifGPcolor
    \GPcolortrue
  }{}%
  \@ifundefined{ifGPblacktext}{%
    \newif\ifGPblacktext
    \GPblacktexttrue
  }{}%
  % define a \g@addto@macro without @ in the name:
  \let\gplgaddtomacro\g@addto@macro
  % define empty templates for all commands taking text:
  \gdef\gplbacktext{}%
  \gdef\gplfronttext{}%
  \makeatother
  \ifGPblacktext
    % no textcolor at all
    \def\colorrgb#1{}%
    \def\colorgray#1{}%
  \else
    % gray or color?
    \ifGPcolor
      \def\colorrgb#1{\color[rgb]{#1}}%
      \def\colorgray#1{\color[gray]{#1}}%
      \expandafter\def\csname LTw\endcsname{\color{white}}%
      \expandafter\def\csname LTb\endcsname{\color{black}}%
      \expandafter\def\csname LTa\endcsname{\color{black}}%
      \expandafter\def\csname LT0\endcsname{\color[rgb]{1,0,0}}%
      \expandafter\def\csname LT1\endcsname{\color[rgb]{0,1,0}}%
      \expandafter\def\csname LT2\endcsname{\color[rgb]{0,0,1}}%
      \expandafter\def\csname LT3\endcsname{\color[rgb]{1,0,1}}%
      \expandafter\def\csname LT4\endcsname{\color[rgb]{0,1,1}}%
      \expandafter\def\csname LT5\endcsname{\color[rgb]{1,1,0}}%
      \expandafter\def\csname LT6\endcsname{\color[rgb]{0,0,0}}%
      \expandafter\def\csname LT7\endcsname{\color[rgb]{1,0.3,0}}%
      \expandafter\def\csname LT8\endcsname{\color[rgb]{0.5,0.5,0.5}}%
    \else
      % gray
      \def\colorrgb#1{\color{black}}%
      \def\colorgray#1{\color[gray]{#1}}%
      \expandafter\def\csname LTw\endcsname{\color{white}}%
      \expandafter\def\csname LTb\endcsname{\color{black}}%
      \expandafter\def\csname LTa\endcsname{\color{black}}%
      \expandafter\def\csname LT0\endcsname{\color{black}}%
      \expandafter\def\csname LT1\endcsname{\color{black}}%
      \expandafter\def\csname LT2\endcsname{\color{black}}%
      \expandafter\def\csname LT3\endcsname{\color{black}}%
      \expandafter\def\csname LT4\endcsname{\color{black}}%
      \expandafter\def\csname LT5\endcsname{\color{black}}%
      \expandafter\def\csname LT6\endcsname{\color{black}}%
      \expandafter\def\csname LT7\endcsname{\color{black}}%
      \expandafter\def\csname LT8\endcsname{\color{black}}%
    \fi
  \fi
    \setlength{\unitlength}{0.0500bp}%
    \ifx\gptboxheight\undefined%
      \newlength{\gptboxheight}%
      \newlength{\gptboxwidth}%
      \newsavebox{\gptboxtext}%
    \fi%
    \setlength{\fboxrule}{0.5pt}%
    \setlength{\fboxsep}{1pt}%
\begin{picture}(7200.00,5040.00)%
    \gplgaddtomacro\gplbacktext{%
      \csname LTb\endcsname%%
      \put(1078,704){\makebox(0,0)[r]{\strut{}$-12.5$}}%
      \put(1078,1292){\makebox(0,0)[r]{\strut{}$-12$}}%
      \put(1078,1880){\makebox(0,0)[r]{\strut{}$-11.5$}}%
      \put(1078,2468){\makebox(0,0)[r]{\strut{}$-11$}}%
      \put(1078,3055){\makebox(0,0)[r]{\strut{}$-10.5$}}%
      \put(1078,3643){\makebox(0,0)[r]{\strut{}$-10$}}%
      \put(1078,4231){\makebox(0,0)[r]{\strut{}$-9.5$}}%
      \put(1078,4819){\makebox(0,0)[r]{\strut{}$-9$}}%
      \put(1210,484){\makebox(0,0){\strut{}$70$}}%
      \put(2142,484){\makebox(0,0){\strut{}$80$}}%
      \put(3074,484){\makebox(0,0){\strut{}$90$}}%
      \put(4007,484){\makebox(0,0){\strut{}$100$}}%
      \put(4939,484){\makebox(0,0){\strut{}$110$}}%
      \put(5871,484){\makebox(0,0){\strut{}$120$}}%
      \put(6803,484){\makebox(0,0){\strut{}$130$}}%
    }%
    \gplgaddtomacro\gplfronttext{%
      \csname LTb\endcsname%%
      \put(198,2761){\rotatebox{-270}{\makebox(0,0){\strut{}$\text{BIC}(t)/M$}}}%
      \put(4006,154){\makebox(0,0){\strut{}$t$}}%
      \csname LTb\endcsname%%
      \put(2145,4298){\makebox(0,0)[r]{\strut{}$M=300$}}%
      \csname LTb\endcsname%%
      \put(2145,4078){\makebox(0,0)[r]{\strut{}$M=500$}}%
      \csname LTb\endcsname%%
      \put(2145,3858){\makebox(0,0)[r]{\strut{}$M=1000$}}%
      \csname LTb\endcsname%%
      \put(2145,3638){\makebox(0,0)[r]{\strut{}$M=2000$}}%
      \csname LTb\endcsname%%
      \put(2145,3418){\makebox(0,0)[r]{\strut{}$M=4000$}}%
      \csname LTb\endcsname%%
      \put(2145,3198){\makebox(0,0)[r]{\strut{}$M=8000$}}%
    }%
    \gplgaddtomacro\gplbacktext{%
      \csname LTb\endcsname%%
      \put(3610,1044){\makebox(0,0)[r]{\strut{}$-3$}}%
      \put(3610,1354){\makebox(0,0)[r]{\strut{}$-2.5$}}%
      \put(3610,1664){\makebox(0,0)[r]{\strut{}$-2$}}%
      \put(3610,1974){\makebox(0,0)[r]{\strut{}$-1.5$}}%
      \put(3610,2284){\makebox(0,0)[r]{\strut{}$-1$}}%
      \put(3610,2594){\makebox(0,0)[r]{\strut{}$-0.5$}}%
      \put(3610,2904){\makebox(0,0)[r]{\strut{}$0$}}%
      \put(3742,824){\makebox(0,0){\strut{}$70$}}%
      \put(4216,824){\makebox(0,0){\strut{}$80$}}%
      \put(4690,824){\makebox(0,0){\strut{}$90$}}%
      \put(5165,824){\makebox(0,0){\strut{}$100$}}%
      \put(5639,824){\makebox(0,0){\strut{}$110$}}%
      \put(6113,824){\makebox(0,0){\strut{}$120$}}%
      \put(6587,824){\makebox(0,0){\strut{}$130$}}%
    }%
    \gplgaddtomacro\gplfronttext{%
      \csname LTb\endcsname%%
      \put(2862,1974){\rotatebox{-270}{\makebox(0,0){\strut{}$\text{BIC}(t)/M$}}}%
      \csname LTb\endcsname%%
      \put(5448,2050){\makebox(0,0)[r]{\strut{}$\beta=0.2$}}%
      \csname LTb\endcsname%%
      \put(5448,1830){\makebox(0,0)[r]{\strut{}$\beta=0.5$}}%
      \csname LTb\endcsname%%
      \put(5448,1610){\makebox(0,0)[r]{\strut{}$\beta=0.8$}}%
      \csname LTb\endcsname%%
      \put(5448,1390){\makebox(0,0)[r]{\strut{}$\beta=1.05$}}%
    }%
    \gplbacktext
    \put(0,0){\includegraphics{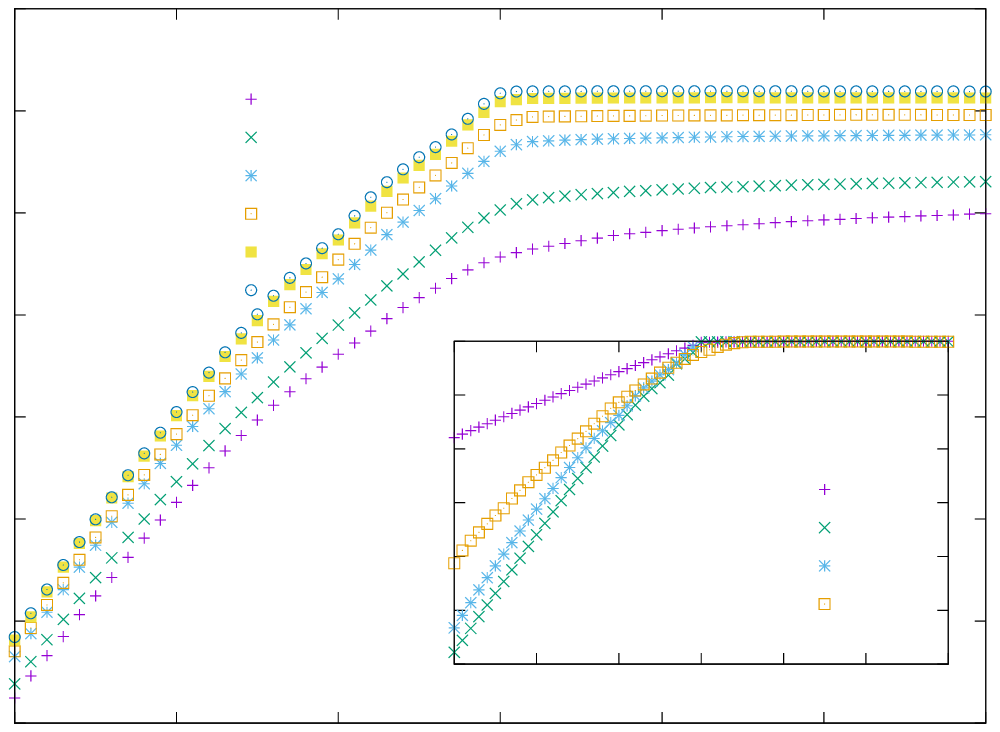}}%
    \gplfronttext
  \end{picture}%
\endgroup

%% file: diff_M_50-RR-05.tex
% GNUPLOT: LaTeX picture with Postscript
\begingroup
  \makeatletter
  \providecommand\color[2][]{%
    \GenericError{(gnuplot) \space\space\space\@spaces}{%
      Package color not loaded in conjunction with
      terminal option `colourtext'%
    }{See the gnuplot documentation for explanation.%
    }{Either use 'blacktext' in gnuplot or load the package
      color.sty in LaTeX.}%
    \renewcommand\color[2][]{}%
  }%
  \providecommand\includegraphics[2][]{%
    \GenericError{(gnuplot) \space\space\space\@spaces}{%
      Package graphicx or graphics not loaded%
    }{See the gnuplot documentation for explanation.%
    }{The gnuplot epslatex terminal needs graphicx.sty or graphics.sty.}%
    \renewcommand\includegraphics[2][]{}%
  }%
  \providecommand\rotatebox[2]{#2}%
  \@ifundefined{ifGPcolor}{%
    \newif\ifGPcolor
    \GPcolortrue
  }{}%
  \@ifundefined{ifGPblacktext}{%
    \newif\ifGPblacktext
    \GPblacktexttrue
  }{}%
  % define a \g@addto@macro without @ in the name:
  \let\gplgaddtomacro\g@addto@macro
  % define empty templates for all commands taking text:
  \gdef\gplbacktext{}%
  \gdef\gplfronttext{}%
  \makeatother
  \ifGPblacktext
    % no textcolor at all
    \def\colorrgb#1{}%
    \def\colorgray#1{}%
  \else
    % gray or color?
    \ifGPcolor
      \def\colorrgb#1{\color[rgb]{#1}}%
      \def\colorgray#1{\color[gray]{#1}}%
      \expandafter\def\csname LTw\endcsname{\color{white}}%
      \expandafter\def\csname LTb\endcsname{\color{black}}%
      \expandafter\def\csname LTa\endcsname{\color{black}}%
      \expandafter\def\csname LT0\endcsname{\color[rgb]{1,0,0}}%
      \expandafter\def\csname LT1\endcsname{\color[rgb]{0,1,0}}%
      \expandafter\def\csname LT2\endcsname{\color[rgb]{0,0,1}}%
      \expandafter\def\csname LT3\endcsname{\color[rgb]{1,0,1}}%
      \expandafter\def\csname LT4\endcsname{\color[rgb]{0,1,1}}%
      \expandafter\def\csname LT5\endcsname{\color[rgb]{1,1,0}}%
      \expandafter\def\csname LT6\endcsname{\color[rgb]{0,0,0}}%
      \expandafter\def\csname LT7\endcsname{\color[rgb]{1,0.3,0}}%
      \expandafter\def\csname LT8\endcsname{\color[rgb]{0.5,0.5,0.5}}%
    \else
      % gray
      \def\colorrgb#1{\color{black}}%
      \def\colorgray#1{\color[gray]{#1}}%
      \expandafter\def\csname LTw\endcsname{\color{white}}%
      \expandafter\def\csname LTb\endcsname{\color{black}}%
      \expandafter\def\csname LTa\endcsname{\color{black}}%
      \expandafter\def\csname LT0\endcsname{\color{black}}%
      \expandafter\def\csname LT1\endcsname{\color{black}}%
      \expandafter\def\csname LT2\endcsname{\color{black}}%
      \expandafter\def\csname LT3\endcsname{\color{black}}%
      \expandafter\def\csname LT4\endcsname{\color{black}}%
      \expandafter\def\csname LT5\endcsname{\color{black}}%
      \expandafter\def\csname LT6\endcsname{\color{black}}%
      \expandafter\def\csname LT7\endcsname{\color{black}}%
      \expandafter\def\csname LT8\endcsname{\color{black}}%
    \fi
  \fi
    \setlength{\unitlength}{0.0500bp}%
    \ifx\gptboxheight\undefined%
      \newlength{\gptboxheight}%
      \newlength{\gptboxwidth}%
      \newsavebox{\gptboxtext}%
    \fi%
    \setlength{\fboxrule}{0.5pt}%
    \setlength{\fboxsep}{1pt}%
\begin{picture}(7200.00,5040.00)%
      \csname LTb\endcsname%%
      \put(3600,4820){\makebox(0,0){\strut{}}}%
    \gplgaddtomacro\gplbacktext{%
      \csname LTb\endcsname%%
      \put(588,2646){\makebox(0,0)[r]{\strut{}$0$}}%
      \put(588,3118){\makebox(0,0)[r]{\strut{}$0.05$}}%
      \put(588,3591){\makebox(0,0)[r]{\strut{}$0.1$}}%
      \put(588,4063){\makebox(0,0)[r]{\strut{}$0.15$}}%
      \put(588,4535){\makebox(0,0)[r]{\strut{}$0.2$}}%
      \put(720,2426){\makebox(0,0){\strut{}}}%
      \put(1131,2426){\makebox(0,0){\strut{}}}%
      \put(1543,2426){\makebox(0,0){\strut{}}}%
      \put(1954,2426){\makebox(0,0){\strut{}}}%
      \put(2365,2426){\makebox(0,0){\strut{}}}%
      \put(2776,2426){\makebox(0,0){\strut{}}}%
      \put(3188,2426){\makebox(0,0){\strut{}}}%
      \put(3599,2426){\makebox(0,0){\strut{}}}%
      \put(1008,2835){\makebox(0,0)[l]{\strut{}$\beta=0.2$}}%
    }%
    \gplgaddtomacro\gplfronttext{%
      \csname LTb\endcsname%%
      \put(-160,3590){\rotatebox{-270}{\makebox(0,0){\strut{}$\Delta \text{BIC}(t) $}}}%
      \csname LTb\endcsname%%
      \put(2715,4368){\makebox(0,0)[r]{\strut{}$M=500$}}%
      \csname LTb\endcsname%%
      \put(2715,4148){\makebox(0,0)[r]{\strut{}$M=1000$}}%
      \csname LTb\endcsname%%
      \put(2715,3928){\makebox(0,0)[r]{\strut{}$M=2000$}}%
      \csname LTb\endcsname%%
      \put(2715,3708){\makebox(0,0)[r]{\strut{}$M=4000$}}%
      \csname LTb\endcsname%%
      \put(2715,3488){\makebox(0,0)[r]{\strut{}$M=8000$}}%
    }%
    \gplgaddtomacro\gplbacktext{%
      \csname LTb\endcsname%%
      \put(588,504){\makebox(0,0)[r]{\strut{}$0$}}%
      \put(588,976){\makebox(0,0)[r]{\strut{}$0.05$}}%
      \put(588,1449){\makebox(0,0)[r]{\strut{}$0.1$}}%
      \put(588,1921){\makebox(0,0)[r]{\strut{}$0.15$}}%
      \put(588,2393){\makebox(0,0)[r]{\strut{}$0.2$}}%
      \put(720,284){\makebox(0,0){\strut{}$0$}}%
      \put(1131,284){\makebox(0,0){\strut{}$20$}}%
      \put(1543,284){\makebox(0,0){\strut{}$40$}}%
      \put(1954,284){\makebox(0,0){\strut{}$60$}}%
      \put(2365,284){\makebox(0,0){\strut{}$80$}}%
      \put(2776,284){\makebox(0,0){\strut{}$100$}}%
      \put(3188,284){\makebox(0,0){\strut{}$120$}}%
      \put(3599,284){\makebox(0,0){\strut{}$140$}}%
      \put(1008,693){\makebox(0,0)[l]{\strut{}$\beta=0.5$}}%
    }%
    \gplgaddtomacro\gplfronttext{%
      \csname LTb\endcsname%%
      \put(-160,1448){\rotatebox{-270}{\makebox(0,0){\strut{}$\Delta \text{BIC}(t) $}}}%
      \put(2159,-46){\makebox(0,0){\strut{}$t$}}%
    }%
    \gplgaddtomacro\gplbacktext{%
      \csname LTb\endcsname%%
      \put(3827,2646){\makebox(0,0)[r]{\strut{} }}%
      \put(3827,3118){\makebox(0,0)[r]{\strut{} }}%
      \put(3827,3591){\makebox(0,0)[r]{\strut{} }}%
      \put(3827,4063){\makebox(0,0)[r]{\strut{} }}%
      \put(3827,4535){\makebox(0,0)[r]{\strut{} }}%
      \put(3959,2426){\makebox(0,0){\strut{}}}%
      \put(4370,2426){\makebox(0,0){\strut{}}}%
      \put(4782,2426){\makebox(0,0){\strut{}}}%
      \put(5193,2426){\makebox(0,0){\strut{}}}%
      \put(5605,2426){\makebox(0,0){\strut{}}}%
      \put(6016,2426){\makebox(0,0){\strut{}}}%
      \put(6428,2426){\makebox(0,0){\strut{}}}%
      \put(6839,2426){\makebox(0,0){\strut{}}}%
      \put(4247,2835){\makebox(0,0)[l]{\strut{}$\beta=0.8$}}%
    }%
    \gplgaddtomacro\gplfronttext{%
    }%
    \gplgaddtomacro\gplbacktext{%
      \csname LTb\endcsname%%
      \put(3827,504){\makebox(0,0)[r]{\strut{} }}%
      \put(3827,976){\makebox(0,0)[r]{\strut{} }}%
      \put(3827,1449){\makebox(0,0)[r]{\strut{} }}%
      \put(3827,1921){\makebox(0,0)[r]{\strut{} }}%
      \put(3827,2393){\makebox(0,0)[r]{\strut{} }}%
      \put(3959,284){\makebox(0,0){\strut{}$0$}}%
      \put(4370,284){\makebox(0,0){\strut{}$20$}}%
      \put(4782,284){\makebox(0,0){\strut{}$40$}}%
      \put(5193,284){\makebox(0,0){\strut{}$60$}}%
      \put(5605,284){\makebox(0,0){\strut{}$80$}}%
      \put(6016,284){\makebox(0,0){\strut{}$100$}}%
      \put(6428,284){\makebox(0,0){\strut{}$120$}}%
      \put(6839,284){\makebox(0,0){\strut{}$140$}}%
      \put(4247,693){\makebox(0,0)[l]{\strut{}$\beta=1.05$}}%
    }%
    \gplgaddtomacro\gplfronttext{%
      \csname LTb\endcsname%%
      \put(5399,-46){\makebox(0,0){\strut{}$t$}}%
    }%
    \gplbacktext
    \put(0,0){\includegraphics{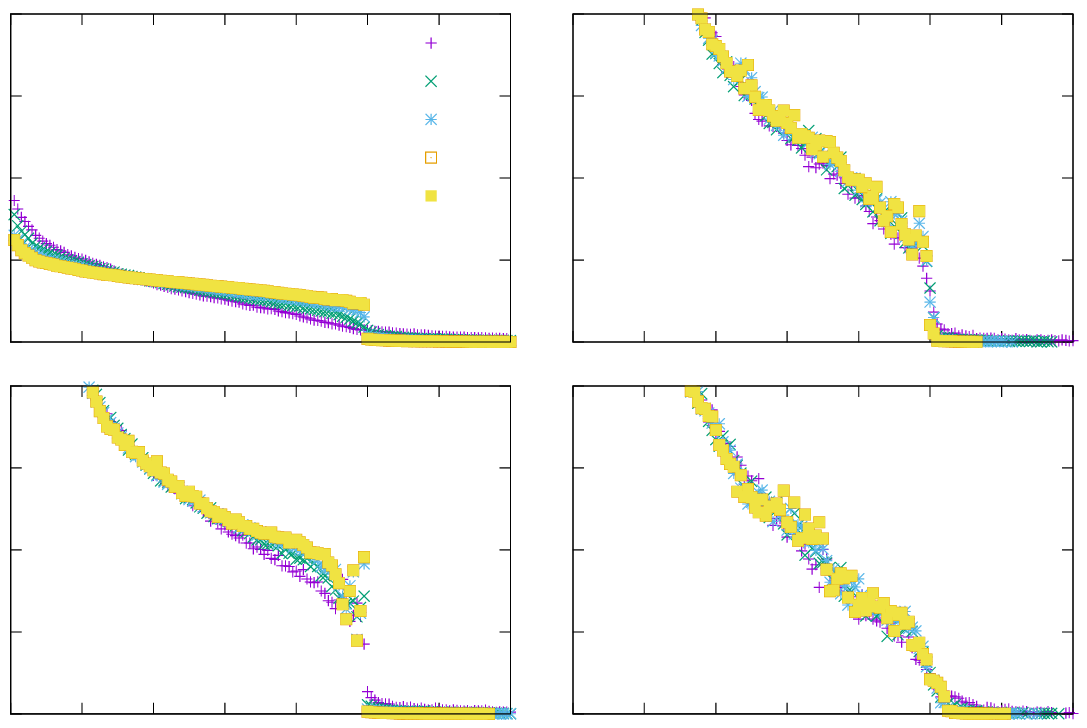}}%
    \gplfronttext
  \end{picture}%
\endgroup

%% file: differentTM-av-44-Dia.tex
% GNUPLOT: LaTeX picture with Postscript
\begingroup
  \makeatletter
  \providecommand\color[2][]{%
    \GenericError{(gnuplot) \space\space\space\@spaces}{%
      Package color not loaded in conjunction with
      terminal option `colourtext'%
    }{See the gnuplot documentation for explanation.%
    }{Either use 'blacktext' in gnuplot or load the package
      color.sty in LaTeX.}%
    \renewcommand\color[2][]{}%
  }%
  \providecommand\includegraphics[2][]{%
    \GenericError{(gnuplot) \space\space\space\@spaces}{%
      Package graphicx or graphics not loaded%
    }{See the gnuplot documentation for explanation.%
    }{The gnuplot epslatex terminal needs graphicx.sty or graphics.sty.}%
    \renewcommand\includegraphics[2][]{}%
  }%
  \providecommand\rotatebox[2]{#2}%
  \@ifundefined{ifGPcolor}{%
    \newif\ifGPcolor
    \GPcolortrue
  }{}%
  \@ifundefined{ifGPblacktext}{%
    \newif\ifGPblacktext
    \GPblacktexttrue
  }{}%
  % define a \g@addto@macro without @ in the name:
  \let\gplgaddtomacro\g@addto@macro
  % define empty templates for all commands taking text:
  \gdef\gplbacktext{}%
  \gdef\gplfronttext{}%
  \makeatother
  \ifGPblacktext
    % no textcolor at all
    \def\colorrgb#1{}%
    \def\colorgray#1{}%
  \else
    % gray or color?
    \ifGPcolor
      \def\colorrgb#1{\color[rgb]{#1}}%
      \def\colorgray#1{\color[gray]{#1}}%
      \expandafter\def\csname LTw\endcsname{\color{white}}%
      \expandafter\def\csname LTb\endcsname{\color{black}}%
      \expandafter\def\csname LTa\endcsname{\color{black}}%
      \expandafter\def\csname LT0\endcsname{\color[rgb]{1,0,0}}%
      \expandafter\def\csname LT1\endcsname{\color[rgb]{0,1,0}}%
      \expandafter\def\csname LT2\endcsname{\color[rgb]{0,0,1}}%
      \expandafter\def\csname LT3\endcsname{\color[rgb]{1,0,1}}%
      \expandafter\def\csname LT4\endcsname{\color[rgb]{0,1,1}}%
      \expandafter\def\csname LT5\endcsname{\color[rgb]{1,1,0}}%
      \expandafter\def\csname LT6\endcsname{\color[rgb]{0,0,0}}%
      \expandafter\def\csname LT7\endcsname{\color[rgb]{1,0.3,0}}%
      \expandafter\def\csname LT8\endcsname{\color[rgb]{0.5,0.5,0.5}}%
    \else
      % gray
      \def\colorrgb#1{\color{black}}%
      \def\colorgray#1{\color[gray]{#1}}%
      \expandafter\def\csname LTw\endcsname{\color{white}}%
      \expandafter\def\csname LTb\endcsname{\color{black}}%
      \expandafter\def\csname LTa\endcsname{\color{black}}%
      \expandafter\def\csname LT0\endcsname{\color{black}}%
      \expandafter\def\csname LT1\endcsname{\color{black}}%
      \expandafter\def\csname LT2\endcsname{\color{black}}%
      \expandafter\def\csname LT3\endcsname{\color{black}}%
      \expandafter\def\csname LT4\endcsname{\color{black}}%
      \expandafter\def\csname LT5\endcsname{\color{black}}%
      \expandafter\def\csname LT6\endcsname{\color{black}}%
      \expandafter\def\csname LT7\endcsname{\color{black}}%
      \expandafter\def\csname LT8\endcsname{\color{black}}%
    \fi
  \fi
    \setlength{\unitlength}{0.0500bp}%
    \ifx\gptboxheight\undefined%
      \newlength{\gptboxheight}%
      \newlength{\gptboxwidth}%
      \newsavebox{\gptboxtext}%
    \fi%
    \setlength{\fboxrule}{0.5pt}%
    \setlength{\fboxsep}{1pt}%
\begin{picture}(7200.00,5040.00)%
    \gplgaddtomacro\gplbacktext{%
      \csname LTb\endcsname%%
      \put(1078,704){\makebox(0,0)[r]{\strut{}$-20.5$}}%
      \put(1078,1292){\makebox(0,0)[r]{\strut{}$-20$}}%
      \put(1078,1880){\makebox(0,0)[r]{\strut{}$-19.5$}}%
      \put(1078,2468){\makebox(0,0)[r]{\strut{}$-19$}}%
      \put(1078,3055){\makebox(0,0)[r]{\strut{}$-18.5$}}%
      \put(1078,3643){\makebox(0,0)[r]{\strut{}$-18$}}%
      \put(1078,4231){\makebox(0,0)[r]{\strut{}$-17.5$}}%
      \put(1078,4819){\makebox(0,0)[r]{\strut{}$-17$}}%
      \put(1210,484){\makebox(0,0){\strut{}$40$}}%
      \put(2329,484){\makebox(0,0){\strut{}$50$}}%
      \put(3447,484){\makebox(0,0){\strut{}$60$}}%
      \put(4566,484){\makebox(0,0){\strut{}$70$}}%
      \put(5684,484){\makebox(0,0){\strut{}$80$}}%
      \put(6803,484){\makebox(0,0){\strut{}$90$}}%
    }%
    \gplgaddtomacro\gplfronttext{%
      \csname LTb\endcsname%%
      \put(198,2761){\rotatebox{-270}{\makebox(0,0){\strut{}$\text{BIC}(t)/M$}}}%
      \put(4006,154){\makebox(0,0){\strut{}$t$}}%
      \csname LTb\endcsname%%
      \put(2145,4298){\makebox(0,0)[r]{\strut{}$M=300$}}%
      \csname LTb\endcsname%%
      \put(2145,4078){\makebox(0,0)[r]{\strut{}$M=500$}}%
      \csname LTb\endcsname%%
      \put(2145,3858){\makebox(0,0)[r]{\strut{}$M=1000$}}%
      \csname LTb\endcsname%%
      \put(2145,3638){\makebox(0,0)[r]{\strut{}$M=2000$}}%
      \csname LTb\endcsname%%
      \put(2145,3418){\makebox(0,0)[r]{\strut{}$M=4000$}}%
      \csname LTb\endcsname%%
      \put(2145,3198){\makebox(0,0)[r]{\strut{}$M=8000$}}%
    }%
    \gplgaddtomacro\gplbacktext{%
      \csname LTb\endcsname%%
      \put(3610,1044){\makebox(0,0)[r]{\strut{}$-3$}}%
      \put(3610,1354){\makebox(0,0)[r]{\strut{}$-2.5$}}%
      \put(3610,1664){\makebox(0,0)[r]{\strut{}$-2$}}%
      \put(3610,1974){\makebox(0,0)[r]{\strut{}$-1.5$}}%
      \put(3610,2284){\makebox(0,0)[r]{\strut{}$-1$}}%
      \put(3610,2594){\makebox(0,0)[r]{\strut{}$-0.5$}}%
      \put(3610,2904){\makebox(0,0)[r]{\strut{}$0$}}%
      \put(3742,824){\makebox(0,0){\strut{}$40$}}%
      \put(4311,824){\makebox(0,0){\strut{}$50$}}%
      \put(4880,824){\makebox(0,0){\strut{}$60$}}%
      \put(5449,824){\makebox(0,0){\strut{}$70$}}%
      \put(6018,824){\makebox(0,0){\strut{}$80$}}%
      \put(6587,824){\makebox(0,0){\strut{}$90$}}%
    }%
    \gplgaddtomacro\gplfronttext{%
      \csname LTb\endcsname%%
      \put(2862,1974){\rotatebox{-270}{\makebox(0,0){\strut{}$\text{BIC}(t)/M$}}}%
      \csname LTb\endcsname%%
      \put(5448,2050){\makebox(0,0)[r]{\strut{}$\beta=0.3$}}%
      \csname LTb\endcsname%%
      \put(5448,1830){\makebox(0,0)[r]{\strut{}$\beta=0.4$}}%
      \csname LTb\endcsname%%
      \put(5448,1610){\makebox(0,0)[r]{\strut{}$\beta=0.5$}}%
      \csname LTb\endcsname%%
      \put(5448,1390){\makebox(0,0)[r]{\strut{}$\beta=0.6$}}%
    }%
    \gplbacktext
    \put(0,0){\includegraphics{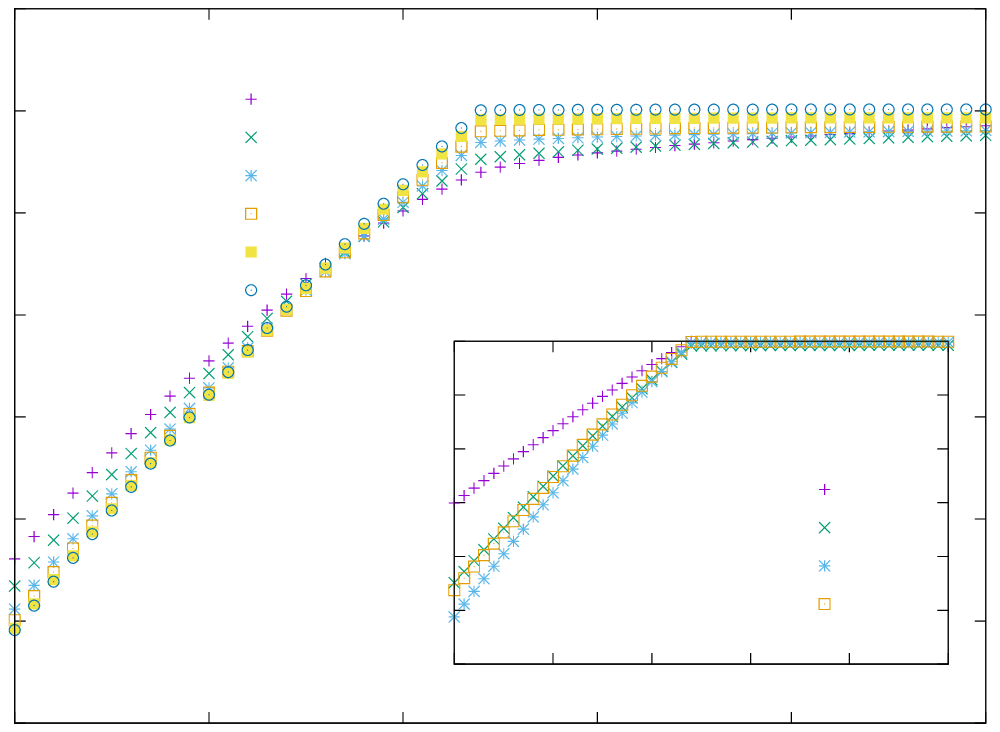}}%
    \gplfronttext
  \end{picture}%
\endgroup

%% file: diff_M_44-Dia.tex
% GNUPLOT: LaTeX picture with Postscript
\begingroup
  \makeatletter
  \providecommand\color[2][]{%
    \GenericError{(gnuplot) \space\space\space\@spaces}{%
      Package color not loaded in conjunction with
      terminal option `colourtext'%
    }{See the gnuplot documentation for explanation.%
    }{Either use 'blacktext' in gnuplot or load the package
      color.sty in LaTeX.}%
    \renewcommand\color[2][]{}%
  }%
  \providecommand\includegraphics[2][]{%
    \GenericError{(gnuplot) \space\space\space\@spaces}{%
      Package graphicx or graphics not loaded%
    }{See the gnuplot documentation for explanation.%
    }{The gnuplot epslatex terminal needs graphicx.sty or graphics.sty.}%
    \renewcommand\includegraphics[2][]{}%
  }%
  \providecommand\rotatebox[2]{#2}%
  \@ifundefined{ifGPcolor}{%
    \newif\ifGPcolor
    \GPcolortrue
  }{}%
  \@ifundefined{ifGPblacktext}{%
    \newif\ifGPblacktext
    \GPblacktexttrue
  }{}%
  % define a \g@addto@macro without @ in the name:
  \let\gplgaddtomacro\g@addto@macro
  % define empty templates for all commands taking text:
  \gdef\gplbacktext{}%
  \gdef\gplfronttext{}%
  \makeatother
  \ifGPblacktext
    % no textcolor at all
    \def\colorrgb#1{}%
    \def\colorgray#1{}%
  \else
    % gray or color?
    \ifGPcolor
      \def\colorrgb#1{\color[rgb]{#1}}%
      \def\colorgray#1{\color[gray]{#1}}%
      \expandafter\def\csname LTw\endcsname{\color{white}}%
      \expandafter\def\csname LTb\endcsname{\color{black}}%
      \expandafter\def\csname LTa\endcsname{\color{black}}%
      \expandafter\def\csname LT0\endcsname{\color[rgb]{1,0,0}}%
      \expandafter\def\csname LT1\endcsname{\color[rgb]{0,1,0}}%
      \expandafter\def\csname LT2\endcsname{\color[rgb]{0,0,1}}%
      \expandafter\def\csname LT3\endcsname{\color[rgb]{1,0,1}}%
      \expandafter\def\csname LT4\endcsname{\color[rgb]{0,1,1}}%
      \expandafter\def\csname LT5\endcsname{\color[rgb]{1,1,0}}%
      \expandafter\def\csname LT6\endcsname{\color[rgb]{0,0,0}}%
      \expandafter\def\csname LT7\endcsname{\color[rgb]{1,0.3,0}}%
      \expandafter\def\csname LT8\endcsname{\color[rgb]{0.5,0.5,0.5}}%
    \else
      % gray
      \def\colorrgb#1{\color{black}}%
      \def\colorgray#1{\color[gray]{#1}}%
      \expandafter\def\csname LTw\endcsname{\color{white}}%
      \expandafter\def\csname LTb\endcsname{\color{black}}%
      \expandafter\def\csname LTa\endcsname{\color{black}}%
      \expandafter\def\csname LT0\endcsname{\color{black}}%
      \expandafter\def\csname LT1\endcsname{\color{black}}%
      \expandafter\def\csname LT2\endcsname{\color{black}}%
      \expandafter\def\csname LT3\endcsname{\color{black}}%
      \expandafter\def\csname LT4\endcsname{\color{black}}%
      \expandafter\def\csname LT5\endcsname{\color{black}}%
      \expandafter\def\csname LT6\endcsname{\color{black}}%
      \expandafter\def\csname LT7\endcsname{\color{black}}%
      \expandafter\def\csname LT8\endcsname{\color{black}}%
    \fi
  \fi
    \setlength{\unitlength}{0.0500bp}%
    \ifx\gptboxheight\undefined%
      \newlength{\gptboxheight}%
      \newlength{\gptboxwidth}%
      \newsavebox{\gptboxtext}%
    \fi%
    \setlength{\fboxrule}{0.5pt}%
    \setlength{\fboxsep}{1pt}%
\begin{picture}(7200.00,5040.00)%
      \csname LTb\endcsname%%
      \put(3600,4820){\makebox(0,0){\strut{}}}%
    \gplgaddtomacro\gplbacktext{%
      \csname LTb\endcsname%%
      \put(588,2646){\makebox(0,0)[r]{\strut{}$0$}}%
      \put(588,2882){\makebox(0,0)[r]{\strut{}$0.05$}}%
      \put(588,3118){\makebox(0,0)[r]{\strut{}$0.1$}}%
      \put(588,3354){\makebox(0,0)[r]{\strut{}$0.15$}}%
      \put(588,3591){\makebox(0,0)[r]{\strut{}$0.2$}}%
      \put(588,3827){\makebox(0,0)[r]{\strut{}$0.25$}}%
      \put(588,4063){\makebox(0,0)[r]{\strut{}$0.3$}}%
      \put(588,4299){\makebox(0,0)[r]{\strut{}$0.35$}}%
      \put(588,4535){\makebox(0,0)[r]{\strut{}$0.4$}}%
      \put(720,2426){\makebox(0,0){\strut{}}}%
      \put(1296,2426){\makebox(0,0){\strut{}}}%
      \put(1872,2426){\makebox(0,0){\strut{}}}%
      \put(2447,2426){\makebox(0,0){\strut{}}}%
      \put(3023,2426){\makebox(0,0){\strut{}}}%
      \put(3599,2426){\makebox(0,0){\strut{}}}%
      \put(1008,2835){\makebox(0,0)[l]{\strut{}$\beta=0.3$}}%
    }%
    \gplgaddtomacro\gplfronttext{%
      \csname LTb\endcsname%%
      \put(-160,3590){\rotatebox{-270}{\makebox(0,0){\strut{}$\Delta \text{BIC}(t) $}}}%
      \csname LTb\endcsname%%
      \put(2612,4362){\makebox(0,0)[r]{\strut{}$M=500$}}%
      \csname LTb\endcsname%%
      \put(2612,4142){\makebox(0,0)[r]{\strut{}$M=1000$}}%
      \csname LTb\endcsname%%
      \put(2612,3922){\makebox(0,0)[r]{\strut{}$M=2000$}}%
      \csname LTb\endcsname%%
      \put(2612,3702){\makebox(0,0)[r]{\strut{}$M=4000$}}%
      \csname LTb\endcsname%%
      \put(2612,3482){\makebox(0,0)[r]{\strut{}$M=8000$}}%
    }%
    \gplgaddtomacro\gplbacktext{%
      \csname LTb\endcsname%%
      \put(588,504){\makebox(0,0)[r]{\strut{}$0$}}%
      \put(588,740){\makebox(0,0)[r]{\strut{}$0.05$}}%
      \put(588,976){\makebox(0,0)[r]{\strut{}$0.1$}}%
      \put(588,1212){\makebox(0,0)[r]{\strut{}$0.15$}}%
      \put(588,1449){\makebox(0,0)[r]{\strut{}$0.2$}}%
      \put(588,1685){\makebox(0,0)[r]{\strut{}$0.25$}}%
      \put(588,1921){\makebox(0,0)[r]{\strut{}$0.3$}}%
      \put(588,2157){\makebox(0,0)[r]{\strut{}$0.35$}}%
      \put(588,2393){\makebox(0,0)[r]{\strut{}$0.4$}}%
      \put(720,284){\makebox(0,0){\strut{}$0$}}%
      \put(1296,284){\makebox(0,0){\strut{}$20$}}%
      \put(1872,284){\makebox(0,0){\strut{}$40$}}%
      \put(2447,284){\makebox(0,0){\strut{}$60$}}%
      \put(3023,284){\makebox(0,0){\strut{}$80$}}%
      \put(3599,284){\makebox(0,0){\strut{}$100$}}%
      \put(1008,693){\makebox(0,0)[l]{\strut{}$\beta=0.4$}}%
    }%
    \gplgaddtomacro\gplfronttext{%
      \csname LTb\endcsname%%
      \put(-160,1448){\rotatebox{-270}{\makebox(0,0){\strut{}$\Delta \text{BIC}(t)$}}}%
      \put(2159,-46){\makebox(0,0){\strut{}$t$}}%
    }%
    \gplgaddtomacro\gplbacktext{%
      \csname LTb\endcsname%%
      \put(3827,2646){\makebox(0,0)[r]{\strut{} }}%
      \put(3827,2882){\makebox(0,0)[r]{\strut{} }}%
      \put(3827,3118){\makebox(0,0)[r]{\strut{} }}%
      \put(3827,3354){\makebox(0,0)[r]{\strut{} }}%
      \put(3827,3591){\makebox(0,0)[r]{\strut{} }}%
      \put(3827,3827){\makebox(0,0)[r]{\strut{} }}%
      \put(3827,4063){\makebox(0,0)[r]{\strut{} }}%
      \put(3827,4299){\makebox(0,0)[r]{\strut{} }}%
      \put(3827,4535){\makebox(0,0)[r]{\strut{} }}%
      \put(3959,2426){\makebox(0,0){\strut{}}}%
      \put(4535,2426){\makebox(0,0){\strut{}}}%
      \put(5111,2426){\makebox(0,0){\strut{}}}%
      \put(5687,2426){\makebox(0,0){\strut{}}}%
      \put(6263,2426){\makebox(0,0){\strut{}}}%
      \put(6839,2426){\makebox(0,0){\strut{}}}%
      \put(4247,2835){\makebox(0,0)[l]{\strut{}$\beta=0.5$}}%
    }%
    \gplgaddtomacro\gplfronttext{%
    }%
    \gplgaddtomacro\gplbacktext{%
      \csname LTb\endcsname%%
      \put(3827,504){\makebox(0,0)[r]{\strut{} }}%
      \put(3827,740){\makebox(0,0)[r]{\strut{} }}%
      \put(3827,976){\makebox(0,0)[r]{\strut{} }}%
      \put(3827,1212){\makebox(0,0)[r]{\strut{} }}%
      \put(3827,1449){\makebox(0,0)[r]{\strut{} }}%
      \put(3827,1685){\makebox(0,0)[r]{\strut{} }}%
      \put(3827,1921){\makebox(0,0)[r]{\strut{} }}%
      \put(3827,2157){\makebox(0,0)[r]{\strut{} }}%
      \put(3827,2393){\makebox(0,0)[r]{\strut{} }}%
      \put(3959,284){\makebox(0,0){\strut{}$0$}}%
      \put(4535,284){\makebox(0,0){\strut{}$20$}}%
      \put(5111,284){\makebox(0,0){\strut{}$40$}}%
      \put(5687,284){\makebox(0,0){\strut{}$60$}}%
      \put(6263,284){\makebox(0,0){\strut{}$80$}}%
      \put(6839,284){\makebox(0,0){\strut{}$100$}}%
      \put(4247,693){\makebox(0,0)[l]{\strut{}$\beta=0.6$}}%
    }%
    \gplgaddtomacro\gplfronttext{%
      \csname LTb\endcsname%%
      \put(5399,-46){\makebox(0,0){\strut{}$t$}}%
    }%
    \gplbacktext
    \put(0,0){\includegraphics{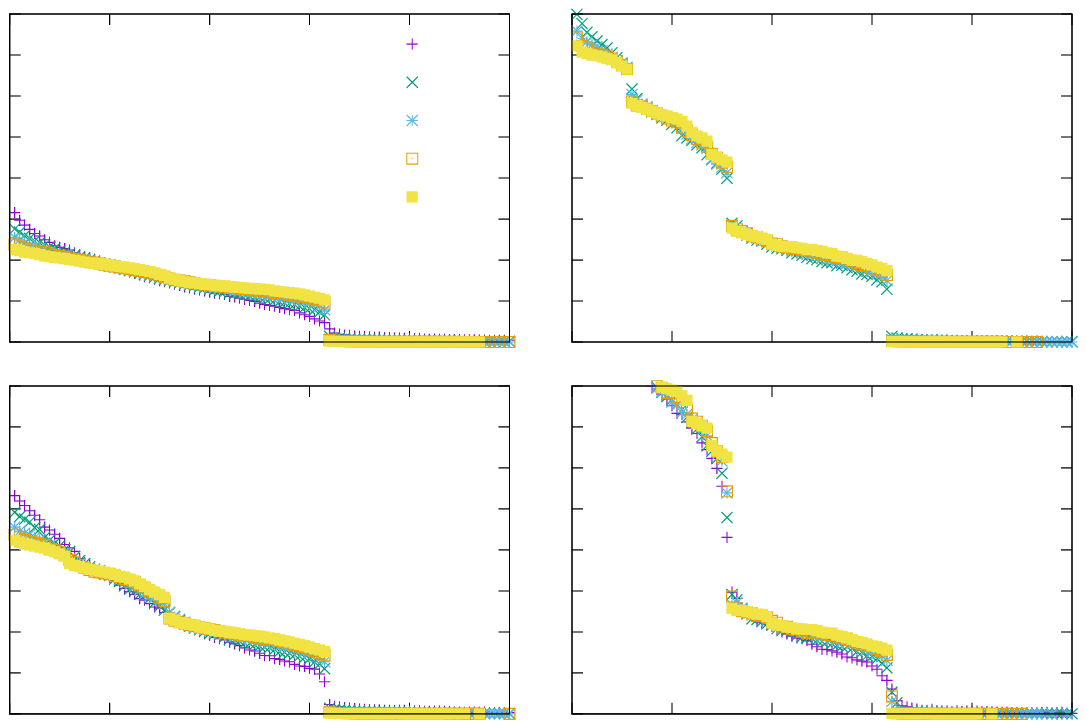}}%
    \gplfronttext
  \end{picture}%
\endgroup

%% file: ROC_M_DiaRR.tex
% GNUPLOT: LaTeX picture with Postscript
\begingroup
  \makeatletter
  \providecommand\color[2][]{%
    \GenericError{(gnuplot) \space\space\space\@spaces}{%
      Package color not loaded in conjunction with
      terminal option `colourtext'%
    }{See the gnuplot documentation for explanation.%
    }{Either use 'blacktext' in gnuplot or load the package
      color.sty in LaTeX.}%
    \renewcommand\color[2][]{}%
  }%
  \providecommand\includegraphics[2][]{%
    \GenericError{(gnuplot) \space\space\space\@spaces}{%
      Package graphicx or graphics not loaded%
    }{See the gnuplot documentation for explanation.%
    }{The gnuplot epslatex terminal needs graphicx.sty or graphics.sty.}%
    \renewcommand\includegraphics[2][]{}%
  }%
  \providecommand\rotatebox[2]{#2}%
  \@ifundefined{ifGPcolor}{%
    \newif\ifGPcolor
    \GPcolortrue
  }{}%
  \@ifundefined{ifGPblacktext}{%
    \newif\ifGPblacktext
    \GPblacktexttrue
  }{}%
  % define a \g@addto@macro without @ in the name:
  \let\gplgaddtomacro\g@addto@macro
  % define empty templates for all commands taking text:
  \gdef\gplbacktext{}%
  \gdef\gplfronttext{}%
  \makeatother
  \ifGPblacktext
    % no textcolor at all
    \def\colorrgb#1{}%
    \def\colorgray#1{}%
  \else
    % gray or color?
    \ifGPcolor
      \def\colorrgb#1{\color[rgb]{#1}}%
      \def\colorgray#1{\color[gray]{#1}}%
      \expandafter\def\csname LTw\endcsname{\color{white}}%
      \expandafter\def\csname LTb\endcsname{\color{black}}%
      \expandafter\def\csname LTa\endcsname{\color{black}}%
      \expandafter\def\csname LT0\endcsname{\color[rgb]{1,0,0}}%
      \expandafter\def\csname LT1\endcsname{\color[rgb]{0,1,0}}%
      \expandafter\def\csname LT2\endcsname{\color[rgb]{0,0,1}}%
      \expandafter\def\csname LT3\endcsname{\color[rgb]{1,0,1}}%
      \expandafter\def\csname LT4\endcsname{\color[rgb]{0,1,1}}%
      \expandafter\def\csname LT5\endcsname{\color[rgb]{1,1,0}}%
      \expandafter\def\csname LT6\endcsname{\color[rgb]{0,0,0}}%
      \expandafter\def\csname LT7\endcsname{\color[rgb]{1,0.3,0}}%
      \expandafter\def\csname LT8\endcsname{\color[rgb]{0.5,0.5,0.5}}%
    \else
      % gray
      \def\colorrgb#1{\color{black}}%
      \def\colorgray#1{\color[gray]{#1}}%
      \expandafter\def\csname LTw\endcsname{\color{white}}%
      \expandafter\def\csname LTb\endcsname{\color{black}}%
      \expandafter\def\csname LTa\endcsname{\color{black}}%
      \expandafter\def\csname LT0\endcsname{\color{black}}%
      \expandafter\def\csname LT1\endcsname{\color{black}}%
      \expandafter\def\csname LT2\endcsname{\color{black}}%
      \expandafter\def\csname LT3\endcsname{\color{black}}%
      \expandafter\def\csname LT4\endcsname{\color{black}}%
      \expandafter\def\csname LT5\endcsname{\color{black}}%
      \expandafter\def\csname LT6\endcsname{\color{black}}%
      \expandafter\def\csname LT7\endcsname{\color{black}}%
      \expandafter\def\csname LT8\endcsname{\color{black}}%
    \fi
  \fi
    \setlength{\unitlength}{0.0500bp}%
    \ifx\gptboxheight\undefined%
      \newlength{\gptboxheight}%
      \newlength{\gptboxwidth}%
      \newsavebox{\gptboxtext}%
    \fi%
    \setlength{\fboxrule}{0.5pt}%
    \setlength{\fboxsep}{1pt}%
\begin{picture}(7200.00,5040.00)%
    \gplgaddtomacro\gplbacktext{%
      \csname LTb\endcsname%%
      \put(588,504){\makebox(0,0)[r]{\strut{}$0.5$}}%
      \put(588,1310){\makebox(0,0)[r]{\strut{}$0.6$}}%
      \put(588,2116){\makebox(0,0)[r]{\strut{}$0.7$}}%
      \put(588,2923){\makebox(0,0)[r]{\strut{}$0.8$}}%
      \put(588,3729){\makebox(0,0)[r]{\strut{}$0.9$}}%
      \put(588,4535){\makebox(0,0)[r]{\strut{}$1$}}%
      \put(720,284){\makebox(0,0){\strut{}$0.9$}}%
      \put(1296,284){\makebox(0,0){\strut{}$0.92$}}%
      \put(1872,284){\makebox(0,0){\strut{}$0.94$}}%
      \put(2447,284){\makebox(0,0){\strut{}$0.96$}}%
      \put(3023,284){\makebox(0,0){\strut{}$0.98$}}%
      \put(3599,284){\makebox(0,0){\strut{}$1$}}%
      \put(1980,4787){\makebox(0,0)[l]{\strut{}\text{(a)}}}%
      \put(5399,4787){\makebox(0,0)[l]{\strut{}\text{(b)}}}%
    }%
    \gplgaddtomacro\gplfronttext{%
      \csname LTb\endcsname%%
      \put(-28,2519){\rotatebox{-270}{\makebox(0,0){\strut{}$\text{TPR}$}}}%
      \put(2159,-46){\makebox(0,0){\strut{}$\text{TNR}$}}%
      \csname LTb\endcsname%%
      \put(1592,2006){\makebox(0,0)[r]{\strut{}$\beta=0.2$}}%
      \csname LTb\endcsname%%
      \put(1592,1786){\makebox(0,0)[r]{\strut{}$\beta=0.5$}}%
      \csname LTb\endcsname%%
      \put(1592,1566){\makebox(0,0)[r]{\strut{}$\beta=0.8$}}%
      \csname LTb\endcsname%%
      \put(1592,1346){\makebox(0,0)[r]{\strut{}$\beta=1.05$}}%
    }%
    \gplgaddtomacro\gplbacktext{%
      \csname LTb\endcsname%%
      \put(3827,504){\makebox(0,0)[r]{\strut{}}}%
      \put(3827,1310){\makebox(0,0)[r]{\strut{}}}%
      \put(3827,2116){\makebox(0,0)[r]{\strut{}}}%
      \put(3827,2923){\makebox(0,0)[r]{\strut{}}}%
      \put(3827,3729){\makebox(0,0)[r]{\strut{}}}%
      \put(3827,4535){\makebox(0,0)[r]{\strut{}}}%
      \put(3959,284){\makebox(0,0){\strut{}$0.9$}}%
      \put(4535,284){\makebox(0,0){\strut{}$0.92$}}%
      \put(5111,284){\makebox(0,0){\strut{}$0.94$}}%
      \put(5687,284){\makebox(0,0){\strut{}$0.96$}}%
      \put(6263,284){\makebox(0,0){\strut{}$0.98$}}%
      \put(6839,284){\makebox(0,0){\strut{}$1$}}%
      \put(1980,4787){\makebox(0,0)[l]{\strut{}\text{(a)}}}%
      \put(5399,4787){\makebox(0,0)[l]{\strut{}\text{(b)}}}%
    }%
    \gplgaddtomacro\gplfronttext{%
      \csname LTb\endcsname%%
      \put(5399,-46){\makebox(0,0){\strut{}$\text{TNR}$”}}%
      \csname LTb\endcsname%%
      \put(4832,2006){\makebox(0,0)[r]{\strut{}$\beta=0.3$}}%
      \csname LTb\endcsname%%
      \put(4832,1786){\makebox(0,0)[r]{\strut{}$\beta=0.4$}}%
      \csname LTb\endcsname%%
      \put(4832,1566){\makebox(0,0)[r]{\strut{}$\beta=0.5$}}%
      \csname LTb\endcsname%%
      \put(4832,1346){\makebox(0,0)[r]{\strut{}$\beta=0.6$}}%
    }%
    \gplbacktext
    \put(0,0){\includegraphics{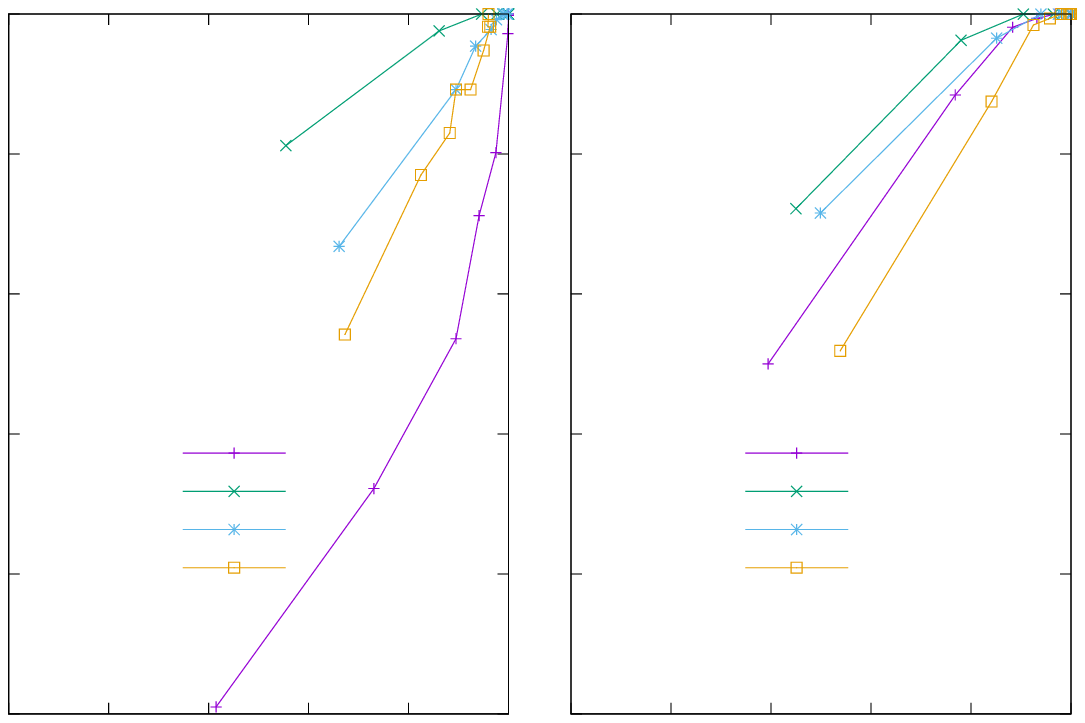}}%
    \gplfronttext
  \end{picture}%
\endgroup

%% file: time-RR.tex
% GNUPLOT: LaTeX picture with Postscript
\begingroup
  \makeatletter
  \providecommand\color[2][]{%
    \GenericError{(gnuplot) \space\space\space\@spaces}{%
      Package color not loaded in conjunction with
      terminal option `colourtext'%
    }{See the gnuplot documentation for explanation.%
    }{Either use 'blacktext' in gnuplot or load the package
      color.sty in LaTeX.}%
    \renewcommand\color[2][]{}%
  }%
  \providecommand\includegraphics[2][]{%
    \GenericError{(gnuplot) \space\space\space\@spaces}{%
      Package graphicx or graphics not loaded%
    }{See the gnuplot documentation for explanation.%
    }{The gnuplot epslatex terminal needs graphicx.sty or graphics.sty.}%
    \renewcommand\includegraphics[2][]{}%
  }%
  \providecommand\rotatebox[2]{#2}%
  \@ifundefined{ifGPcolor}{%
    \newif\ifGPcolor
    \GPcolortrue
  }{}%
  \@ifundefined{ifGPblacktext}{%
    \newif\ifGPblacktext
    \GPblacktexttrue
  }{}%
  % define a \g@addto@macro without @ in the name:
  \let\gplgaddtomacro\g@addto@macro
  % define empty templates for all commands taking text:
  \gdef\gplbacktext{}%
  \gdef\gplfronttext{}%
  \makeatother
  \ifGPblacktext
    % no textcolor at all
    \def\colorrgb#1{}%
    \def\colorgray#1{}%
  \else
    % gray or color?
    \ifGPcolor
      \def\colorrgb#1{\color[rgb]{#1}}%
      \def\colorgray#1{\color[gray]{#1}}%
      \expandafter\def\csname LTw\endcsname{\color{white}}%
      \expandafter\def\csname LTb\endcsname{\color{black}}%
      \expandafter\def\csname LTa\endcsname{\color{black}}%
      \expandafter\def\csname LT0\endcsname{\color[rgb]{1,0,0}}%
      \expandafter\def\csname LT1\endcsname{\color[rgb]{0,1,0}}%
      \expandafter\def\csname LT2\endcsname{\color[rgb]{0,0,1}}%
      \expandafter\def\csname LT3\endcsname{\color[rgb]{1,0,1}}%
      \expandafter\def\csname LT4\endcsname{\color[rgb]{0,1,1}}%
      \expandafter\def\csname LT5\endcsname{\color[rgb]{1,1,0}}%
      \expandafter\def\csname LT6\endcsname{\color[rgb]{0,0,0}}%
      \expandafter\def\csname LT7\endcsname{\color[rgb]{1,0.3,0}}%
      \expandafter\def\csname LT8\endcsname{\color[rgb]{0.5,0.5,0.5}}%
    \else
      % gray
      \def\colorrgb#1{\color{black}}%
      \def\colorgray#1{\color[gray]{#1}}%
      \expandafter\def\csname LTw\endcsname{\color{white}}%
      \expandafter\def\csname LTb\endcsname{\color{black}}%
      \expandafter\def\csname LTa\endcsname{\color{black}}%
      \expandafter\def\csname LT0\endcsname{\color{black}}%
      \expandafter\def\csname LT1\endcsname{\color{black}}%
      \expandafter\def\csname LT2\endcsname{\color{black}}%
      \expandafter\def\csname LT3\endcsname{\color{black}}%
      \expandafter\def\csname LT4\endcsname{\color{black}}%
      \expandafter\def\csname LT5\endcsname{\color{black}}%
      \expandafter\def\csname LT6\endcsname{\color{black}}%
      \expandafter\def\csname LT7\endcsname{\color{black}}%
      \expandafter\def\csname LT8\endcsname{\color{black}}%
    \fi
  \fi
    \setlength{\unitlength}{0.0500bp}%
    \ifx\gptboxheight\undefined%
      \newlength{\gptboxheight}%
      \newlength{\gptboxwidth}%
      \newsavebox{\gptboxtext}%
    \fi%
    \setlength{\fboxrule}{0.5pt}%
    \setlength{\fboxsep}{1pt}%
\begin{picture}(7200.00,5040.00)%
    \gplgaddtomacro\gplbacktext{%
      \csname LTb\endcsname%%
      \put(588,504){\makebox(0,0)[r]{\strut{}$0.5$}}%
      \put(588,907){\makebox(0,0)[r]{\strut{}$1$}}%
      \put(588,1310){\makebox(0,0)[r]{\strut{}$1.5$}}%
      \put(588,1713){\makebox(0,0)[r]{\strut{}$2$}}%
      \put(588,2116){\makebox(0,0)[r]{\strut{}$2.5$}}%
      \put(588,2520){\makebox(0,0)[r]{\strut{}$3$}}%
      \put(588,2923){\makebox(0,0)[r]{\strut{}$3.5$}}%
      \put(588,3326){\makebox(0,0)[r]{\strut{}$4$}}%
      \put(588,3729){\makebox(0,0)[r]{\strut{}$4.5$}}%
      \put(588,4132){\makebox(0,0)[r]{\strut{}$5$}}%
      \put(588,4535){\makebox(0,0)[r]{\strut{}$5.5$}}%
      \put(720,284){\makebox(0,0){\strut{}$2.8$}}%
      \put(990,284){\makebox(0,0){\strut{}$3$}}%
      \put(1260,284){\makebox(0,0){\strut{}$3.2$}}%
      \put(1530,284){\makebox(0,0){\strut{}$3.4$}}%
      \put(1800,284){\makebox(0,0){\strut{}$3.6$}}%
      \put(2070,284){\makebox(0,0){\strut{}$3.8$}}%
      \put(2339,284){\makebox(0,0){\strut{}$4$}}%
      \put(2609,284){\makebox(0,0){\strut{}$4.2$}}%
      \put(2879,284){\makebox(0,0){\strut{}$4.4$}}%
      \put(3149,284){\makebox(0,0){\strut{}$4.6$}}%
      \put(3419,284){\makebox(0,0){\strut{}$4.8$}}%
      \put(1980,4787){\makebox(0,0)[l]{\strut{}\text{(a)}}}%
      \put(5399,4787){\makebox(0,0)[l]{\strut{}\text{(b)}}}%
    }%
    \gplgaddtomacro\gplfronttext{%
      \csname LTb\endcsname%%
      \put(-28,2519){\rotatebox{-270}{\makebox(0,0){\strut{}$\log[\text{time(s)}]$}}}%
      \put(2069,-46){\makebox(0,0){\strut{}$\log(N)$}}%
      \csname LTb\endcsname%%
      \put(2429,1402){\makebox(0,0)[r]{\strut{}$\beta=0.3$}}%
      \csname LTb\endcsname%%
      \put(2429,1182){\makebox(0,0)[r]{\strut{}$\beta=0.5$}}%
      \csname LTb\endcsname%%
      \put(2429,962){\makebox(0,0)[r]{\strut{}$\beta=0.9$}}%
      \csname LTb\endcsname%%
      \put(2429,742){\makebox(0,0)[r]{\strut{}$\beta=1.1$}}%
    }%
    \gplgaddtomacro\gplbacktext{%
      \csname LTb\endcsname%%
      \put(4008,504){\makebox(0,0)[r]{\strut{}$0.5$}}%
      \put(4008,870){\makebox(0,0)[r]{\strut{}$1$}}%
      \put(4008,1237){\makebox(0,0)[r]{\strut{}$1.5$}}%
      \put(4008,1603){\makebox(0,0)[r]{\strut{}$2$}}%
      \put(4008,1970){\makebox(0,0)[r]{\strut{}$2.5$}}%
      \put(4008,2336){\makebox(0,0)[r]{\strut{}$3$}}%
      \put(4008,2703){\makebox(0,0)[r]{\strut{}$3.5$}}%
      \put(4008,3069){\makebox(0,0)[r]{\strut{}$4$}}%
      \put(4008,3436){\makebox(0,0)[r]{\strut{}$4.5$}}%
      \put(4008,3802){\makebox(0,0)[r]{\strut{}$5$}}%
      \put(4008,4169){\makebox(0,0)[r]{\strut{}$5.5$}}%
      \put(4008,4535){\makebox(0,0)[r]{\strut{}$6$}}%
      \put(4140,284){\makebox(0,0){\strut{}$2.8$}}%
      \put(4410,284){\makebox(0,0){\strut{}$3$}}%
      \put(4680,284){\makebox(0,0){\strut{}$3.2$}}%
      \put(4950,284){\makebox(0,0){\strut{}$3.4$}}%
      \put(5220,284){\makebox(0,0){\strut{}$3.6$}}%
      \put(5490,284){\makebox(0,0){\strut{}$3.8$}}%
      \put(5759,284){\makebox(0,0){\strut{}$4$}}%
      \put(6029,284){\makebox(0,0){\strut{}$4.2$}}%
      \put(6299,284){\makebox(0,0){\strut{}$4.4$}}%
      \put(6569,284){\makebox(0,0){\strut{}$4.6$}}%
      \put(6839,284){\makebox(0,0){\strut{}$4.8$}}%
      \put(1980,4787){\makebox(0,0)[l]{\strut{}\text{(a)}}}%
      \put(5399,4787){\makebox(0,0)[l]{\strut{}\text{(b)}}}%
    }%
    \gplgaddtomacro\gplfronttext{%
      \csname LTb\endcsname%%
      \put(5489,-46){\makebox(0,0){\strut{}$\log(N)$}}%
      \csname LTb\endcsname%%
      \put(5849,1402){\makebox(0,0)[r]{\strut{}$\beta=0.3$}}%
      \csname LTb\endcsname%%
      \put(5849,1182){\makebox(0,0)[r]{\strut{}$\beta=0.5$}}%
      \csname LTb\endcsname%%
      \put(5849,962){\makebox(0,0)[r]{\strut{}$\beta=0.9$}}%
      \csname LTb\endcsname%%
      \put(5849,742){\makebox(0,0)[r]{\strut{}$\beta=1.1$}}%
    }%
    \gplbacktext
    \put(0,0){\includegraphics{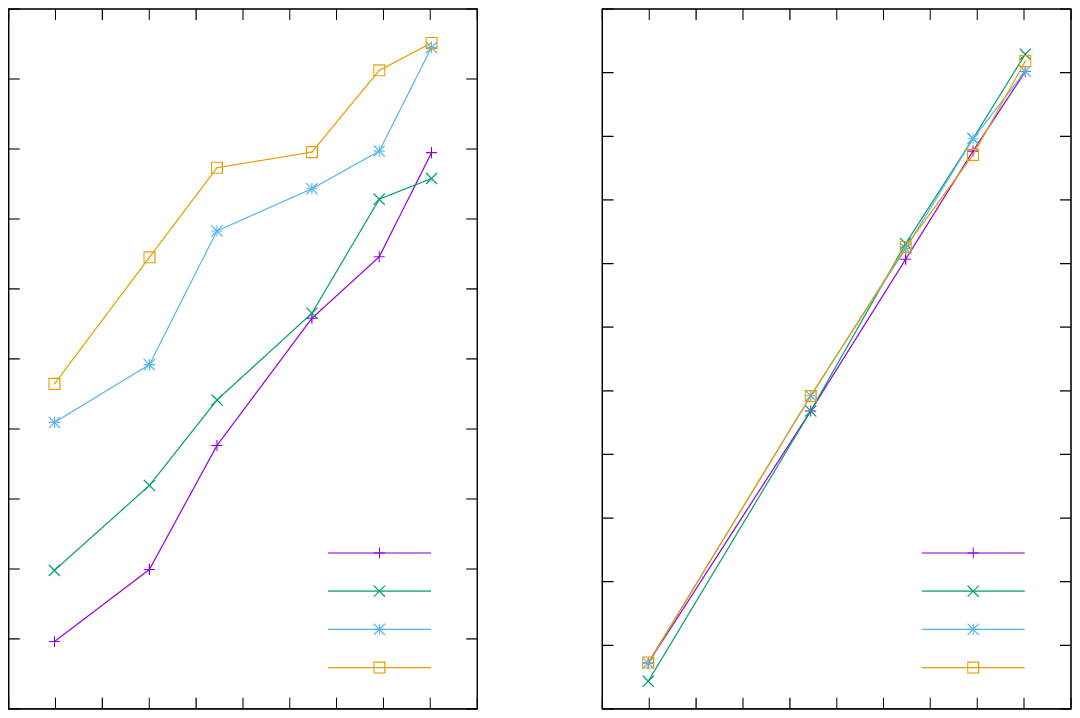}}%
    \gplfronttext
  \end{picture}%
\endgroup

%% file: eps-RR.tex
% GNUPLOT: LaTeX picture with Postscript
\begingroup
  \makeatletter
  \providecommand\color[2][]{%
    \GenericError{(gnuplot) \space\space\space\@spaces}{%
      Package color not loaded in conjunction with
      terminal option `colourtext'%
    }{See the gnuplot documentation for explanation.%
    }{Either use 'blacktext' in gnuplot or load the package
      color.sty in LaTeX.}%
    \renewcommand\color[2][]{}%
  }%
  \providecommand\includegraphics[2][]{%
    \GenericError{(gnuplot) \space\space\space\@spaces}{%
      Package graphicx or graphics not loaded%
    }{See the gnuplot documentation for explanation.%
    }{The gnuplot epslatex terminal needs graphicx.sty or graphics.sty.}%
    \renewcommand\includegraphics[2][]{}%
  }%
  \providecommand\rotatebox[2]{#2}%
  \@ifundefined{ifGPcolor}{%
    \newif\ifGPcolor
    \GPcolortrue
  }{}%
  \@ifundefined{ifGPblacktext}{%
    \newif\ifGPblacktext
    \GPblacktexttrue
  }{}%
  % define a \g@addto@macro without @ in the name:
  \let\gplgaddtomacro\g@addto@macro
  % define empty templates for all commands taking text:
  \gdef\gplbacktext{}%
  \gdef\gplfronttext{}%
  \makeatother
  \ifGPblacktext
    % no textcolor at all
    \def\colorrgb#1{}%
    \def\colorgray#1{}%
  \else
    % gray or color?
    \ifGPcolor
      \def\colorrgb#1{\color[rgb]{#1}}%
      \def\colorgray#1{\color[gray]{#1}}%
      \expandafter\def\csname LTw\endcsname{\color{white}}%
      \expandafter\def\csname LTb\endcsname{\color{black}}%
      \expandafter\def\csname LTa\endcsname{\color{black}}%
      \expandafter\def\csname LT0\endcsname{\color[rgb]{1,0,0}}%
      \expandafter\def\csname LT1\endcsname{\color[rgb]{0,1,0}}%
      \expandafter\def\csname LT2\endcsname{\color[rgb]{0,0,1}}%
      \expandafter\def\csname LT3\endcsname{\color[rgb]{1,0,1}}%
      \expandafter\def\csname LT4\endcsname{\color[rgb]{0,1,1}}%
      \expandafter\def\csname LT5\endcsname{\color[rgb]{1,1,0}}%
      \expandafter\def\csname LT6\endcsname{\color[rgb]{0,0,0}}%
      \expandafter\def\csname LT7\endcsname{\color[rgb]{1,0.3,0}}%
      \expandafter\def\csname LT8\endcsname{\color[rgb]{0.5,0.5,0.5}}%
    \else
      % gray
      \def\colorrgb#1{\color{black}}%
      \def\colorgray#1{\color[gray]{#1}}%
      \expandafter\def\csname LTw\endcsname{\color{white}}%
      \expandafter\def\csname LTb\endcsname{\color{black}}%
      \expandafter\def\csname LTa\endcsname{\color{black}}%
      \expandafter\def\csname LT0\endcsname{\color{black}}%
      \expandafter\def\csname LT1\endcsname{\color{black}}%
      \expandafter\def\csname LT2\endcsname{\color{black}}%
      \expandafter\def\csname LT3\endcsname{\color{black}}%
      \expandafter\def\csname LT4\endcsname{\color{black}}%
      \expandafter\def\csname LT5\endcsname{\color{black}}%
      \expandafter\def\csname LT6\endcsname{\color{black}}%
      \expandafter\def\csname LT7\endcsname{\color{black}}%
      \expandafter\def\csname LT8\endcsname{\color{black}}%
    \fi
  \fi
    \setlength{\unitlength}{0.0500bp}%
    \ifx\gptboxheight\undefined%
      \newlength{\gptboxheight}%
      \newlength{\gptboxwidth}%
      \newsavebox{\gptboxtext}%
    \fi%
    \setlength{\fboxrule}{0.5pt}%
    \setlength{\fboxsep}{1pt}%
\begin{picture}(7200.00,5040.00)%
    \gplgaddtomacro\gplbacktext{%
      \csname LTb\endcsname%%
      \put(588,504){\makebox(0,0)[r]{\strut{}$0$}}%
      \put(588,1310){\makebox(0,0)[r]{\strut{}$0.2$}}%
      \put(588,2116){\makebox(0,0)[r]{\strut{}$0.4$}}%
      \put(588,2923){\makebox(0,0)[r]{\strut{}$0.6$}}%
      \put(588,3729){\makebox(0,0)[r]{\strut{}$0.8$}}%
      \put(588,4535){\makebox(0,0)[r]{\strut{}$1$}}%
      \put(720,284){\makebox(0,0){\strut{}$20$}}%
      \put(1057,284){\makebox(0,0){\strut{}$30$}}%
      \put(1395,284){\makebox(0,0){\strut{}$40$}}%
      \put(1732,284){\makebox(0,0){\strut{}$50$}}%
      \put(2070,284){\makebox(0,0){\strut{}$60$}}%
      \put(2407,284){\makebox(0,0){\strut{}$70$}}%
      \put(2744,284){\makebox(0,0){\strut{}$80$}}%
      \put(3082,284){\makebox(0,0){\strut{}$90$}}%
      \put(3419,284){\makebox(0,0){\strut{}$100$}}%
      \put(1980,4787){\makebox(0,0)[l]{\strut{}\text{(a)}}}%
      \put(5399,4787){\makebox(0,0)[l]{\strut{}\text{(b)}}}%
    }%
    \gplgaddtomacro\gplfronttext{%
      \csname LTb\endcsname%%
      \put(-28,2519){\rotatebox{-270}{\makebox(0,0){\strut{}$\epsilon$}}}%
      \put(2069,-46){\makebox(0,0){\strut{}$N$}}%
      \csname LTb\endcsname%%
      \put(1484,4183){\makebox(0,0)[r]{\strut{}$\beta=0.3$}}%
      \csname LTb\endcsname%%
      \put(1484,3963){\makebox(0,0)[r]{\strut{}$\beta=0.5$}}%
      \csname LTb\endcsname%%
      \put(1484,3743){\makebox(0,0)[r]{\strut{}$\beta=0.9$}}%
      \csname LTb\endcsname%%
      \put(1484,3523){\makebox(0,0)[r]{\strut{}$\beta=1.1$}}%
    }%
    \gplgaddtomacro\gplbacktext{%
      \csname LTb\endcsname%%
      \put(4008,504){\makebox(0,0)[r]{\strut{}$0$}}%
      \put(4008,1310){\makebox(0,0)[r]{\strut{}$0.2$}}%
      \put(4008,2116){\makebox(0,0)[r]{\strut{}$0.4$}}%
      \put(4008,2923){\makebox(0,0)[r]{\strut{}$0.6$}}%
      \put(4008,3729){\makebox(0,0)[r]{\strut{}$0.8$}}%
      \put(4008,4535){\makebox(0,0)[r]{\strut{}$1$}}%
      \put(4140,284){\makebox(0,0){\strut{}$20$}}%
      \put(4477,284){\makebox(0,0){\strut{}$30$}}%
      \put(4815,284){\makebox(0,0){\strut{}$40$}}%
      \put(5152,284){\makebox(0,0){\strut{}$50$}}%
      \put(5490,284){\makebox(0,0){\strut{}$60$}}%
      \put(5827,284){\makebox(0,0){\strut{}$70$}}%
      \put(6164,284){\makebox(0,0){\strut{}$80$}}%
      \put(6502,284){\makebox(0,0){\strut{}$90$}}%
      \put(6839,284){\makebox(0,0){\strut{}$100$}}%
      \put(1980,4787){\makebox(0,0)[l]{\strut{}\text{(a)}}}%
      \put(5399,4787){\makebox(0,0)[l]{\strut{}\text{(b)}}}%
    }%
    \gplgaddtomacro\gplfronttext{%
      \csname LTb\endcsname%%
      \put(5489,-46){\makebox(0,0){\strut{}$N$}}%
      \csname LTb\endcsname%%
      \put(4904,4183){\makebox(0,0)[r]{\strut{}$\beta=0.3$}}%
      \csname LTb\endcsname%%
      \put(4904,3963){\makebox(0,0)[r]{\strut{}$\beta=0.5$}}%
      \csname LTb\endcsname%%
      \put(4904,3743){\makebox(0,0)[r]{\strut{}$\beta=0.9$}}%
      \csname LTb\endcsname%%
      \put(4904,3523){\makebox(0,0)[r]{\strut{}$\beta=1.1$}}%
    }%
    \gplbacktext
    \put(0,0){\includegraphics{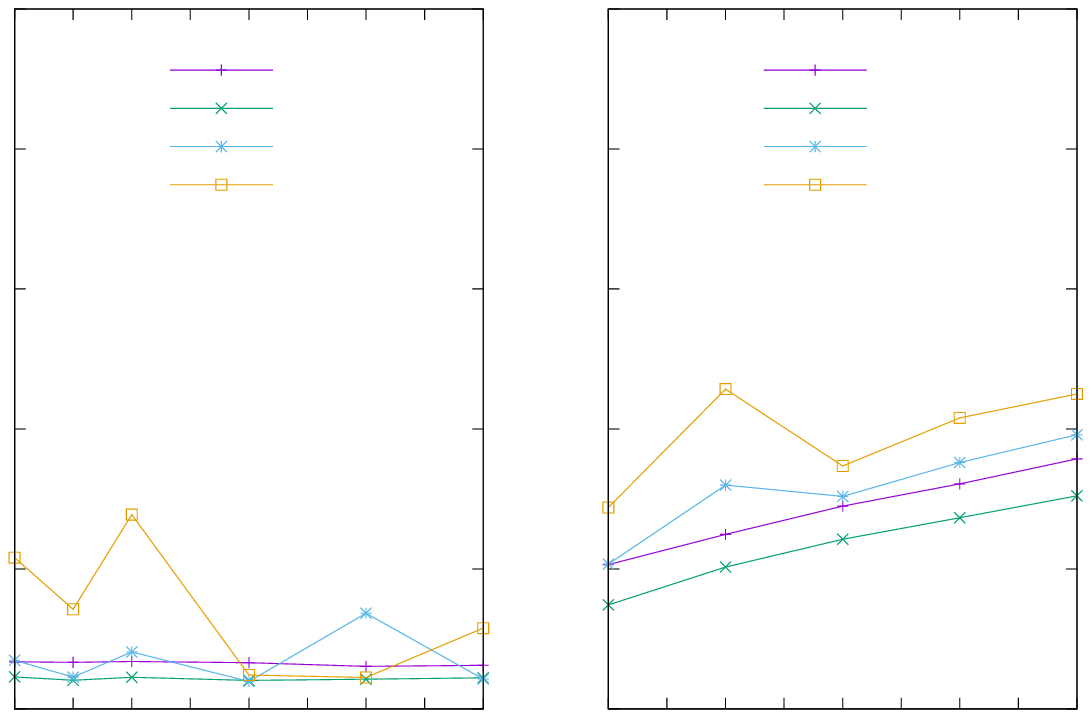}}%
    \gplfronttext
  \end{picture}%
\endgroup

%% file: Decimation-36-2D-FB.tex
% GNUPLOT: LaTeX picture with Postscript
\begingroup
  \makeatletter
  \providecommand\color[2][]{%
    \GenericError{(gnuplot) \space\space\space\@spaces}{%
      Package color not loaded in conjunction with
      terminal option `colourtext'%
    }{See the gnuplot documentation for explanation.%
    }{Either use 'blacktext' in gnuplot or load the package
      color.sty in LaTeX.}%
    \renewcommand\color[2][]{}%
  }%
  \providecommand\includegraphics[2][]{%
    \GenericError{(gnuplot) \space\space\space\@spaces}{%
      Package graphicx or graphics not loaded%
    }{See the gnuplot documentation for explanation.%
    }{The gnuplot epslatex terminal needs graphicx.sty or graphics.sty.}%
    \renewcommand\includegraphics[2][]{}%
  }%
  \providecommand\rotatebox[2]{#2}%
  \@ifundefined{ifGPcolor}{%
    \newif\ifGPcolor
    \GPcolortrue
  }{}%
  \@ifundefined{ifGPblacktext}{%
    \newif\ifGPblacktext
    \GPblacktexttrue
  }{}%
  % define a \g@addto@macro without @ in the name:
  \let\gplgaddtomacro\g@addto@macro
  % define empty templates for all commands taking text:
  \gdef\gplbacktext{}%
  \gdef\gplfronttext{}%
  \makeatother
  \ifGPblacktext
    % no textcolor at all
    \def\colorrgb#1{}%
    \def\colorgray#1{}%
  \else
    % gray or color?
    \ifGPcolor
      \def\colorrgb#1{\color[rgb]{#1}}%
      \def\colorgray#1{\color[gray]{#1}}%
      \expandafter\def\csname LTw\endcsname{\color{white}}%
      \expandafter\def\csname LTb\endcsname{\color{black}}%
      \expandafter\def\csname LTa\endcsname{\color{black}}%
      \expandafter\def\csname LT0\endcsname{\color[rgb]{1,0,0}}%
      \expandafter\def\csname LT1\endcsname{\color[rgb]{0,1,0}}%
      \expandafter\def\csname LT2\endcsname{\color[rgb]{0,0,1}}%
      \expandafter\def\csname LT3\endcsname{\color[rgb]{1,0,1}}%
      \expandafter\def\csname LT4\endcsname{\color[rgb]{0,1,1}}%
      \expandafter\def\csname LT5\endcsname{\color[rgb]{1,1,0}}%
      \expandafter\def\csname LT6\endcsname{\color[rgb]{0,0,0}}%
      \expandafter\def\csname LT7\endcsname{\color[rgb]{1,0.3,0}}%
      \expandafter\def\csname LT8\endcsname{\color[rgb]{0.5,0.5,0.5}}%
    \else
      % gray
      \def\colorrgb#1{\color{black}}%
      \def\colorgray#1{\color[gray]{#1}}%
      \expandafter\def\csname LTw\endcsname{\color{white}}%
      \expandafter\def\csname LTb\endcsname{\color{black}}%
      \expandafter\def\csname LTa\endcsname{\color{black}}%
      \expandafter\def\csname LT0\endcsname{\color{black}}%
      \expandafter\def\csname LT1\endcsname{\color{black}}%
      \expandafter\def\csname LT2\endcsname{\color{black}}%
      \expandafter\def\csname LT3\endcsname{\color{black}}%
      \expandafter\def\csname LT4\endcsname{\color{black}}%
      \expandafter\def\csname LT5\endcsname{\color{black}}%
      \expandafter\def\csname LT6\endcsname{\color{black}}%
      \expandafter\def\csname LT7\endcsname{\color{black}}%
      \expandafter\def\csname LT8\endcsname{\color{black}}%
    \fi
  \fi
    \setlength{\unitlength}{0.0500bp}%
    \ifx\gptboxheight\undefined%
      \newlength{\gptboxheight}%
      \newlength{\gptboxwidth}%
      \newsavebox{\gptboxtext}%
    \fi%
    \setlength{\fboxrule}{0.5pt}%
    \setlength{\fboxsep}{1pt}%
\begin{picture}(7200.00,5040.00)%
    \gplgaddtomacro\gplbacktext{%
      \csname LTb\endcsname%%
      \put(462,704){\makebox(0,0)[r]{\strut{}$0$}}%
      \put(462,1495){\makebox(0,0)[r]{\strut{}$5$}}%
      \put(462,2287){\makebox(0,0)[r]{\strut{}$10$}}%
      \put(462,3078){\makebox(0,0)[r]{\strut{}$15$}}%
      \put(462,3869){\makebox(0,0)[r]{\strut{}$20$}}%
      \put(462,4661){\makebox(0,0)[r]{\strut{}$25$}}%
      \put(594,484){\makebox(0,0){\strut{}$0$}}%
      \put(1629,484){\makebox(0,0){\strut{}$0.05$}}%
      \put(2664,484){\makebox(0,0){\strut{}$0.1$}}%
      \put(3699,484){\makebox(0,0){\strut{}$0.15$}}%
      \put(4733,484){\makebox(0,0){\strut{}$0.2$}}%
      \put(5768,484){\makebox(0,0){\strut{}$0.25$}}%
      \put(6803,484){\makebox(0,0){\strut{}$0.3$}}%
    }%
    \gplgaddtomacro\gplfronttext{%
      \csname LTb\endcsname%%
      \put(3698,154){\makebox(0,0){\strut{}$x$}}%
      \csname LTb\endcsname%%
      \put(5327,1829){\makebox(0,0)[r]{\strut{}$\mathcal{S}_t(x)$}}%
      \csname LTb\endcsname%%
      \put(5327,1609){\makebox(0,0)[r]{\strut{}$10\epsilon(x)$}}%
    }%
    \gplgaddtomacro\gplbacktext{%
      \csname LTb\endcsname%%
      \put(3550,3173){\makebox(0,0)[r]{\strut{}$0.5$}}%
      \put(3550,3391){\makebox(0,0)[r]{\strut{}$0.6$}}%
      \put(3550,3610){\makebox(0,0)[r]{\strut{}$0.7$}}%
      \put(3550,3828){\makebox(0,0)[r]{\strut{}$0.8$}}%
      \put(3550,4047){\makebox(0,0)[r]{\strut{}$0.9$}}%
      \put(3550,4265){\makebox(0,0)[r]{\strut{}$1$}}%
      \put(3682,2953){\makebox(0,0){\strut{}$0$}}%
      \put(4133,2953){\makebox(0,0){\strut{}$0.2$}}%
      \put(4585,2953){\makebox(0,0){\strut{}$0.4$}}%
      \put(5036,2953){\makebox(0,0){\strut{}$0.6$}}%
      \put(5488,2953){\makebox(0,0){\strut{}$0.8$}}%
      \put(5939,2953){\makebox(0,0){\strut{}$1$}}%
    }%
    \gplgaddtomacro\gplfronttext{%
      \csname LTb\endcsname%%
      \put(2934,3719){\rotatebox{-270}{\makebox(0,0){\strut{}$\text{TPR}$}}}%
      \put(4810,2623){\makebox(0,0){\strut{}$\text{TNR}$}}%
      \csname LTb\endcsname%%
      \put(4633,3937){\makebox(0,0)[r]{\strut{}$\text{ROC}$}}%
      \csname LTb\endcsname%%
      \put(4633,3717){\makebox(0,0)[r]{\strut{}$\max[\mathcal{S}_t(x)]$}}%
      \csname LTb\endcsname%%
      \put(4633,3497){\makebox(0,0)[r]{\strut{}$\min[\epsilon(x)]$}}%
    }%
    \gplbacktext
    \put(0,0){\includegraphics{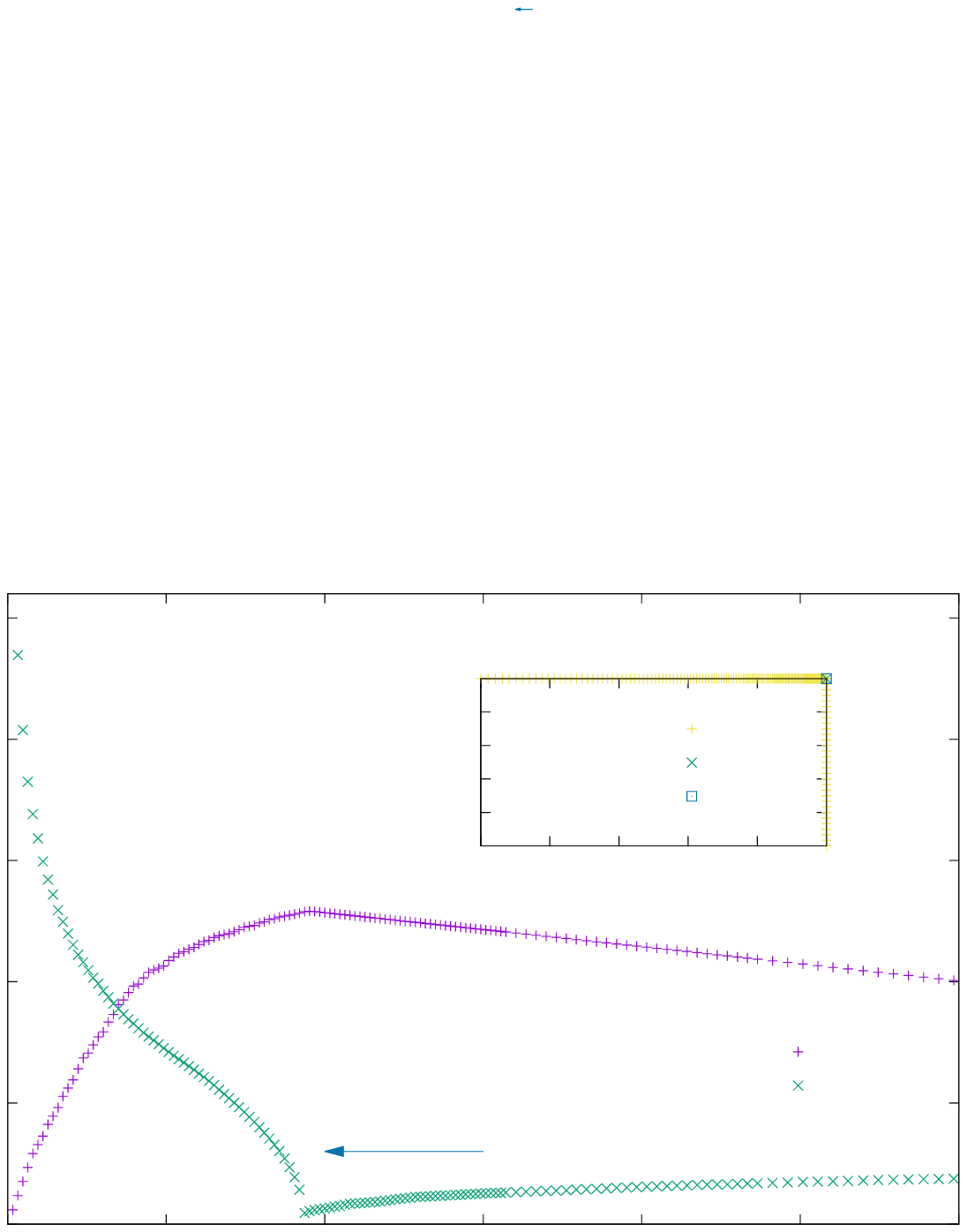}}%
    \gplfronttext
  \end{picture}%
\endgroup

%% file: Av-49-2D-FB_05.tex
% GNUPLOT: LaTeX picture with Postscript
\begingroup
  \makeatletter
  \providecommand\color[2][]{%
    \GenericError{(gnuplot) \space\space\space\@spaces}{%
      Package color not loaded in conjunction with
      terminal option `colourtext'%
    }{See the gnuplot documentation for explanation.%
    }{Either use 'blacktext' in gnuplot or load the package
      color.sty in LaTeX.}%
    \renewcommand\color[2][]{}%
  }%
  \providecommand\includegraphics[2][]{%
    \GenericError{(gnuplot) \space\space\space\@spaces}{%
      Package graphicx or graphics not loaded%
    }{See the gnuplot documentation for explanation.%
    }{The gnuplot epslatex terminal needs graphicx.sty or graphics.sty.}%
    \renewcommand\includegraphics[2][]{}%
  }%
  \providecommand\rotatebox[2]{#2}%
  \@ifundefined{ifGPcolor}{%
    \newif\ifGPcolor
    \GPcolortrue
  }{}%
  \@ifundefined{ifGPblacktext}{%
    \newif\ifGPblacktext
    \GPblacktexttrue
  }{}%
  % define a \g@addto@macro without @ in the name:
  \let\gplgaddtomacro\g@addto@macro
  % define empty templates for all commands taking text:
  \gdef\gplbacktext{}%
  \gdef\gplfronttext{}%
  \makeatother
  \ifGPblacktext
    % no textcolor at all
    \def\colorrgb#1{}%
    \def\colorgray#1{}%
  \else
    % gray or color?
    \ifGPcolor
      \def\colorrgb#1{\color[rgb]{#1}}%
      \def\colorgray#1{\color[gray]{#1}}%
      \expandafter\def\csname LTw\endcsname{\color{white}}%
      \expandafter\def\csname LTb\endcsname{\color{black}}%
      \expandafter\def\csname LTa\endcsname{\color{black}}%
      \expandafter\def\csname LT0\endcsname{\color[rgb]{1,0,0}}%
      \expandafter\def\csname LT1\endcsname{\color[rgb]{0,1,0}}%
      \expandafter\def\csname LT2\endcsname{\color[rgb]{0,0,1}}%
      \expandafter\def\csname LT3\endcsname{\color[rgb]{1,0,1}}%
      \expandafter\def\csname LT4\endcsname{\color[rgb]{0,1,1}}%
      \expandafter\def\csname LT5\endcsname{\color[rgb]{1,1,0}}%
      \expandafter\def\csname LT6\endcsname{\color[rgb]{0,0,0}}%
      \expandafter\def\csname LT7\endcsname{\color[rgb]{1,0.3,0}}%
      \expandafter\def\csname LT8\endcsname{\color[rgb]{0.5,0.5,0.5}}%
    \else
      % gray
      \def\colorrgb#1{\color{black}}%
      \def\colorgray#1{\color[gray]{#1}}%
      \expandafter\def\csname LTw\endcsname{\color{white}}%
      \expandafter\def\csname LTb\endcsname{\color{black}}%
      \expandafter\def\csname LTa\endcsname{\color{black}}%
      \expandafter\def\csname LT0\endcsname{\color{black}}%
      \expandafter\def\csname LT1\endcsname{\color{black}}%
      \expandafter\def\csname LT2\endcsname{\color{black}}%
      \expandafter\def\csname LT3\endcsname{\color{black}}%
      \expandafter\def\csname LT4\endcsname{\color{black}}%
      \expandafter\def\csname LT5\endcsname{\color{black}}%
      \expandafter\def\csname LT6\endcsname{\color{black}}%
      \expandafter\def\csname LT7\endcsname{\color{black}}%
      \expandafter\def\csname LT8\endcsname{\color{black}}%
    \fi
  \fi
    \setlength{\unitlength}{0.0500bp}%
    \ifx\gptboxheight\undefined%
      \newlength{\gptboxheight}%
      \newlength{\gptboxwidth}%
      \newsavebox{\gptboxtext}%
    \fi%
    \setlength{\fboxrule}{0.5pt}%
    \setlength{\fboxsep}{1pt}%
\begin{picture}(7200.00,5040.00)%
    \gplgaddtomacro\gplbacktext{%
      \csname LTb\endcsname%%
      \put(858,1488){\makebox(0,0)[r]{\strut{}$-18$}}%
      \put(858,2468){\makebox(0,0)[r]{\strut{}$-17.5$}}%
      \put(858,3447){\makebox(0,0)[r]{\strut{}$-17$}}%
      \put(858,4427){\makebox(0,0)[r]{\strut{}$-16.5$}}%
      \put(990,484){\makebox(0,0){\strut{}$70$}}%
      \put(1959,484){\makebox(0,0){\strut{}$80$}}%
      \put(2928,484){\makebox(0,0){\strut{}$90$}}%
      \put(3897,484){\makebox(0,0){\strut{}$100$}}%
      \put(4865,484){\makebox(0,0){\strut{}$110$}}%
      \put(5834,484){\makebox(0,0){\strut{}$120$}}%
      \put(6803,484){\makebox(0,0){\strut{}$130$}}%
    }%
    \gplgaddtomacro\gplfronttext{%
      \csname LTb\endcsname%%
      \put(3896,154){\makebox(0,0){\strut{}$t$}}%
      \csname LTb\endcsname%%
      \put(5367,3475){\makebox(0,0)[r]{\strut{}$\text{BIC}(t)/M$”}}%
    }%
    \gplgaddtomacro\gplbacktext{%
      \csname LTb\endcsname%%
      \put(3114,1655){\makebox(0,0)[r]{\strut{}$0$}}%
      \put(3114,1945){\makebox(0,0)[r]{\strut{}$0.1$}}%
      \put(3114,2235){\makebox(0,0)[r]{\strut{}$0.2$}}%
      \put(3114,2526){\makebox(0,0)[r]{\strut{}$0.3$}}%
      \put(3114,2816){\makebox(0,0)[r]{\strut{}$0.4$}}%
      \put(3114,3106){\makebox(0,0)[r]{\strut{}$0.5$}}%
      \put(3246,1290){\makebox(0,0){\strut{}$70$}}%
      \put(3779,1290){\makebox(0,0){\strut{}$80$}}%
      \put(4312,1290){\makebox(0,0){\strut{}$90$}}%
      \put(4845,1290){\makebox(0,0){\strut{}$100$}}%
      \put(5377,1290){\makebox(0,0){\strut{}$110$}}%
      \put(5910,1290){\makebox(0,0){\strut{}$120$}}%
      \put(6443,1290){\makebox(0,0){\strut{}$130$}}%
    }%
    \gplgaddtomacro\gplfronttext{%
      \csname LTb\endcsname%%
      \put(4844,960){\makebox(0,0){\strut{}$t$}}%
      \csname LTb\endcsname%%
      \put(4949,2677){\makebox(0,0)[r]{\strut{}$\epsilon(t)$}}%
    }%
    \gplbacktext
    \put(0,0){\includegraphics{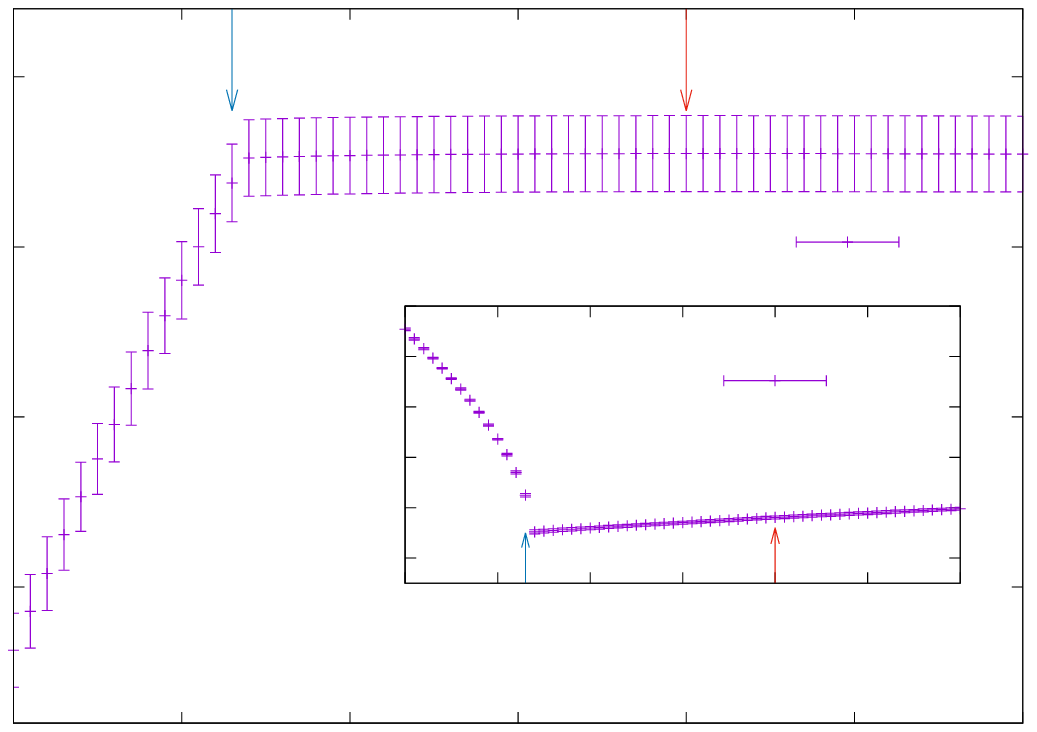}}%
    \gplfronttext
  \end{picture}%
\endgroup

%% file: Av-100-RR-05.tex
% GNUPLOT: LaTeX picture with Postscript
\begingroup
  \makeatletter
  \providecommand\color[2][]{%
    \GenericError{(gnuplot) \space\space\space\@spaces}{%
      Package color not loaded in conjunction with
      terminal option `colourtext'%
    }{See the gnuplot documentation for explanation.%
    }{Either use 'blacktext' in gnuplot or load the package
      color.sty in LaTeX.}%
    \renewcommand\color[2][]{}%
  }%
  \providecommand\includegraphics[2][]{%
    \GenericError{(gnuplot) \space\space\space\@spaces}{%
      Package graphicx or graphics not loaded%
    }{See the gnuplot documentation for explanation.%
    }{The gnuplot epslatex terminal needs graphicx.sty or graphics.sty.}%
    \renewcommand\includegraphics[2][]{}%
  }%
  \providecommand\rotatebox[2]{#2}%
  \@ifundefined{ifGPcolor}{%
    \newif\ifGPcolor
    \GPcolortrue
  }{}%
  \@ifundefined{ifGPblacktext}{%
    \newif\ifGPblacktext
    \GPblacktexttrue
  }{}%
  % define a \g@addto@macro without @ in the name:
  \let\gplgaddtomacro\g@addto@macro
  % define empty templates for all commands taking text:
  \gdef\gplbacktext{}%
  \gdef\gplfronttext{}%
  \makeatother
  \ifGPblacktext
    % no textcolor at all
    \def\colorrgb#1{}%
    \def\colorgray#1{}%
  \else
    % gray or color?
    \ifGPcolor
      \def\colorrgb#1{\color[rgb]{#1}}%
      \def\colorgray#1{\color[gray]{#1}}%
      \expandafter\def\csname LTw\endcsname{\color{white}}%
      \expandafter\def\csname LTb\endcsname{\color{black}}%
      \expandafter\def\csname LTa\endcsname{\color{black}}%
      \expandafter\def\csname LT0\endcsname{\color[rgb]{1,0,0}}%
      \expandafter\def\csname LT1\endcsname{\color[rgb]{0,1,0}}%
      \expandafter\def\csname LT2\endcsname{\color[rgb]{0,0,1}}%
      \expandafter\def\csname LT3\endcsname{\color[rgb]{1,0,1}}%
      \expandafter\def\csname LT4\endcsname{\color[rgb]{0,1,1}}%
      \expandafter\def\csname LT5\endcsname{\color[rgb]{1,1,0}}%
      \expandafter\def\csname LT6\endcsname{\color[rgb]{0,0,0}}%
      \expandafter\def\csname LT7\endcsname{\color[rgb]{1,0.3,0}}%
      \expandafter\def\csname LT8\endcsname{\color[rgb]{0.5,0.5,0.5}}%
    \else
      % gray
      \def\colorrgb#1{\color{black}}%
      \def\colorgray#1{\color[gray]{#1}}%
      \expandafter\def\csname LTw\endcsname{\color{white}}%
      \expandafter\def\csname LTb\endcsname{\color{black}}%
      \expandafter\def\csname LTa\endcsname{\color{black}}%
      \expandafter\def\csname LT0\endcsname{\color{black}}%
      \expandafter\def\csname LT1\endcsname{\color{black}}%
      \expandafter\def\csname LT2\endcsname{\color{black}}%
      \expandafter\def\csname LT3\endcsname{\color{black}}%
      \expandafter\def\csname LT4\endcsname{\color{black}}%
      \expandafter\def\csname LT5\endcsname{\color{black}}%
      \expandafter\def\csname LT6\endcsname{\color{black}}%
      \expandafter\def\csname LT7\endcsname{\color{black}}%
      \expandafter\def\csname LT8\endcsname{\color{black}}%
    \fi
  \fi
    \setlength{\unitlength}{0.0500bp}%
    \ifx\gptboxheight\undefined%
      \newlength{\gptboxheight}%
      \newlength{\gptboxwidth}%
      \newsavebox{\gptboxtext}%
    \fi%
    \setlength{\fboxrule}{0.5pt}%
    \setlength{\fboxsep}{1pt}%
\begin{picture}(7200.00,5040.00)%
    \gplgaddtomacro\gplbacktext{%
      \csname LTb\endcsname%%
      \put(858,704){\makebox(0,0)[r]{\strut{}$-20$}}%
      \put(858,1439){\makebox(0,0)[r]{\strut{}$-19.5$}}%
      \put(858,2174){\makebox(0,0)[r]{\strut{}$-19$}}%
      \put(858,2908){\makebox(0,0)[r]{\strut{}$-18.5$}}%
      \put(858,3643){\makebox(0,0)[r]{\strut{}$-18$}}%
      \put(858,4378){\makebox(0,0)[r]{\strut{}$-17.5$}}%
      \put(1571,484){\makebox(0,0){\strut{}$180$}}%
      \put(2734,484){\makebox(0,0){\strut{}$200$}}%
      \put(3897,484){\makebox(0,0){\strut{}$220$}}%
      \put(5059,484){\makebox(0,0){\strut{}$240$}}%
      \put(6222,484){\makebox(0,0){\strut{}$260$}}%
    }%
    \gplgaddtomacro\gplfronttext{%
      \csname LTb\endcsname%%
      \put(3896,154){\makebox(0,0){\strut{}$t$}}%
      \csname LTb\endcsname%%
      \put(5367,3475){\makebox(0,0)[r]{\strut{}$\text{BIC}(t)/M$}}%
    }%
    \gplgaddtomacro\gplbacktext{%
      \csname LTb\endcsname%%
      \put(3114,1655){\makebox(0,0)[r]{\strut{}$0$}}%
      \put(3114,1945){\makebox(0,0)[r]{\strut{}$0.1$}}%
      \put(3114,2235){\makebox(0,0)[r]{\strut{}$0.2$}}%
      \put(3114,2526){\makebox(0,0)[r]{\strut{}$0.3$}}%
      \put(3114,2816){\makebox(0,0)[r]{\strut{}$0.4$}}%
      \put(3114,3106){\makebox(0,0)[r]{\strut{}$0.5$}}%
      \put(3566,1290){\makebox(0,0){\strut{}$180$}}%
      \put(4205,1290){\makebox(0,0){\strut{}$200$}}%
      \put(4845,1290){\makebox(0,0){\strut{}$220$}}%
      \put(5484,1290){\makebox(0,0){\strut{}$240$}}%
      \put(6123,1290){\makebox(0,0){\strut{}$260$}}%
    }%
    \gplgaddtomacro\gplfronttext{%
      \csname LTb\endcsname%%
      \put(4844,960){\makebox(0,0){\strut{}$t$}}%
      \csname LTb\endcsname%%
      \put(4949,2677){\makebox(0,0)[r]{\strut{}$\epsilon(t)$}}%
    }%
    \gplbacktext
    \put(0,0){\includegraphics{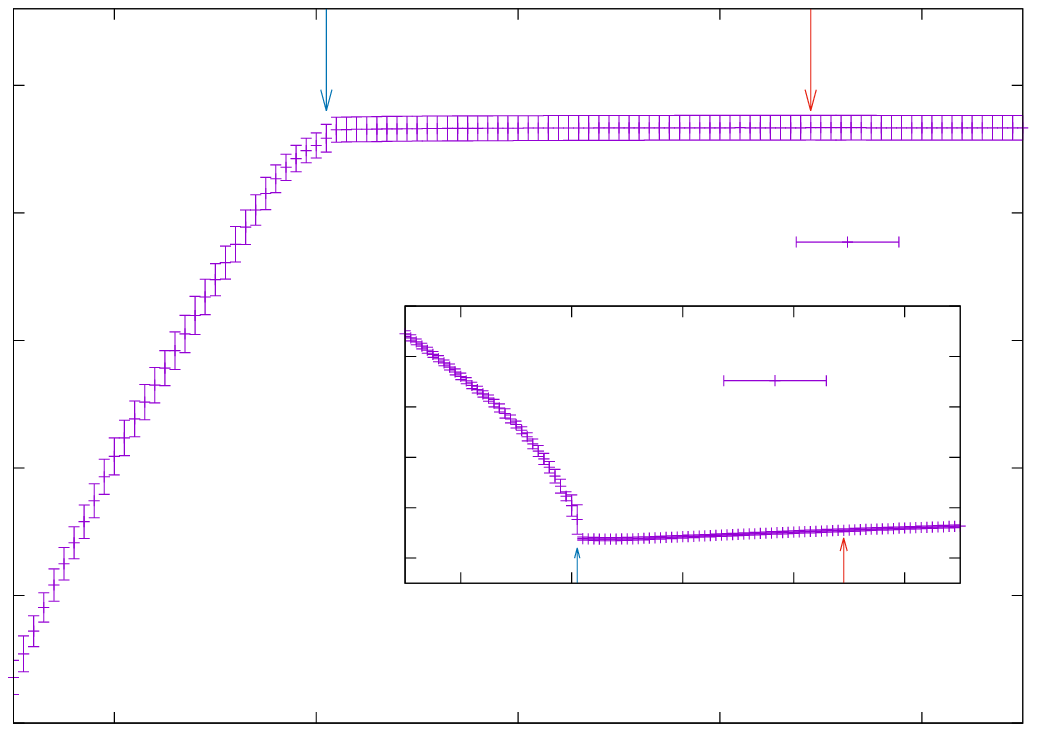}}%
    \gplfronttext
  \end{picture}%
\endgroup

%% file: ROCvsM-2D-PB.tex
% GNUPLOT: LaTeX picture with Postscript
\begingroup
  \makeatletter
  \providecommand\color[2][]{%
    \GenericError{(gnuplot) \space\space\space\@spaces}{%
      Package color not loaded in conjunction with
      terminal option `colourtext'%
    }{See the gnuplot documentation for explanation.%
    }{Either use 'blacktext' in gnuplot or load the package
      color.sty in LaTeX.}%
    \renewcommand\color[2][]{}%
  }%
  \providecommand\includegraphics[2][]{%
    \GenericError{(gnuplot) \space\space\space\@spaces}{%
      Package graphicx or graphics not loaded%
    }{See the gnuplot documentation for explanation.%
    }{The gnuplot epslatex terminal needs graphicx.sty or graphics.sty.}%
    \renewcommand\includegraphics[2][]{}%
  }%
  \providecommand\rotatebox[2]{#2}%
  \@ifundefined{ifGPcolor}{%
    \newif\ifGPcolor
    \GPcolortrue
  }{}%
  \@ifundefined{ifGPblacktext}{%
    \newif\ifGPblacktext
    \GPblacktexttrue
  }{}%
  % define a \g@addto@macro without @ in the name:
  \let\gplgaddtomacro\g@addto@macro
  % define empty templates for all commands taking text:
  \gdef\gplbacktext{}%
  \gdef\gplfronttext{}%
  \makeatother
  \ifGPblacktext
    % no textcolor at all
    \def\colorrgb#1{}%
    \def\colorgray#1{}%
  \else
    % gray or color?
    \ifGPcolor
      \def\colorrgb#1{\color[rgb]{#1}}%
      \def\colorgray#1{\color[gray]{#1}}%
      \expandafter\def\csname LTw\endcsname{\color{white}}%
      \expandafter\def\csname LTb\endcsname{\color{black}}%
      \expandafter\def\csname LTa\endcsname{\color{black}}%
      \expandafter\def\csname LT0\endcsname{\color[rgb]{1,0,0}}%
      \expandafter\def\csname LT1\endcsname{\color[rgb]{0,1,0}}%
      \expandafter\def\csname LT2\endcsname{\color[rgb]{0,0,1}}%
      \expandafter\def\csname LT3\endcsname{\color[rgb]{1,0,1}}%
      \expandafter\def\csname LT4\endcsname{\color[rgb]{0,1,1}}%
      \expandafter\def\csname LT5\endcsname{\color[rgb]{1,1,0}}%
      \expandafter\def\csname LT6\endcsname{\color[rgb]{0,0,0}}%
      \expandafter\def\csname LT7\endcsname{\color[rgb]{1,0.3,0}}%
      \expandafter\def\csname LT8\endcsname{\color[rgb]{0.5,0.5,0.5}}%
    \else
      % gray
      \def\colorrgb#1{\color{black}}%
      \def\colorgray#1{\color[gray]{#1}}%
      \expandafter\def\csname LTw\endcsname{\color{white}}%
      \expandafter\def\csname LTb\endcsname{\color{black}}%
      \expandafter\def\csname LTa\endcsname{\color{black}}%
      \expandafter\def\csname LT0\endcsname{\color{black}}%
      \expandafter\def\csname LT1\endcsname{\color{black}}%
      \expandafter\def\csname LT2\endcsname{\color{black}}%
      \expandafter\def\csname LT3\endcsname{\color{black}}%
      \expandafter\def\csname LT4\endcsname{\color{black}}%
      \expandafter\def\csname LT5\endcsname{\color{black}}%
      \expandafter\def\csname LT6\endcsname{\color{black}}%
      \expandafter\def\csname LT7\endcsname{\color{black}}%
      \expandafter\def\csname LT8\endcsname{\color{black}}%
    \fi
  \fi
    \setlength{\unitlength}{0.0500bp}%
    \ifx\gptboxheight\undefined%
      \newlength{\gptboxheight}%
      \newlength{\gptboxwidth}%
      \newsavebox{\gptboxtext}%
    \fi%
    \setlength{\fboxrule}{0.5pt}%
    \setlength{\fboxsep}{1pt}%
\begin{picture}(7200.00,5040.00)%
      \csname LTb\endcsname%%
      \put(3600,4820){\makebox(0,0){\strut{}}}%
    \gplgaddtomacro\gplbacktext{%
      \csname LTb\endcsname%%
      \put(588,3360){\makebox(0,0)[r]{\strut{}-8}}%
      \put(588,3654){\makebox(0,0)[r]{\strut{}-6}}%
      \put(588,3948){\makebox(0,0)[r]{\strut{}-4}}%
      \put(588,4241){\makebox(0,0)[r]{\strut{}-2}}%
      \put(588,4535){\makebox(0,0)[r]{\strut{}0}}%
      \put(1584,3140){\makebox(0,0){\strut{}}}%
      \put(2447,3140){\makebox(0,0){\strut{}}}%
      \put(3311,3140){\makebox(0,0){\strut{}}}%
      \put(2735,3478){\makebox(0,0)[l]{\strut{}$\beta=0.2$}}%
    }%
    \gplgaddtomacro\gplfronttext{%
    }%
    \gplgaddtomacro\gplbacktext{%
      \csname LTb\endcsname%%
      \put(588,1932){\makebox(0,0)[r]{\strut{}-8}}%
      \put(588,2226){\makebox(0,0)[r]{\strut{}-6}}%
      \put(588,2520){\makebox(0,0)[r]{\strut{}-4}}%
      \put(588,2813){\makebox(0,0)[r]{\strut{}-2}}%
      \put(588,3107){\makebox(0,0)[r]{\strut{}0}}%
      \put(720,1712){\makebox(0,0){\strut{}0}}%
      \put(1584,1712){\makebox(0,0){\strut{}1}}%
      \put(2447,1712){\makebox(0,0){\strut{}2}}%
      \put(3311,1712){\makebox(0,0){\strut{}3}}%
      \put(3596,1712){\makebox(0,0){\strut{}e-3}}%
      \put(2735,2050){\makebox(0,0)[l]{\strut{}$\beta=0.3$}}%
    }%
    \gplgaddtomacro\gplfronttext{%
      \csname LTb\endcsname%%
      \put(2159,1382){\makebox(0,0){\strut{}$1/M$}}%
    }%
    \gplgaddtomacro\gplbacktext{%
    }%
    \gplgaddtomacro\gplfronttext{%
      \csname LTb\endcsname%%
      \put(2326,1053){\makebox(0,0)[r]{\strut{}TNR N=36}}%
      \csname LTb\endcsname%%
      \put(2326,833){\makebox(0,0)[r]{\strut{}TNR N=49}}%
      \csname LTb\endcsname%%
      \put(2326,613){\makebox(0,0)[r]{\strut{}TNR N=64}}%
      \csname LTb\endcsname%%
      \put(2326,393){\makebox(0,0)[r]{\strut{}TNR N=81}}%
      \csname LTb\endcsname%%
      \put(2326,173){\makebox(0,0)[r]{\strut{}TNR N=100}}%
    }%
    \gplgaddtomacro\gplbacktext{%
      \csname LTb\endcsname%%
      \put(3827,3360){\makebox(0,0)[r]{\strut{} }}%
      \put(3827,3507){\makebox(0,0)[r]{\strut{} }}%
      \put(3827,3654){\makebox(0,0)[r]{\strut{} }}%
      \put(3827,3801){\makebox(0,0)[r]{\strut{} }}%
      \put(3827,3948){\makebox(0,0)[r]{\strut{} }}%
      \put(3827,4094){\makebox(0,0)[r]{\strut{} }}%
      \put(3827,4241){\makebox(0,0)[r]{\strut{} }}%
      \put(3827,4388){\makebox(0,0)[r]{\strut{} }}%
      \put(3827,4535){\makebox(0,0)[r]{\strut{} }}%
      \put(4823,3140){\makebox(0,0){\strut{}}}%
      \put(5687,3140){\makebox(0,0){\strut{}}}%
      \put(6551,3140){\makebox(0,0){\strut{}}}%
      \put(5975,3478){\makebox(0,0)[l]{\strut{}$\beta=0.5$}}%
    }%
    \gplgaddtomacro\gplfronttext{%
    }%
    \gplgaddtomacro\gplbacktext{%
      \csname LTb\endcsname%%
      \put(3827,1932){\makebox(0,0)[r]{\strut{} }}%
      \put(3827,2079){\makebox(0,0)[r]{\strut{} }}%
      \put(3827,2226){\makebox(0,0)[r]{\strut{} }}%
      \put(3827,2373){\makebox(0,0)[r]{\strut{} }}%
      \put(3827,2520){\makebox(0,0)[r]{\strut{} }}%
      \put(3827,2666){\makebox(0,0)[r]{\strut{} }}%
      \put(3827,2813){\makebox(0,0)[r]{\strut{} }}%
      \put(3827,2960){\makebox(0,0)[r]{\strut{} }}%
      \put(3827,3107){\makebox(0,0)[r]{\strut{} }}%
      \put(3959,1712){\makebox(0,0){\strut{}0}}%
      \put(4823,1712){\makebox(0,0){\strut{}1}}%
      \put(5687,1712){\makebox(0,0){\strut{}2}}%
      \put(6551,1712){\makebox(0,0){\strut{}3}}%
      \put(6836,1712){\makebox(0,0){\strut{}e-3}}%
      \put(5975,2050){\makebox(0,0)[l]{\strut{}$\beta=0.4$}}%
    }%
    \gplgaddtomacro\gplfronttext{%
      \csname LTb\endcsname%%
      \put(5399,1382){\makebox(0,0){\strut{}$1/M$}}%
    }%
    \gplgaddtomacro\gplbacktext{%
    }%
    \gplgaddtomacro\gplfronttext{%
      \csname LTb\endcsname%%
      \put(5565,1053){\makebox(0,0)[r]{\strut{}TPR N=36}}%
      \csname LTb\endcsname%%
      \put(5565,833){\makebox(0,0)[r]{\strut{}TPR N=49}}%
      \csname LTb\endcsname%%
      \put(5565,613){\makebox(0,0)[r]{\strut{}TPR N=64}}%
      \csname LTb\endcsname%%
      \put(5565,393){\makebox(0,0)[r]{\strut{}TPR N=81}}%
      \csname LTb\endcsname%%
      \put(5565,173){\makebox(0,0)[r]{\strut{}TPR N=100}}%
    }%
    \gplbacktext
    \put(0,0){\includegraphics{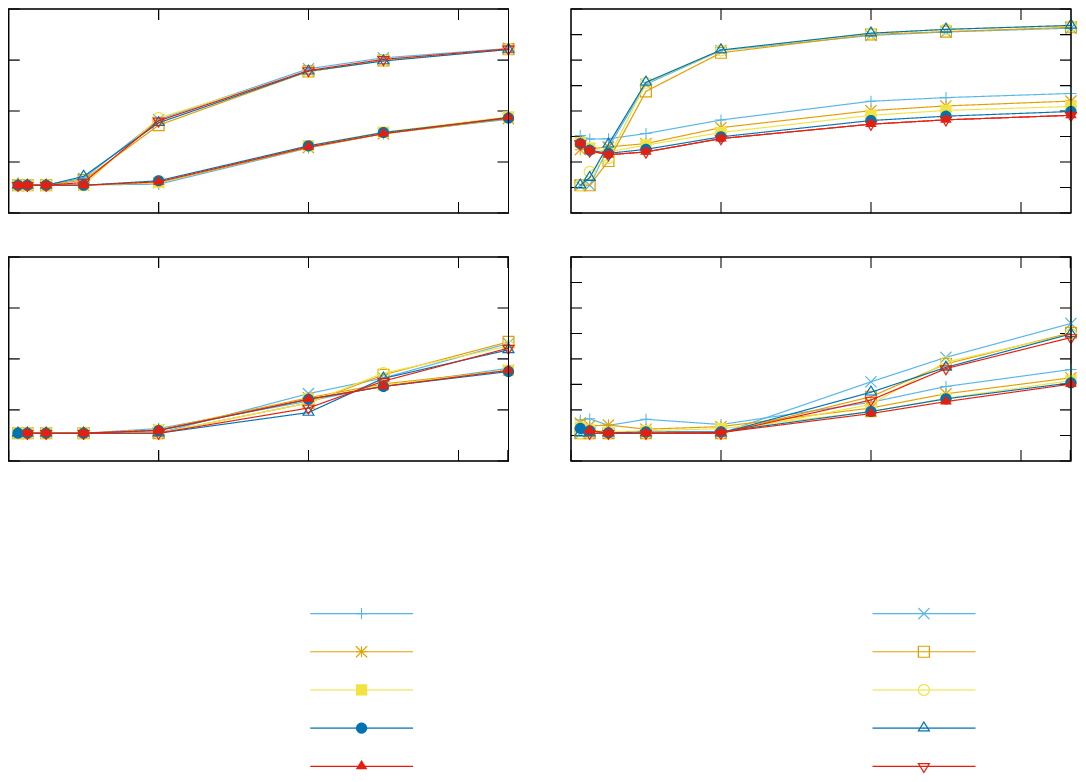}}%
    \gplfronttext
  \end{picture}%
\endgroup

%% file: ROCvsM-RR.tex
% GNUPLOT: LaTeX picture with Postscript
\begingroup
  \makeatletter
  \providecommand\color[2][]{%
    \GenericError{(gnuplot) \space\space\space\@spaces}{%
      Package color not loaded in conjunction with
      terminal option `colourtext'%
    }{See the gnuplot documentation for explanation.%
    }{Either use 'blacktext' in gnuplot or load the package
      color.sty in LaTeX.}%
    \renewcommand\color[2][]{}%
  }%
  \providecommand\includegraphics[2][]{%
    \GenericError{(gnuplot) \space\space\space\@spaces}{%
      Package graphicx or graphics not loaded%
    }{See the gnuplot documentation for explanation.%
    }{The gnuplot epslatex terminal needs graphicx.sty or graphics.sty.}%
    \renewcommand\includegraphics[2][]{}%
  }%
  \providecommand\rotatebox[2]{#2}%
  \@ifundefined{ifGPcolor}{%
    \newif\ifGPcolor
    \GPcolortrue
  }{}%
  \@ifundefined{ifGPblacktext}{%
    \newif\ifGPblacktext
    \GPblacktexttrue
  }{}%
  % define a \g@addto@macro without @ in the name:
  \let\gplgaddtomacro\g@addto@macro
  % define empty templates for all commands taking text:
  \gdef\gplbacktext{}%
  \gdef\gplfronttext{}%
  \makeatother
  \ifGPblacktext
    % no textcolor at all
    \def\colorrgb#1{}%
    \def\colorgray#1{}%
  \else
    % gray or color?
    \ifGPcolor
      \def\colorrgb#1{\color[rgb]{#1}}%
      \def\colorgray#1{\color[gray]{#1}}%
      \expandafter\def\csname LTw\endcsname{\color{white}}%
      \expandafter\def\csname LTb\endcsname{\color{black}}%
      \expandafter\def\csname LTa\endcsname{\color{black}}%
      \expandafter\def\csname LT0\endcsname{\color[rgb]{1,0,0}}%
      \expandafter\def\csname LT1\endcsname{\color[rgb]{0,1,0}}%
      \expandafter\def\csname LT2\endcsname{\color[rgb]{0,0,1}}%
      \expandafter\def\csname LT3\endcsname{\color[rgb]{1,0,1}}%
      \expandafter\def\csname LT4\endcsname{\color[rgb]{0,1,1}}%
      \expandafter\def\csname LT5\endcsname{\color[rgb]{1,1,0}}%
      \expandafter\def\csname LT6\endcsname{\color[rgb]{0,0,0}}%
      \expandafter\def\csname LT7\endcsname{\color[rgb]{1,0.3,0}}%
      \expandafter\def\csname LT8\endcsname{\color[rgb]{0.5,0.5,0.5}}%
    \else
      % gray
      \def\colorrgb#1{\color{black}}%
      \def\colorgray#1{\color[gray]{#1}}%
      \expandafter\def\csname LTw\endcsname{\color{white}}%
      \expandafter\def\csname LTb\endcsname{\color{black}}%
      \expandafter\def\csname LTa\endcsname{\color{black}}%
      \expandafter\def\csname LT0\endcsname{\color{black}}%
      \expandafter\def\csname LT1\endcsname{\color{black}}%
      \expandafter\def\csname LT2\endcsname{\color{black}}%
      \expandafter\def\csname LT3\endcsname{\color{black}}%
      \expandafter\def\csname LT4\endcsname{\color{black}}%
      \expandafter\def\csname LT5\endcsname{\color{black}}%
      \expandafter\def\csname LT6\endcsname{\color{black}}%
      \expandafter\def\csname LT7\endcsname{\color{black}}%
      \expandafter\def\csname LT8\endcsname{\color{black}}%
    \fi
  \fi
    \setlength{\unitlength}{0.0500bp}%
    \ifx\gptboxheight\undefined%
      \newlength{\gptboxheight}%
      \newlength{\gptboxwidth}%
      \newsavebox{\gptboxtext}%
    \fi%
    \setlength{\fboxrule}{0.5pt}%
    \setlength{\fboxsep}{1pt}%
\begin{picture}(7200.00,5040.00)%
      \csname LTb\endcsname%%
      \put(3600,4820){\makebox(0,0){\strut{}}}%
    \gplgaddtomacro\gplbacktext{%
      \csname LTb\endcsname%%
      \put(588,3360){\makebox(0,0)[r]{\strut{}-8}}%
      \put(588,3654){\makebox(0,0)[r]{\strut{}-6}}%
      \put(588,3948){\makebox(0,0)[r]{\strut{}-4}}%
      \put(588,4241){\makebox(0,0)[r]{\strut{}-2}}%
      \put(588,4535){\makebox(0,0)[r]{\strut{}0}}%
      \put(1584,3140){\makebox(0,0){\strut{}}}%
      \put(2447,3140){\makebox(0,0){\strut{}}}%
      \put(3311,3140){\makebox(0,0){\strut{}}}%
      \put(778,4300){\makebox(0,0)[l]{\strut{}$\beta=0.3$}}%
    }%
    \gplgaddtomacro\gplfronttext{%
    }%
    \gplgaddtomacro\gplbacktext{%
      \csname LTb\endcsname%%
      \put(588,1932){\makebox(0,0)[r]{\strut{}-8}}%
      \put(588,2226){\makebox(0,0)[r]{\strut{}-6}}%
      \put(588,2520){\makebox(0,0)[r]{\strut{}-4}}%
      \put(588,2813){\makebox(0,0)[r]{\strut{}-2}}%
      \put(588,3107){\makebox(0,0)[r]{\strut{}0}}%
      \put(720,1712){\makebox(0,0){\strut{}0}}%
      \put(1584,1712){\makebox(0,0){\strut{}1}}%
      \put(2447,1712){\makebox(0,0){\strut{}2}}%
      \put(3311,1712){\makebox(0,0){\strut{}3}}%
      \put(3596,1712){\makebox(0,0){\strut{}e-3}}%
      \put(778,2872){\makebox(0,0)[l]{\strut{}$\beta=0.9$}}%
    }%
    \gplgaddtomacro\gplfronttext{%
      \csname LTb\endcsname%%
      \put(2159,1382){\makebox(0,0){\strut{}$1/M$}}%
    }%
    \gplgaddtomacro\gplbacktext{%
    }%
    \gplgaddtomacro\gplfronttext{%
      \csname LTb\endcsname%%
      \put(2326,1053){\makebox(0,0)[r]{\strut{}TNR N=30}}%
      \csname LTb\endcsname%%
      \put(2326,833){\makebox(0,0)[r]{\strut{}TNR N=40}}%
      \csname LTb\endcsname%%
      \put(2326,613){\makebox(0,0)[r]{\strut{}TNR N=60}}%
      \csname LTb\endcsname%%
      \put(2326,393){\makebox(0,0)[r]{\strut{}TNR N=80}}%
      \csname LTb\endcsname%%
      \put(2326,173){\makebox(0,0)[r]{\strut{}TNR N=100}}%
    }%
    \gplgaddtomacro\gplbacktext{%
      \csname LTb\endcsname%%
      \put(3827,3360){\makebox(0,0)[r]{\strut{} }}%
      \put(3827,3507){\makebox(0,0)[r]{\strut{} }}%
      \put(3827,3654){\makebox(0,0)[r]{\strut{} }}%
      \put(3827,3801){\makebox(0,0)[r]{\strut{} }}%
      \put(3827,3948){\makebox(0,0)[r]{\strut{} }}%
      \put(3827,4094){\makebox(0,0)[r]{\strut{} }}%
      \put(3827,4241){\makebox(0,0)[r]{\strut{} }}%
      \put(3827,4388){\makebox(0,0)[r]{\strut{} }}%
      \put(3827,4535){\makebox(0,0)[r]{\strut{} }}%
      \put(4823,3140){\makebox(0,0){\strut{}}}%
      \put(5687,3140){\makebox(0,0){\strut{}}}%
      \put(6551,3140){\makebox(0,0){\strut{}}}%
      \put(4017,4300){\makebox(0,0)[l]{\strut{}$\beta=0.5$}}%
    }%
    \gplgaddtomacro\gplfronttext{%
    }%
    \gplgaddtomacro\gplbacktext{%
      \csname LTb\endcsname%%
      \put(3827,1932){\makebox(0,0)[r]{\strut{} }}%
      \put(3827,2079){\makebox(0,0)[r]{\strut{} }}%
      \put(3827,2226){\makebox(0,0)[r]{\strut{} }}%
      \put(3827,2373){\makebox(0,0)[r]{\strut{} }}%
      \put(3827,2520){\makebox(0,0)[r]{\strut{} }}%
      \put(3827,2666){\makebox(0,0)[r]{\strut{} }}%
      \put(3827,2813){\makebox(0,0)[r]{\strut{} }}%
      \put(3827,2960){\makebox(0,0)[r]{\strut{} }}%
      \put(3827,3107){\makebox(0,0)[r]{\strut{} }}%
      \put(3959,1712){\makebox(0,0){\strut{}0}}%
      \put(4823,1712){\makebox(0,0){\strut{}1}}%
      \put(5687,1712){\makebox(0,0){\strut{}2}}%
      \put(6551,1712){\makebox(0,0){\strut{}3}}%
      \put(6836,1712){\makebox(0,0){\strut{}e-3}}%
      \put(4017,2872){\makebox(0,0)[l]{\strut{}$\beta=1.1$}}%
    }%
    \gplgaddtomacro\gplfronttext{%
      \csname LTb\endcsname%%
      \put(5399,1382){\makebox(0,0){\strut{}$1/M$}}%
    }%
    \gplgaddtomacro\gplbacktext{%
    }%
    \gplgaddtomacro\gplfronttext{%
      \csname LTb\endcsname%%
      \put(5565,1053){\makebox(0,0)[r]{\strut{}TPR N=30}}%
      \csname LTb\endcsname%%
      \put(5565,833){\makebox(0,0)[r]{\strut{}TPR N=40}}%
      \csname LTb\endcsname%%
      \put(5565,613){\makebox(0,0)[r]{\strut{}TPR N=60}}%
      \csname LTb\endcsname%%
      \put(5565,393){\makebox(0,0)[r]{\strut{}TPR N=80}}%
      \csname LTb\endcsname%%
      \put(5565,173){\makebox(0,0)[r]{\strut{}TPR N=100}}%
    }%
    \gplbacktext
    \put(0,0){\includegraphics{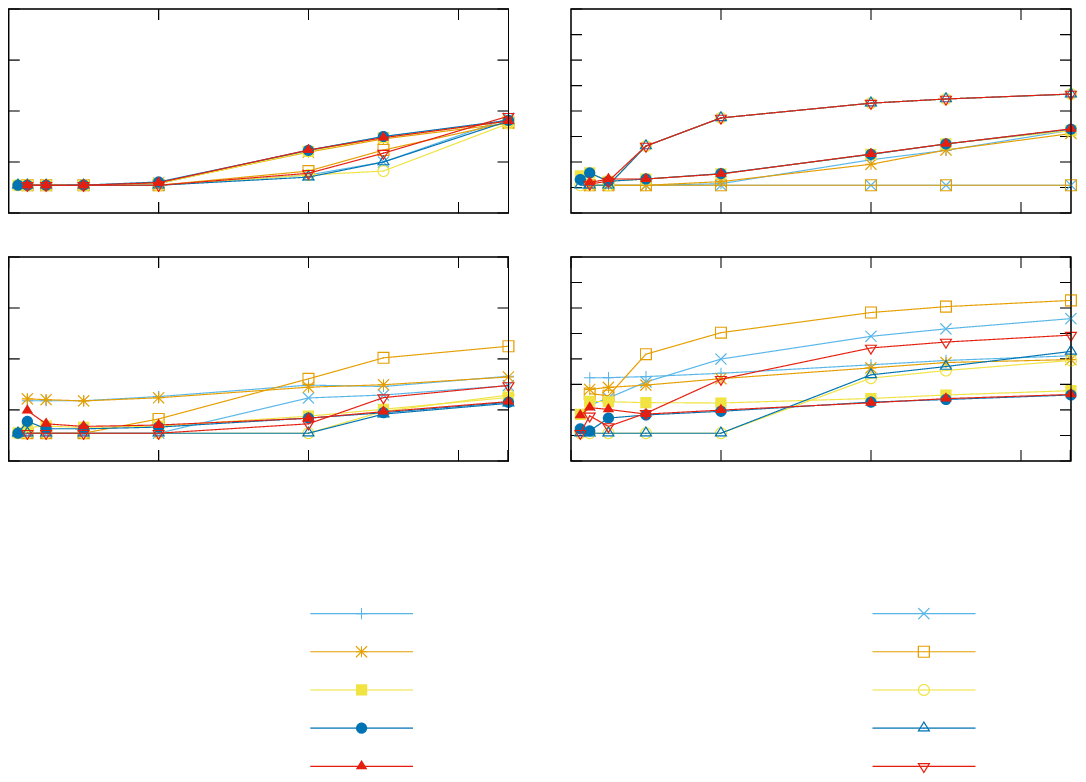}}%
    \gplfronttext
  \end{picture}%
\endgroup

%% file: hist_RR_03.tex
% GNUPLOT: LaTeX picture with Postscript
\begingroup
  \makeatletter
  \providecommand\color[2][]{%
    \GenericError{(gnuplot) \space\space\space\@spaces}{%
      Package color not loaded in conjunction with
      terminal option `colourtext'%
    }{See the gnuplot documentation for explanation.%
    }{Either use 'blacktext' in gnuplot or load the package
      color.sty in LaTeX.}%
    \renewcommand\color[2][]{}%
  }%
  \providecommand\includegraphics[2][]{%
    \GenericError{(gnuplot) \space\space\space\@spaces}{%
      Package graphicx or graphics not loaded%
    }{See the gnuplot documentation for explanation.%
    }{The gnuplot epslatex terminal needs graphicx.sty or graphics.sty.}%
    \renewcommand\includegraphics[2][]{}%
  }%
  \providecommand\rotatebox[2]{#2}%
  \@ifundefined{ifGPcolor}{%
    \newif\ifGPcolor
    \GPcolortrue
  }{}%
  \@ifundefined{ifGPblacktext}{%
    \newif\ifGPblacktext
    \GPblacktexttrue
  }{}%
  % define a \g@addto@macro without @ in the name:
  \let\gplgaddtomacro\g@addto@macro
  % define empty templates for all commands taking text:
  \gdef\gplbacktext{}%
  \gdef\gplfronttext{}%
  \makeatother
  \ifGPblacktext
    % no textcolor at all
    \def\colorrgb#1{}%
    \def\colorgray#1{}%
  \else
    % gray or color?
    \ifGPcolor
      \def\colorrgb#1{\color[rgb]{#1}}%
      \def\colorgray#1{\color[gray]{#1}}%
      \expandafter\def\csname LTw\endcsname{\color{white}}%
      \expandafter\def\csname LTb\endcsname{\color{black}}%
      \expandafter\def\csname LTa\endcsname{\color{black}}%
      \expandafter\def\csname LT0\endcsname{\color[rgb]{1,0,0}}%
      \expandafter\def\csname LT1\endcsname{\color[rgb]{0,1,0}}%
      \expandafter\def\csname LT2\endcsname{\color[rgb]{0,0,1}}%
      \expandafter\def\csname LT3\endcsname{\color[rgb]{1,0,1}}%
      \expandafter\def\csname LT4\endcsname{\color[rgb]{0,1,1}}%
      \expandafter\def\csname LT5\endcsname{\color[rgb]{1,1,0}}%
      \expandafter\def\csname LT6\endcsname{\color[rgb]{0,0,0}}%
      \expandafter\def\csname LT7\endcsname{\color[rgb]{1,0.3,0}}%
      \expandafter\def\csname LT8\endcsname{\color[rgb]{0.5,0.5,0.5}}%
    \else
      % gray
      \def\colorrgb#1{\color{black}}%
      \def\colorgray#1{\color[gray]{#1}}%
      \expandafter\def\csname LTw\endcsname{\color{white}}%
      \expandafter\def\csname LTb\endcsname{\color{black}}%
      \expandafter\def\csname LTa\endcsname{\color{black}}%
      \expandafter\def\csname LT0\endcsname{\color{black}}%
      \expandafter\def\csname LT1\endcsname{\color{black}}%
      \expandafter\def\csname LT2\endcsname{\color{black}}%
      \expandafter\def\csname LT3\endcsname{\color{black}}%
      \expandafter\def\csname LT4\endcsname{\color{black}}%
      \expandafter\def\csname LT5\endcsname{\color{black}}%
      \expandafter\def\csname LT6\endcsname{\color{black}}%
      \expandafter\def\csname LT7\endcsname{\color{black}}%
      \expandafter\def\csname LT8\endcsname{\color{black}}%
    \fi
  \fi
    \setlength{\unitlength}{0.0500bp}%
    \ifx\gptboxheight\undefined%
      \newlength{\gptboxheight}%
      \newlength{\gptboxwidth}%
      \newsavebox{\gptboxtext}%
    \fi%
    \setlength{\fboxrule}{0.5pt}%
    \setlength{\fboxsep}{1pt}%
\begin{picture}(7200.00,5040.00)%
    \gplgaddtomacro\gplbacktext{%
      \csname LTb\endcsname%%
      \put(462,704){\makebox(0,0)[r]{\strut{}$0$}}%
      \put(462,1733){\makebox(0,0)[r]{\strut{}$5$}}%
      \put(462,2762){\makebox(0,0)[r]{\strut{}$10$}}%
      \put(462,3790){\makebox(0,0)[r]{\strut{}$15$}}%
      \put(462,4819){\makebox(0,0)[r]{\strut{}$20$}}%
      \put(594,484){\makebox(0,0){\strut{}$0$}}%
      \put(1481,484){\makebox(0,0){\strut{}$100$}}%
      \put(2368,484){\makebox(0,0){\strut{}$200$}}%
      \put(3255,484){\makebox(0,0){\strut{}$300$}}%
      \put(4142,484){\makebox(0,0){\strut{}$400$}}%
      \put(5029,484){\makebox(0,0){\strut{}$500$}}%
      \put(5916,484){\makebox(0,0){\strut{}$600$}}%
      \put(6803,484){\makebox(0,0){\strut{}$700$}}%
    }%
    \gplgaddtomacro\gplfronttext{%
      \csname LTb\endcsname%%
      \put(3698,154){\makebox(0,0){\strut{}$t$}}%
      \csname LTb\endcsname%%
      \put(1726,4298){\makebox(0,0)[r]{\strut{}$\epsilon$}}%
      \csname LTb\endcsname%%
      \put(1726,4078){\makebox(0,0)[r]{\strut{}$100 \delta K$}}%
    }%
    \gplgaddtomacro\gplbacktext{%
      \csname LTb\endcsname%%
      \put(3258,2316){\makebox(0,0)[r]{\strut{}$0$}}%
      \put(3258,2736){\makebox(0,0)[r]{\strut{}$0.1$}}%
      \put(3258,3156){\makebox(0,0)[r]{\strut{}$0.2$}}%
      \put(3258,3576){\makebox(0,0)[r]{\strut{}$0.3$}}%
      \put(3258,3996){\makebox(0,0)[r]{\strut{}$0.4$}}%
      \put(3258,4416){\makebox(0,0)[r]{\strut{}$0.5$}}%
      \put(3390,2096){\makebox(0,0){\strut{}$-1$}}%
      \put(4189,2096){\makebox(0,0){\strut{}$-0.5$}}%
      \put(4989,2096){\makebox(0,0){\strut{}$0$}}%
      \put(5788,2096){\makebox(0,0){\strut{}$0.5$}}%
      \put(6587,2096){\makebox(0,0){\strut{}$1$}}%
    }%
    \gplgaddtomacro\gplfronttext{%
      \csname LTb\endcsname%%
      \put(4988,1766){\makebox(0,0){\strut{}$J$}}%
      \csname LTb\endcsname%%
      \put(4134,3886){\makebox(0,0)[r]{\strut{}$t=100$}}%
      \csname LTb\endcsname%%
      \put(4134,3666){\makebox(0,0)[r]{\strut{}$t=300$}}%
      \csname LTb\endcsname%%
      \put(4134,3446){\makebox(0,0)[r]{\strut{}$t=500$}}%
      \csname LTb\endcsname%%
      \put(4134,3226){\makebox(0,0)[r]{\strut{}$t=700$}}%
    }%
    \gplbacktext
    \put(0,0){\includegraphics{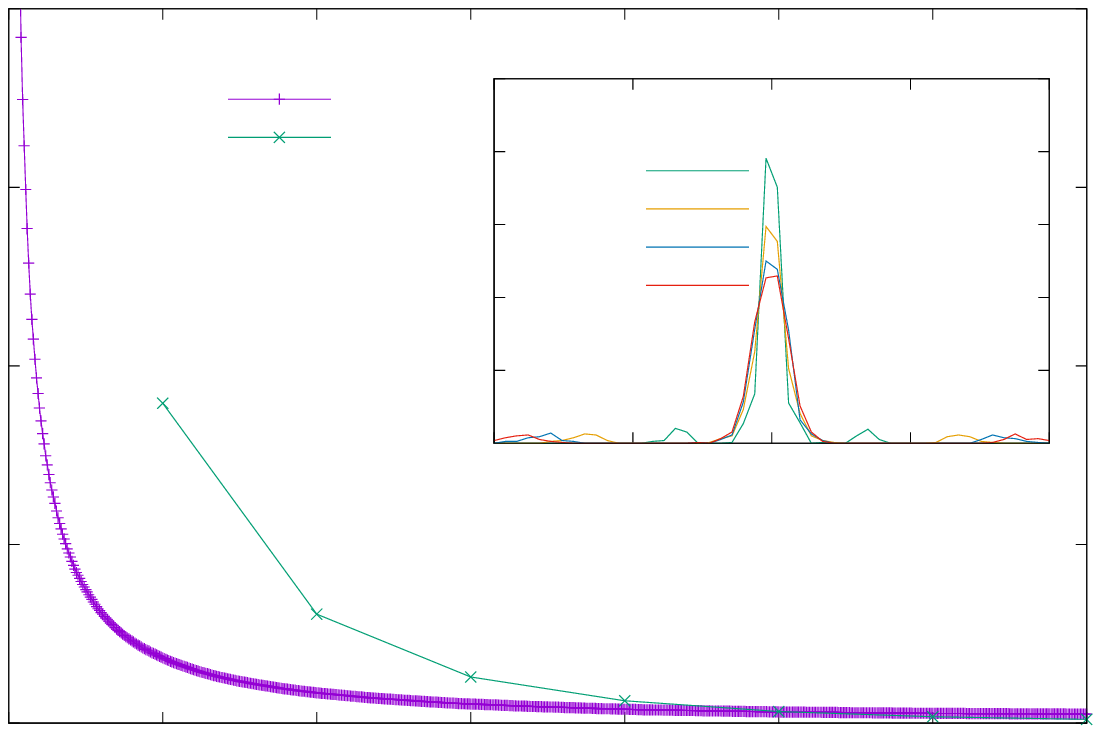}}%
    \gplfronttext
  \end{picture}%
\endgroup

%% file: hist_RR_11.tex
% GNUPLOT: LaTeX picture with Postscript
\begingroup
  \makeatletter
  \providecommand\color[2][]{%
    \GenericError{(gnuplot) \space\space\space\@spaces}{%
      Package color not loaded in conjunction with
      terminal option `colourtext'%
    }{See the gnuplot documentation for explanation.%
    }{Either use 'blacktext' in gnuplot or load the package
      color.sty in LaTeX.}%
    \renewcommand\color[2][]{}%
  }%
  \providecommand\includegraphics[2][]{%
    \GenericError{(gnuplot) \space\space\space\@spaces}{%
      Package graphicx or graphics not loaded%
    }{See the gnuplot documentation for explanation.%
    }{The gnuplot epslatex terminal needs graphicx.sty or graphics.sty.}%
    \renewcommand\includegraphics[2][]{}%
  }%
  \providecommand\rotatebox[2]{#2}%
  \@ifundefined{ifGPcolor}{%
    \newif\ifGPcolor
    \GPcolortrue
  }{}%
  \@ifundefined{ifGPblacktext}{%
    \newif\ifGPblacktext
    \GPblacktexttrue
  }{}%
  % define a \g@addto@macro without @ in the name:
  \let\gplgaddtomacro\g@addto@macro
  % define empty templates for all commands taking text:
  \gdef\gplbacktext{}%
  \gdef\gplfronttext{}%
  \makeatother
  \ifGPblacktext
    % no textcolor at all
    \def\colorrgb#1{}%
    \def\colorgray#1{}%
  \else
    % gray or color?
    \ifGPcolor
      \def\colorrgb#1{\color[rgb]{#1}}%
      \def\colorgray#1{\color[gray]{#1}}%
      \expandafter\def\csname LTw\endcsname{\color{white}}%
      \expandafter\def\csname LTb\endcsname{\color{black}}%
      \expandafter\def\csname LTa\endcsname{\color{black}}%
      \expandafter\def\csname LT0\endcsname{\color[rgb]{1,0,0}}%
      \expandafter\def\csname LT1\endcsname{\color[rgb]{0,1,0}}%
      \expandafter\def\csname LT2\endcsname{\color[rgb]{0,0,1}}%
      \expandafter\def\csname LT3\endcsname{\color[rgb]{1,0,1}}%
      \expandafter\def\csname LT4\endcsname{\color[rgb]{0,1,1}}%
      \expandafter\def\csname LT5\endcsname{\color[rgb]{1,1,0}}%
      \expandafter\def\csname LT6\endcsname{\color[rgb]{0,0,0}}%
      \expandafter\def\csname LT7\endcsname{\color[rgb]{1,0.3,0}}%
      \expandafter\def\csname LT8\endcsname{\color[rgb]{0.5,0.5,0.5}}%
    \else
      % gray
      \def\colorrgb#1{\color{black}}%
      \def\colorgray#1{\color[gray]{#1}}%
      \expandafter\def\csname LTw\endcsname{\color{white}}%
      \expandafter\def\csname LTb\endcsname{\color{black}}%
      \expandafter\def\csname LTa\endcsname{\color{black}}%
      \expandafter\def\csname LT0\endcsname{\color{black}}%
      \expandafter\def\csname LT1\endcsname{\color{black}}%
      \expandafter\def\csname LT2\endcsname{\color{black}}%
      \expandafter\def\csname LT3\endcsname{\color{black}}%
      \expandafter\def\csname LT4\endcsname{\color{black}}%
      \expandafter\def\csname LT5\endcsname{\color{black}}%
      \expandafter\def\csname LT6\endcsname{\color{black}}%
      \expandafter\def\csname LT7\endcsname{\color{black}}%
      \expandafter\def\csname LT8\endcsname{\color{black}}%
    \fi
  \fi
    \setlength{\unitlength}{0.0500bp}%
    \ifx\gptboxheight\undefined%
      \newlength{\gptboxheight}%
      \newlength{\gptboxwidth}%
      \newsavebox{\gptboxtext}%
    \fi%
    \setlength{\fboxrule}{0.5pt}%
    \setlength{\fboxsep}{1pt}%
\begin{picture}(7200.00,5040.00)%
    \gplgaddtomacro\gplbacktext{%
      \csname LTb\endcsname%%
      \put(462,704){\makebox(0,0)[r]{\strut{}$0$}}%
      \put(462,1733){\makebox(0,0)[r]{\strut{}$5$}}%
      \put(462,2762){\makebox(0,0)[r]{\strut{}$10$}}%
      \put(462,3790){\makebox(0,0)[r]{\strut{}$15$}}%
      \put(462,4819){\makebox(0,0)[r]{\strut{}$20$}}%
      \put(594,484){\makebox(0,0){\strut{}$0$}}%
      \put(1481,484){\makebox(0,0){\strut{}$100$}}%
      \put(2368,484){\makebox(0,0){\strut{}$200$}}%
      \put(3255,484){\makebox(0,0){\strut{}$300$}}%
      \put(4142,484){\makebox(0,0){\strut{}$400$}}%
      \put(5029,484){\makebox(0,0){\strut{}$500$}}%
      \put(5916,484){\makebox(0,0){\strut{}$600$}}%
      \put(6803,484){\makebox(0,0){\strut{}$700$}}%
    }%
    \gplgaddtomacro\gplfronttext{%
      \csname LTb\endcsname%%
      \put(3698,154){\makebox(0,0){\strut{}$t$}}%
      \csname LTb\endcsname%%
      \put(1726,4298){\makebox(0,0)[r]{\strut{}$\epsilon$}}%
      \csname LTb\endcsname%%
      \put(1726,4078){\makebox(0,0)[r]{\strut{}$100 \delta K$}}%
    }%
    \gplgaddtomacro\gplbacktext{%
      \csname LTb\endcsname%%
      \put(3390,2316){\makebox(0,0)[r]{\strut{}$0$}}%
      \put(3390,2666){\makebox(0,0)[r]{\strut{}$0.05$}}%
      \put(3390,3016){\makebox(0,0)[r]{\strut{}$0.1$}}%
      \put(3390,3366){\makebox(0,0)[r]{\strut{}$0.15$}}%
      \put(3390,3716){\makebox(0,0)[r]{\strut{}$0.2$}}%
      \put(3390,4066){\makebox(0,0)[r]{\strut{}$0.25$}}%
      \put(3390,4416){\makebox(0,0)[r]{\strut{}$0.3$}}%
      \put(3522,2096){\makebox(0,0){\strut{}$-1$}}%
      \put(4288,2096){\makebox(0,0){\strut{}$-0.5$}}%
      \put(5055,2096){\makebox(0,0){\strut{}$0$}}%
      \put(5821,2096){\makebox(0,0){\strut{}$0.5$}}%
      \put(6587,2096){\makebox(0,0){\strut{}$1$}}%
    }%
    \gplgaddtomacro\gplfronttext{%
      \csname LTb\endcsname%%
      \put(5054,1766){\makebox(0,0){\strut{}$J$}}%
      \csname LTb\endcsname%%
      \put(4200,3886){\makebox(0,0)[r]{\strut{}$t=100$}}%
      \csname LTb\endcsname%%
      \put(4200,3666){\makebox(0,0)[r]{\strut{}$t=300$}}%
      \csname LTb\endcsname%%
      \put(4200,3446){\makebox(0,0)[r]{\strut{}$t=500$}}%
      \csname LTb\endcsname%%
      \put(4200,3226){\makebox(0,0)[r]{\strut{}$t=700$}}%
    }%
    \gplbacktext
    \put(0,0){\includegraphics{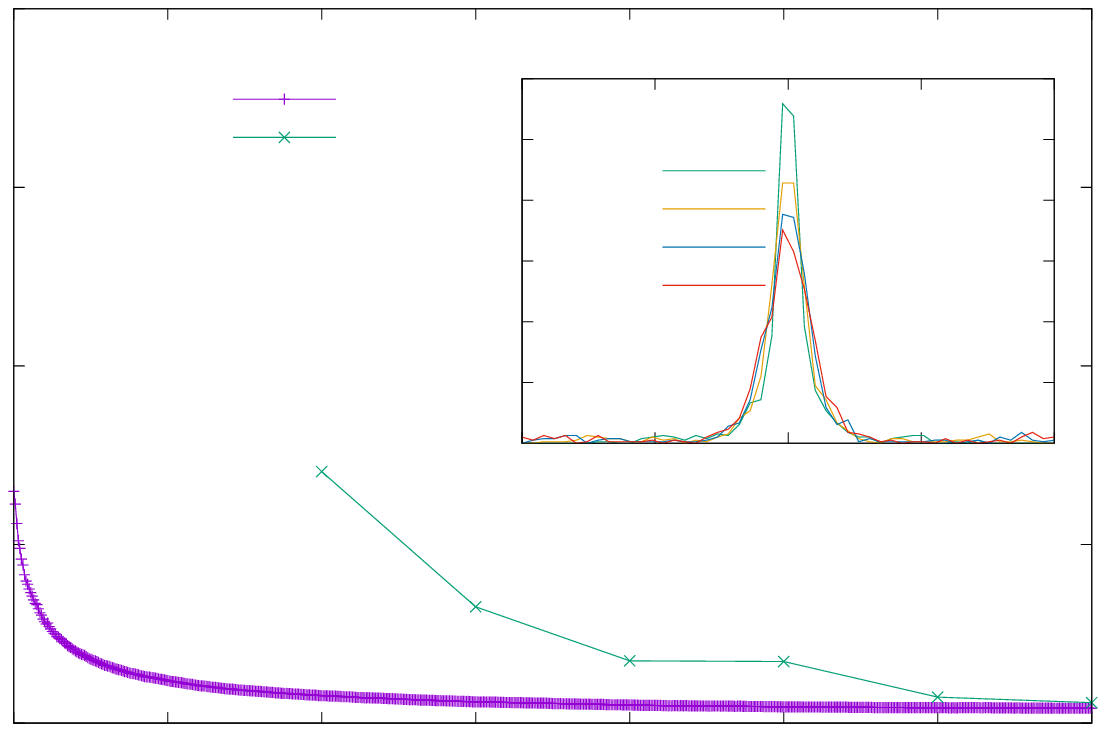}}%
    \gplfronttext
  \end{picture}%
\endgroup

%% file: hist_2D_PB_02.tex
% GNUPLOT: LaTeX picture with Postscript
\begingroup
  \makeatletter
  \providecommand\color[2][]{%
    \GenericError{(gnuplot) \space\space\space\@spaces}{%
      Package color not loaded in conjunction with
      terminal option `colourtext'%
    }{See the gnuplot documentation for explanation.%
    }{Either use 'blacktext' in gnuplot or load the package
      color.sty in LaTeX.}%
    \renewcommand\color[2][]{}%
  }%
  \providecommand\includegraphics[2][]{%
    \GenericError{(gnuplot) \space\space\space\@spaces}{%
      Package graphicx or graphics not loaded%
    }{See the gnuplot documentation for explanation.%
    }{The gnuplot epslatex terminal needs graphicx.sty or graphics.sty.}%
    \renewcommand\includegraphics[2][]{}%
  }%
  \providecommand\rotatebox[2]{#2}%
  \@ifundefined{ifGPcolor}{%
    \newif\ifGPcolor
    \GPcolortrue
  }{}%
  \@ifundefined{ifGPblacktext}{%
    \newif\ifGPblacktext
    \GPblacktexttrue
  }{}%
  % define a \g@addto@macro without @ in the name:
  \let\gplgaddtomacro\g@addto@macro
  % define empty templates for all commands taking text:
  \gdef\gplbacktext{}%
  \gdef\gplfronttext{}%
  \makeatother
  \ifGPblacktext
    % no textcolor at all
    \def\colorrgb#1{}%
    \def\colorgray#1{}%
  \else
    % gray or color?
    \ifGPcolor
      \def\colorrgb#1{\color[rgb]{#1}}%
      \def\colorgray#1{\color[gray]{#1}}%
      \expandafter\def\csname LTw\endcsname{\color{white}}%
      \expandafter\def\csname LTb\endcsname{\color{black}}%
      \expandafter\def\csname LTa\endcsname{\color{black}}%
      \expandafter\def\csname LT0\endcsname{\color[rgb]{1,0,0}}%
      \expandafter\def\csname LT1\endcsname{\color[rgb]{0,1,0}}%
      \expandafter\def\csname LT2\endcsname{\color[rgb]{0,0,1}}%
      \expandafter\def\csname LT3\endcsname{\color[rgb]{1,0,1}}%
      \expandafter\def\csname LT4\endcsname{\color[rgb]{0,1,1}}%
      \expandafter\def\csname LT5\endcsname{\color[rgb]{1,1,0}}%
      \expandafter\def\csname LT6\endcsname{\color[rgb]{0,0,0}}%
      \expandafter\def\csname LT7\endcsname{\color[rgb]{1,0.3,0}}%
      \expandafter\def\csname LT8\endcsname{\color[rgb]{0.5,0.5,0.5}}%
    \else
      % gray
      \def\colorrgb#1{\color{black}}%
      \def\colorgray#1{\color[gray]{#1}}%
      \expandafter\def\csname LTw\endcsname{\color{white}}%
      \expandafter\def\csname LTb\endcsname{\color{black}}%
      \expandafter\def\csname LTa\endcsname{\color{black}}%
      \expandafter\def\csname LT0\endcsname{\color{black}}%
      \expandafter\def\csname LT1\endcsname{\color{black}}%
      \expandafter\def\csname LT2\endcsname{\color{black}}%
      \expandafter\def\csname LT3\endcsname{\color{black}}%
      \expandafter\def\csname LT4\endcsname{\color{black}}%
      \expandafter\def\csname LT5\endcsname{\color{black}}%
      \expandafter\def\csname LT6\endcsname{\color{black}}%
      \expandafter\def\csname LT7\endcsname{\color{black}}%
      \expandafter\def\csname LT8\endcsname{\color{black}}%
    \fi
  \fi
    \setlength{\unitlength}{0.0500bp}%
    \ifx\gptboxheight\undefined%
      \newlength{\gptboxheight}%
      \newlength{\gptboxwidth}%
      \newsavebox{\gptboxtext}%
    \fi%
    \setlength{\fboxrule}{0.5pt}%
    \setlength{\fboxsep}{1pt}%
\begin{picture}(7200.00,5040.00)%
    \gplgaddtomacro\gplbacktext{%
      \csname LTb\endcsname%%
      \put(462,704){\makebox(0,0)[r]{\strut{}$0$}}%
      \put(462,1733){\makebox(0,0)[r]{\strut{}$5$}}%
      \put(462,2762){\makebox(0,0)[r]{\strut{}$10$}}%
      \put(462,3790){\makebox(0,0)[r]{\strut{}$15$}}%
      \put(462,4819){\makebox(0,0)[r]{\strut{}$20$}}%
      \put(594,484){\makebox(0,0){\strut{}$0$}}%
      \put(1481,484){\makebox(0,0){\strut{}$100$}}%
      \put(2368,484){\makebox(0,0){\strut{}$200$}}%
      \put(3255,484){\makebox(0,0){\strut{}$300$}}%
      \put(4142,484){\makebox(0,0){\strut{}$400$}}%
      \put(5029,484){\makebox(0,0){\strut{}$500$}}%
      \put(5916,484){\makebox(0,0){\strut{}$600$}}%
      \put(6803,484){\makebox(0,0){\strut{}$700$}}%
    }%
    \gplgaddtomacro\gplfronttext{%
      \csname LTb\endcsname%%
      \put(3698,154){\makebox(0,0){\strut{}$t$}}%
      \csname LTb\endcsname%%
      \put(1726,4298){\makebox(0,0)[r]{\strut{}$\epsilon$}}%
      \csname LTb\endcsname%%
      \put(1726,4078){\makebox(0,0)[r]{\strut{}$200 \delta K$}}%
    }%
    \gplgaddtomacro\gplbacktext{%
      \csname LTb\endcsname%%
      \put(3258,2316){\makebox(0,0)[r]{\strut{}$0$}}%
      \put(3258,2736){\makebox(0,0)[r]{\strut{}$0.1$}}%
      \put(3258,3156){\makebox(0,0)[r]{\strut{}$0.2$}}%
      \put(3258,3576){\makebox(0,0)[r]{\strut{}$0.3$}}%
      \put(3258,3996){\makebox(0,0)[r]{\strut{}$0.4$}}%
      \put(3258,4416){\makebox(0,0)[r]{\strut{}$0.5$}}%
      \put(3636,2096){\makebox(0,0){\strut{}$-0.2$}}%
      \put(4128,2096){\makebox(0,0){\strut{}$0$}}%
      \put(4620,2096){\makebox(0,0){\strut{}$0.2$}}%
      \put(5111,2096){\makebox(0,0){\strut{}$0.4$}}%
      \put(5603,2096){\makebox(0,0){\strut{}$0.6$}}%
      \put(6095,2096){\makebox(0,0){\strut{}$0.8$}}%
      \put(6587,2096){\makebox(0,0){\strut{}$1$}}%
    }%
    \gplgaddtomacro\gplfronttext{%
      \csname LTb\endcsname%%
      \put(4988,1766){\makebox(0,0){\strut{}$J$}}%
      \csname LTb\endcsname%%
      \put(5093,3886){\makebox(0,0)[r]{\strut{}$t=100$}}%
      \csname LTb\endcsname%%
      \put(5093,3666){\makebox(0,0)[r]{\strut{}$t=300$}}%
      \csname LTb\endcsname%%
      \put(5093,3446){\makebox(0,0)[r]{\strut{}$t=500$}}%
      \csname LTb\endcsname%%
      \put(5093,3226){\makebox(0,0)[r]{\strut{}$t=700$}}%
    }%
    \gplbacktext
    \put(0,0){\includegraphics{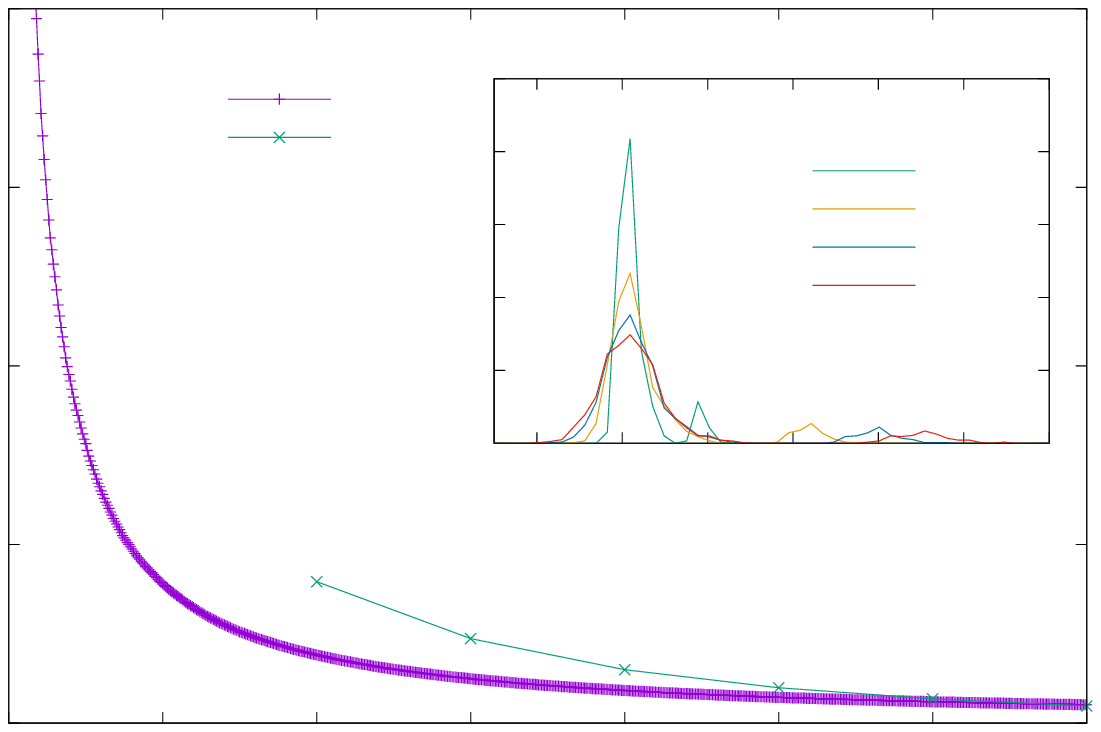}}%
    \gplfronttext
  \end{picture}%
\endgroup

%% file: hist_2D_PB_05.tex
% GNUPLOT: LaTeX picture with Postscript
\begingroup
  \makeatletter
  \providecommand\color[2][]{%
    \GenericError{(gnuplot) \space\space\space\@spaces}{%
      Package color not loaded in conjunction with
      terminal option `colourtext'%
    }{See the gnuplot documentation for explanation.%
    }{Either use 'blacktext' in gnuplot or load the package
      color.sty in LaTeX.}%
    \renewcommand\color[2][]{}%
  }%
  \providecommand\includegraphics[2][]{%
    \GenericError{(gnuplot) \space\space\space\@spaces}{%
      Package graphicx or graphics not loaded%
    }{See the gnuplot documentation for explanation.%
    }{The gnuplot epslatex terminal needs graphicx.sty or graphics.sty.}%
    \renewcommand\includegraphics[2][]{}%
  }%
  \providecommand\rotatebox[2]{#2}%
  \@ifundefined{ifGPcolor}{%
    \newif\ifGPcolor
    \GPcolortrue
  }{}%
  \@ifundefined{ifGPblacktext}{%
    \newif\ifGPblacktext
    \GPblacktexttrue
  }{}%
  % define a \g@addto@macro without @ in the name:
  \let\gplgaddtomacro\g@addto@macro
  % define empty templates for all commands taking text:
  \gdef\gplbacktext{}%
  \gdef\gplfronttext{}%
  \makeatother
  \ifGPblacktext
    % no textcolor at all
    \def\colorrgb#1{}%
    \def\colorgray#1{}%
  \else
    % gray or color?
    \ifGPcolor
      \def\colorrgb#1{\color[rgb]{#1}}%
      \def\colorgray#1{\color[gray]{#1}}%
      \expandafter\def\csname LTw\endcsname{\color{white}}%
      \expandafter\def\csname LTb\endcsname{\color{black}}%
      \expandafter\def\csname LTa\endcsname{\color{black}}%
      \expandafter\def\csname LT0\endcsname{\color[rgb]{1,0,0}}%
      \expandafter\def\csname LT1\endcsname{\color[rgb]{0,1,0}}%
      \expandafter\def\csname LT2\endcsname{\color[rgb]{0,0,1}}%
      \expandafter\def\csname LT3\endcsname{\color[rgb]{1,0,1}}%
      \expandafter\def\csname LT4\endcsname{\color[rgb]{0,1,1}}%
      \expandafter\def\csname LT5\endcsname{\color[rgb]{1,1,0}}%
      \expandafter\def\csname LT6\endcsname{\color[rgb]{0,0,0}}%
      \expandafter\def\csname LT7\endcsname{\color[rgb]{1,0.3,0}}%
      \expandafter\def\csname LT8\endcsname{\color[rgb]{0.5,0.5,0.5}}%
    \else
      % gray
      \def\colorrgb#1{\color{black}}%
      \def\colorgray#1{\color[gray]{#1}}%
      \expandafter\def\csname LTw\endcsname{\color{white}}%
      \expandafter\def\csname LTb\endcsname{\color{black}}%
      \expandafter\def\csname LTa\endcsname{\color{black}}%
      \expandafter\def\csname LT0\endcsname{\color{black}}%
      \expandafter\def\csname LT1\endcsname{\color{black}}%
      \expandafter\def\csname LT2\endcsname{\color{black}}%
      \expandafter\def\csname LT3\endcsname{\color{black}}%
      \expandafter\def\csname LT4\endcsname{\color{black}}%
      \expandafter\def\csname LT5\endcsname{\color{black}}%
      \expandafter\def\csname LT6\endcsname{\color{black}}%
      \expandafter\def\csname LT7\endcsname{\color{black}}%
      \expandafter\def\csname LT8\endcsname{\color{black}}%
    \fi
  \fi
    \setlength{\unitlength}{0.0500bp}%
    \ifx\gptboxheight\undefined%
      \newlength{\gptboxheight}%
      \newlength{\gptboxwidth}%
      \newsavebox{\gptboxtext}%
    \fi%
    \setlength{\fboxrule}{0.5pt}%
    \setlength{\fboxsep}{1pt}%
\begin{picture}(7200.00,5040.00)%
    \gplgaddtomacro\gplbacktext{%
      \csname LTb\endcsname%%
      \put(462,704){\makebox(0,0)[r]{\strut{}$0$}}%
      \put(462,1733){\makebox(0,0)[r]{\strut{}$5$}}%
      \put(462,2762){\makebox(0,0)[r]{\strut{}$10$}}%
      \put(462,3790){\makebox(0,0)[r]{\strut{}$15$}}%
      \put(462,4819){\makebox(0,0)[r]{\strut{}$20$}}%
      \put(594,484){\makebox(0,0){\strut{}$0$}}%
      \put(1481,484){\makebox(0,0){\strut{}$100$}}%
      \put(2368,484){\makebox(0,0){\strut{}$200$}}%
      \put(3255,484){\makebox(0,0){\strut{}$300$}}%
      \put(4142,484){\makebox(0,0){\strut{}$400$}}%
      \put(5029,484){\makebox(0,0){\strut{}$500$}}%
      \put(5916,484){\makebox(0,0){\strut{}$600$}}%
      \put(6803,484){\makebox(0,0){\strut{}$700$}}%
    }%
    \gplgaddtomacro\gplfronttext{%
      \csname LTb\endcsname%%
      \put(3698,154){\makebox(0,0){\strut{}$t$}}%
      \csname LTb\endcsname%%
      \put(1726,4298){\makebox(0,0)[r]{\strut{}$\epsilon$}}%
      \csname LTb\endcsname%%
      \put(1726,4078){\makebox(0,0)[r]{\strut{}$200 \delta K$}}%
    }%
    \gplgaddtomacro\gplbacktext{%
      \csname LTb\endcsname%%
      \put(3258,2316){\makebox(0,0)[r]{\strut{}$0$}}%
      \put(3258,2736){\makebox(0,0)[r]{\strut{}$0.1$}}%
      \put(3258,3156){\makebox(0,0)[r]{\strut{}$0.2$}}%
      \put(3258,3576){\makebox(0,0)[r]{\strut{}$0.3$}}%
      \put(3258,3996){\makebox(0,0)[r]{\strut{}$0.4$}}%
      \put(3258,4416){\makebox(0,0)[r]{\strut{}$0.5$}}%
      \put(3636,2096){\makebox(0,0){\strut{}$-0.2$}}%
      \put(4128,2096){\makebox(0,0){\strut{}$0$}}%
      \put(4620,2096){\makebox(0,0){\strut{}$0.2$}}%
      \put(5111,2096){\makebox(0,0){\strut{}$0.4$}}%
      \put(5603,2096){\makebox(0,0){\strut{}$0.6$}}%
      \put(6095,2096){\makebox(0,0){\strut{}$0.8$}}%
      \put(6587,2096){\makebox(0,0){\strut{}$1$}}%
    }%
    \gplgaddtomacro\gplfronttext{%
      \csname LTb\endcsname%%
      \put(4988,1766){\makebox(0,0){\strut{}$J$}}%
      \csname LTb\endcsname%%
      \put(5093,3886){\makebox(0,0)[r]{\strut{}$t=100$}}%
      \csname LTb\endcsname%%
      \put(5093,3666){\makebox(0,0)[r]{\strut{}$t=300$}}%
      \csname LTb\endcsname%%
      \put(5093,3446){\makebox(0,0)[r]{\strut{}$t=500$}}%
      \csname LTb\endcsname%%
      \put(5093,3226){\makebox(0,0)[r]{\strut{}$t=700$}}%
    }%
    \gplbacktext
    \put(0,0){\includegraphics{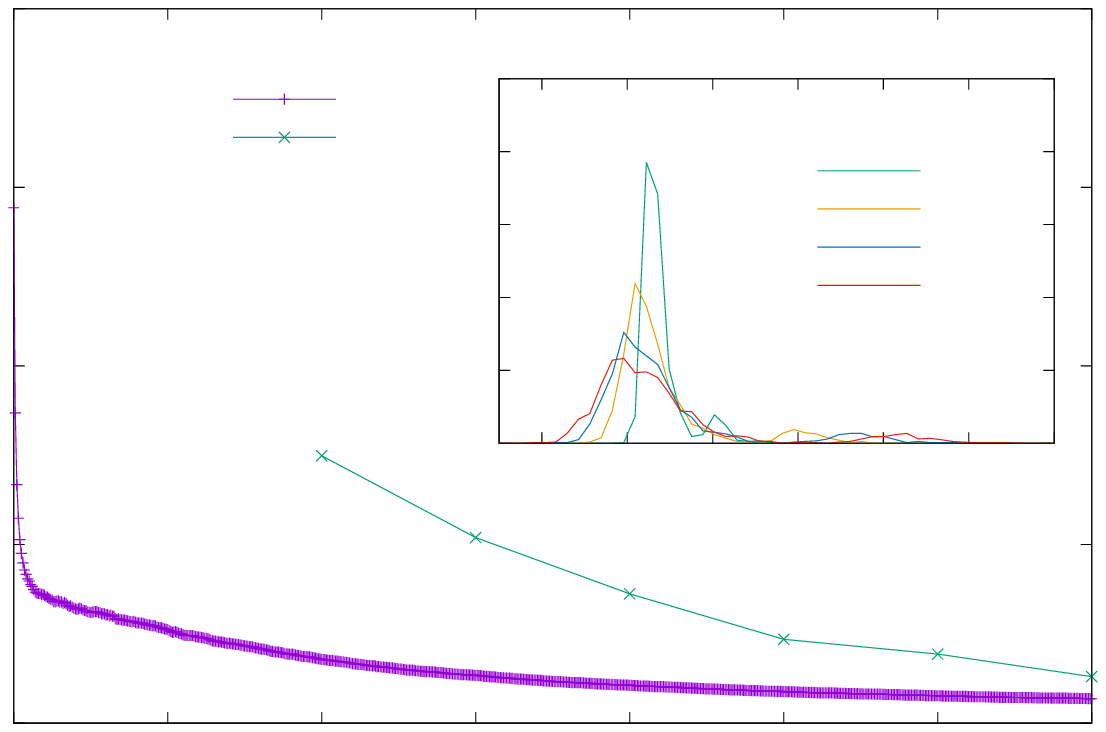}}%
    \gplfronttext
  \end{picture}%
\endgroup

%% file: time-2D-PB.tex
% GNUPLOT: LaTeX picture with Postscript
\begingroup
  \makeatletter
  \providecommand\color[2][]{%
    \GenericError{(gnuplot) \space\space\space\@spaces}{%
      Package color not loaded in conjunction with
      terminal option `colourtext'%
    }{See the gnuplot documentation for explanation.%
    }{Either use 'blacktext' in gnuplot or load the package
      color.sty in LaTeX.}%
    \renewcommand\color[2][]{}%
  }%
  \providecommand\includegraphics[2][]{%
    \GenericError{(gnuplot) \space\space\space\@spaces}{%
      Package graphicx or graphics not loaded%
    }{See the gnuplot documentation for explanation.%
    }{The gnuplot epslatex terminal needs graphicx.sty or graphics.sty.}%
    \renewcommand\includegraphics[2][]{}%
  }%
  \providecommand\rotatebox[2]{#2}%
  \@ifundefined{ifGPcolor}{%
    \newif\ifGPcolor
    \GPcolortrue
  }{}%
  \@ifundefined{ifGPblacktext}{%
    \newif\ifGPblacktext
    \GPblacktexttrue
  }{}%
  % define a \g@addto@macro without @ in the name:
  \let\gplgaddtomacro\g@addto@macro
  % define empty templates for all commands taking text:
  \gdef\gplbacktext{}%
  \gdef\gplfronttext{}%
  \makeatother
  \ifGPblacktext
    % no textcolor at all
    \def\colorrgb#1{}%
    \def\colorgray#1{}%
  \else
    % gray or color?
    \ifGPcolor
      \def\colorrgb#1{\color[rgb]{#1}}%
      \def\colorgray#1{\color[gray]{#1}}%
      \expandafter\def\csname LTw\endcsname{\color{white}}%
      \expandafter\def\csname LTb\endcsname{\color{black}}%
      \expandafter\def\csname LTa\endcsname{\color{black}}%
      \expandafter\def\csname LT0\endcsname{\color[rgb]{1,0,0}}%
      \expandafter\def\csname LT1\endcsname{\color[rgb]{0,1,0}}%
      \expandafter\def\csname LT2\endcsname{\color[rgb]{0,0,1}}%
      \expandafter\def\csname LT3\endcsname{\color[rgb]{1,0,1}}%
      \expandafter\def\csname LT4\endcsname{\color[rgb]{0,1,1}}%
      \expandafter\def\csname LT5\endcsname{\color[rgb]{1,1,0}}%
      \expandafter\def\csname LT6\endcsname{\color[rgb]{0,0,0}}%
      \expandafter\def\csname LT7\endcsname{\color[rgb]{1,0.3,0}}%
      \expandafter\def\csname LT8\endcsname{\color[rgb]{0.5,0.5,0.5}}%
    \else
      % gray
      \def\colorrgb#1{\color{black}}%
      \def\colorgray#1{\color[gray]{#1}}%
      \expandafter\def\csname LTw\endcsname{\color{white}}%
      \expandafter\def\csname LTb\endcsname{\color{black}}%
      \expandafter\def\csname LTa\endcsname{\color{black}}%
      \expandafter\def\csname LT0\endcsname{\color{black}}%
      \expandafter\def\csname LT1\endcsname{\color{black}}%
      \expandafter\def\csname LT2\endcsname{\color{black}}%
      \expandafter\def\csname LT3\endcsname{\color{black}}%
      \expandafter\def\csname LT4\endcsname{\color{black}}%
      \expandafter\def\csname LT5\endcsname{\color{black}}%
      \expandafter\def\csname LT6\endcsname{\color{black}}%
      \expandafter\def\csname LT7\endcsname{\color{black}}%
      \expandafter\def\csname LT8\endcsname{\color{black}}%
    \fi
  \fi
    \setlength{\unitlength}{0.0500bp}%
    \ifx\gptboxheight\undefined%
      \newlength{\gptboxheight}%
      \newlength{\gptboxwidth}%
      \newsavebox{\gptboxtext}%
    \fi%
    \setlength{\fboxrule}{0.5pt}%
    \setlength{\fboxsep}{1pt}%
\begin{picture}(7200.00,5040.00)%
    \gplgaddtomacro\gplbacktext{%
      \csname LTb\endcsname%%
      \put(588,504){\makebox(0,0)[r]{\strut{}$1.5$}}%
      \put(588,907){\makebox(0,0)[r]{\strut{}$2$}}%
      \put(588,1310){\makebox(0,0)[r]{\strut{}$2.5$}}%
      \put(588,1713){\makebox(0,0)[r]{\strut{}$3$}}%
      \put(588,2116){\makebox(0,0)[r]{\strut{}$3.5$}}%
      \put(588,2520){\makebox(0,0)[r]{\strut{}$4$}}%
      \put(588,2923){\makebox(0,0)[r]{\strut{}$4.5$}}%
      \put(588,3326){\makebox(0,0)[r]{\strut{}$5$}}%
      \put(588,3729){\makebox(0,0)[r]{\strut{}$5.5$}}%
      \put(588,4132){\makebox(0,0)[r]{\strut{}$6$}}%
      \put(588,4535){\makebox(0,0)[r]{\strut{}$6.5$}}%
      \put(720,284){\makebox(0,0){\strut{}$3.2$}}%
      \put(1057,284){\makebox(0,0){\strut{}$3.4$}}%
      \put(1395,284){\makebox(0,0){\strut{}$3.6$}}%
      \put(1732,284){\makebox(0,0){\strut{}$3.8$}}%
      \put(2070,284){\makebox(0,0){\strut{}$4$}}%
      \put(2407,284){\makebox(0,0){\strut{}$4.2$}}%
      \put(2744,284){\makebox(0,0){\strut{}$4.4$}}%
      \put(3082,284){\makebox(0,0){\strut{}$4.6$}}%
      \put(3419,284){\makebox(0,0){\strut{}$4.8$}}%
      \put(1980,4787){\makebox(0,0)[l]{\strut{}\text{(a)}}}%
      \put(5399,4787){\makebox(0,0)[l]{\strut{}\text{(b)}}}%
    }%
    \gplgaddtomacro\gplfronttext{%
      \csname LTb\endcsname%%
      \put(-28,2519){\rotatebox{-270}{\makebox(0,0){\strut{}$\log[\text{time(s)}]$}}}%
      \put(2069,-46){\makebox(0,0){\strut{}$\log(N)$}}%
      \csname LTb\endcsname%%
      \put(2429,1402){\makebox(0,0)[r]{\strut{}$\beta=0.2$}}%
      \csname LTb\endcsname%%
      \put(2429,1182){\makebox(0,0)[r]{\strut{}$\beta=0.3$}}%
      \csname LTb\endcsname%%
      \put(2429,962){\makebox(0,0)[r]{\strut{}$\beta=0.4$}}%
      \csname LTb\endcsname%%
      \put(2429,742){\makebox(0,0)[r]{\strut{}$\beta=0.5$}}%
    }%
    \gplgaddtomacro\gplbacktext{%
      \csname LTb\endcsname%%
      \put(4008,504){\makebox(0,0)[r]{\strut{}$1$}}%
      \put(4008,907){\makebox(0,0)[r]{\strut{}$1.5$}}%
      \put(4008,1310){\makebox(0,0)[r]{\strut{}$2$}}%
      \put(4008,1713){\makebox(0,0)[r]{\strut{}$2.5$}}%
      \put(4008,2116){\makebox(0,0)[r]{\strut{}$3$}}%
      \put(4008,2520){\makebox(0,0)[r]{\strut{}$3.5$}}%
      \put(4008,2923){\makebox(0,0)[r]{\strut{}$4$}}%
      \put(4008,3326){\makebox(0,0)[r]{\strut{}$4.5$}}%
      \put(4008,3729){\makebox(0,0)[r]{\strut{}$5$}}%
      \put(4008,4132){\makebox(0,0)[r]{\strut{}$5.5$}}%
      \put(4008,4535){\makebox(0,0)[r]{\strut{}$6$}}%
      \put(4140,284){\makebox(0,0){\strut{}$3.2$}}%
      \put(4477,284){\makebox(0,0){\strut{}$3.4$}}%
      \put(4815,284){\makebox(0,0){\strut{}$3.6$}}%
      \put(5152,284){\makebox(0,0){\strut{}$3.8$}}%
      \put(5490,284){\makebox(0,0){\strut{}$4$}}%
      \put(5827,284){\makebox(0,0){\strut{}$4.2$}}%
      \put(6164,284){\makebox(0,0){\strut{}$4.4$}}%
      \put(6502,284){\makebox(0,0){\strut{}$4.6$}}%
      \put(6839,284){\makebox(0,0){\strut{}$4.8$}}%
      \put(1980,4787){\makebox(0,0)[l]{\strut{}\text{(a)}}}%
      \put(5399,4787){\makebox(0,0)[l]{\strut{}\text{(b)}}}%
    }%
    \gplgaddtomacro\gplfronttext{%
      \csname LTb\endcsname%%
      \put(5489,-46){\makebox(0,0){\strut{}$\log(N)$}}%
      \csname LTb\endcsname%%
      \put(5849,1402){\makebox(0,0)[r]{\strut{}$\beta=0.2$}}%
      \csname LTb\endcsname%%
      \put(5849,1182){\makebox(0,0)[r]{\strut{}$\beta=0.3$}}%
      \csname LTb\endcsname%%
      \put(5849,962){\makebox(0,0)[r]{\strut{}$\beta=0.4$}}%
      \csname LTb\endcsname%%
      \put(5849,742){\makebox(0,0)[r]{\strut{}$\beta=0.5$}}%
    }%
    \gplbacktext
    \put(0,0){\includegraphics{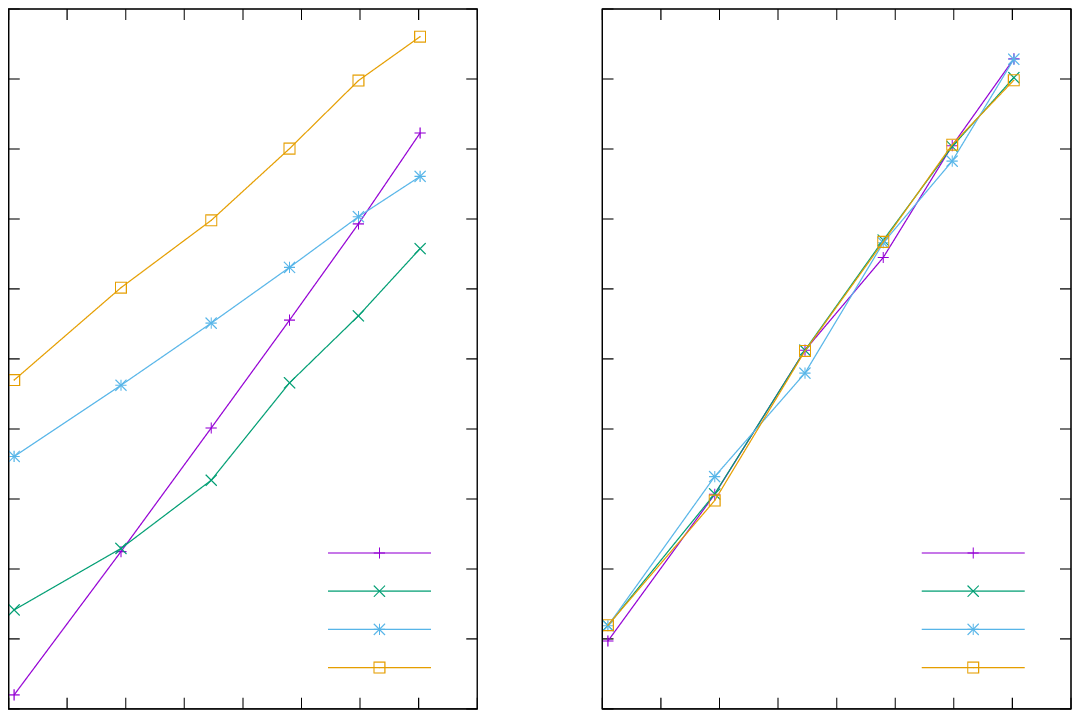}}%
    \gplfronttext
  \end{picture}%
\endgroup

%% file: eps-2D-PB.tex
% GNUPLOT: LaTeX picture with Postscript
\begingroup
  \makeatletter
  \providecommand\color[2][]{%
    \GenericError{(gnuplot) \space\space\space\@spaces}{%
      Package color not loaded in conjunction with
      terminal option `colourtext'%
    }{See the gnuplot documentation for explanation.%
    }{Either use 'blacktext' in gnuplot or load the package
      color.sty in LaTeX.}%
    \renewcommand\color[2][]{}%
  }%
  \providecommand\includegraphics[2][]{%
    \GenericError{(gnuplot) \space\space\space\@spaces}{%
      Package graphicx or graphics not loaded%
    }{See the gnuplot documentation for explanation.%
    }{The gnuplot epslatex terminal needs graphicx.sty or graphics.sty.}%
    \renewcommand\includegraphics[2][]{}%
  }%
  \providecommand\rotatebox[2]{#2}%
  \@ifundefined{ifGPcolor}{%
    \newif\ifGPcolor
    \GPcolortrue
  }{}%
  \@ifundefined{ifGPblacktext}{%
    \newif\ifGPblacktext
    \GPblacktexttrue
  }{}%
  % define a \g@addto@macro without @ in the name:
  \let\gplgaddtomacro\g@addto@macro
  % define empty templates for all commands taking text:
  \gdef\gplbacktext{}%
  \gdef\gplfronttext{}%
  \makeatother
  \ifGPblacktext
    % no textcolor at all
    \def\colorrgb#1{}%
    \def\colorgray#1{}%
  \else
    % gray or color?
    \ifGPcolor
      \def\colorrgb#1{\color[rgb]{#1}}%
      \def\colorgray#1{\color[gray]{#1}}%
      \expandafter\def\csname LTw\endcsname{\color{white}}%
      \expandafter\def\csname LTb\endcsname{\color{black}}%
      \expandafter\def\csname LTa\endcsname{\color{black}}%
      \expandafter\def\csname LT0\endcsname{\color[rgb]{1,0,0}}%
      \expandafter\def\csname LT1\endcsname{\color[rgb]{0,1,0}}%
      \expandafter\def\csname LT2\endcsname{\color[rgb]{0,0,1}}%
      \expandafter\def\csname LT3\endcsname{\color[rgb]{1,0,1}}%
      \expandafter\def\csname LT4\endcsname{\color[rgb]{0,1,1}}%
      \expandafter\def\csname LT5\endcsname{\color[rgb]{1,1,0}}%
      \expandafter\def\csname LT6\endcsname{\color[rgb]{0,0,0}}%
      \expandafter\def\csname LT7\endcsname{\color[rgb]{1,0.3,0}}%
      \expandafter\def\csname LT8\endcsname{\color[rgb]{0.5,0.5,0.5}}%
    \else
      % gray
      \def\colorrgb#1{\color{black}}%
      \def\colorgray#1{\color[gray]{#1}}%
      \expandafter\def\csname LTw\endcsname{\color{white}}%
      \expandafter\def\csname LTb\endcsname{\color{black}}%
      \expandafter\def\csname LTa\endcsname{\color{black}}%
      \expandafter\def\csname LT0\endcsname{\color{black}}%
      \expandafter\def\csname LT1\endcsname{\color{black}}%
      \expandafter\def\csname LT2\endcsname{\color{black}}%
      \expandafter\def\csname LT3\endcsname{\color{black}}%
      \expandafter\def\csname LT4\endcsname{\color{black}}%
      \expandafter\def\csname LT5\endcsname{\color{black}}%
      \expandafter\def\csname LT6\endcsname{\color{black}}%
      \expandafter\def\csname LT7\endcsname{\color{black}}%
      \expandafter\def\csname LT8\endcsname{\color{black}}%
    \fi
  \fi
    \setlength{\unitlength}{0.0500bp}%
    \ifx\gptboxheight\undefined%
      \newlength{\gptboxheight}%
      \newlength{\gptboxwidth}%
      \newsavebox{\gptboxtext}%
    \fi%
    \setlength{\fboxrule}{0.5pt}%
    \setlength{\fboxsep}{1pt}%
\begin{picture}(7200.00,5040.00)%
    \gplgaddtomacro\gplbacktext{%
      \csname LTb\endcsname%%
      \put(588,504){\makebox(0,0)[r]{\strut{}$0$}}%
      \put(588,1310){\makebox(0,0)[r]{\strut{}$0.2$}}%
      \put(588,2116){\makebox(0,0)[r]{\strut{}$0.4$}}%
      \put(588,2923){\makebox(0,0)[r]{\strut{}$0.6$}}%
      \put(588,3729){\makebox(0,0)[r]{\strut{}$0.8$}}%
      \put(588,4535){\makebox(0,0)[r]{\strut{}$1$}}%
      \put(720,284){\makebox(0,0){\strut{}$20$}}%
      \put(1057,284){\makebox(0,0){\strut{}$30$}}%
      \put(1395,284){\makebox(0,0){\strut{}$40$}}%
      \put(1732,284){\makebox(0,0){\strut{}$50$}}%
      \put(2070,284){\makebox(0,0){\strut{}$60$}}%
      \put(2407,284){\makebox(0,0){\strut{}$70$}}%
      \put(2744,284){\makebox(0,0){\strut{}$80$}}%
      \put(3082,284){\makebox(0,0){\strut{}$90$}}%
      \put(3419,284){\makebox(0,0){\strut{}$100$}}%
      \put(1980,4787){\makebox(0,0)[l]{\strut{}\text{(a)}}}%
      \put(5399,4787){\makebox(0,0)[l]{\strut{}\text{(b)}}}%
    }%
    \gplgaddtomacro\gplfronttext{%
      \csname LTb\endcsname%%
      \put(-28,2519){\rotatebox{-270}{\makebox(0,0){\strut{}$\epsilon$}}}%
      \put(2069,-46){\makebox(0,0){\strut{}$N$}}%
      \csname LTb\endcsname%%
      \put(1484,4183){\makebox(0,0)[r]{\strut{}$\beta=0.2$}}%
      \csname LTb\endcsname%%
      \put(1484,3963){\makebox(0,0)[r]{\strut{}$\beta=0.3$}}%
      \csname LTb\endcsname%%
      \put(1484,3743){\makebox(0,0)[r]{\strut{}$\beta=0.4$}}%
      \csname LTb\endcsname%%
      \put(1484,3523){\makebox(0,0)[r]{\strut{}$\beta=0.5$}}%
    }%
    \gplgaddtomacro\gplbacktext{%
      \csname LTb\endcsname%%
      \put(4008,504){\makebox(0,0)[r]{\strut{}$0$}}%
      \put(4008,1310){\makebox(0,0)[r]{\strut{}$0.2$}}%
      \put(4008,2116){\makebox(0,0)[r]{\strut{}$0.4$}}%
      \put(4008,2923){\makebox(0,0)[r]{\strut{}$0.6$}}%
      \put(4008,3729){\makebox(0,0)[r]{\strut{}$0.8$}}%
      \put(4008,4535){\makebox(0,0)[r]{\strut{}$1$}}%
      \put(4140,284){\makebox(0,0){\strut{}$20$}}%
      \put(4477,284){\makebox(0,0){\strut{}$30$}}%
      \put(4815,284){\makebox(0,0){\strut{}$40$}}%
      \put(5152,284){\makebox(0,0){\strut{}$50$}}%
      \put(5490,284){\makebox(0,0){\strut{}$60$}}%
      \put(5827,284){\makebox(0,0){\strut{}$70$}}%
      \put(6164,284){\makebox(0,0){\strut{}$80$}}%
      \put(6502,284){\makebox(0,0){\strut{}$90$}}%
      \put(6839,284){\makebox(0,0){\strut{}$100$}}%
      \put(1980,4787){\makebox(0,0)[l]{\strut{}\text{(a)}}}%
      \put(5399,4787){\makebox(0,0)[l]{\strut{}\text{(b)}}}%
    }%
    \gplgaddtomacro\gplfronttext{%
      \csname LTb\endcsname%%
      \put(5489,-46){\makebox(0,0){\strut{}$N$}}%
      \csname LTb\endcsname%%
      \put(4904,4183){\makebox(0,0)[r]{\strut{}$\beta=0.2$}}%
      \csname LTb\endcsname%%
      \put(4904,3963){\makebox(0,0)[r]{\strut{}$\beta=0.3$}}%
      \csname LTb\endcsname%%
      \put(4904,3743){\makebox(0,0)[r]{\strut{}$\beta=0.4$}}%
      \csname LTb\endcsname%%
      \put(4904,3523){\makebox(0,0)[r]{\strut{}$\beta=0.5$}}%
    }%
    \gplbacktext
    \put(0,0){\includegraphics{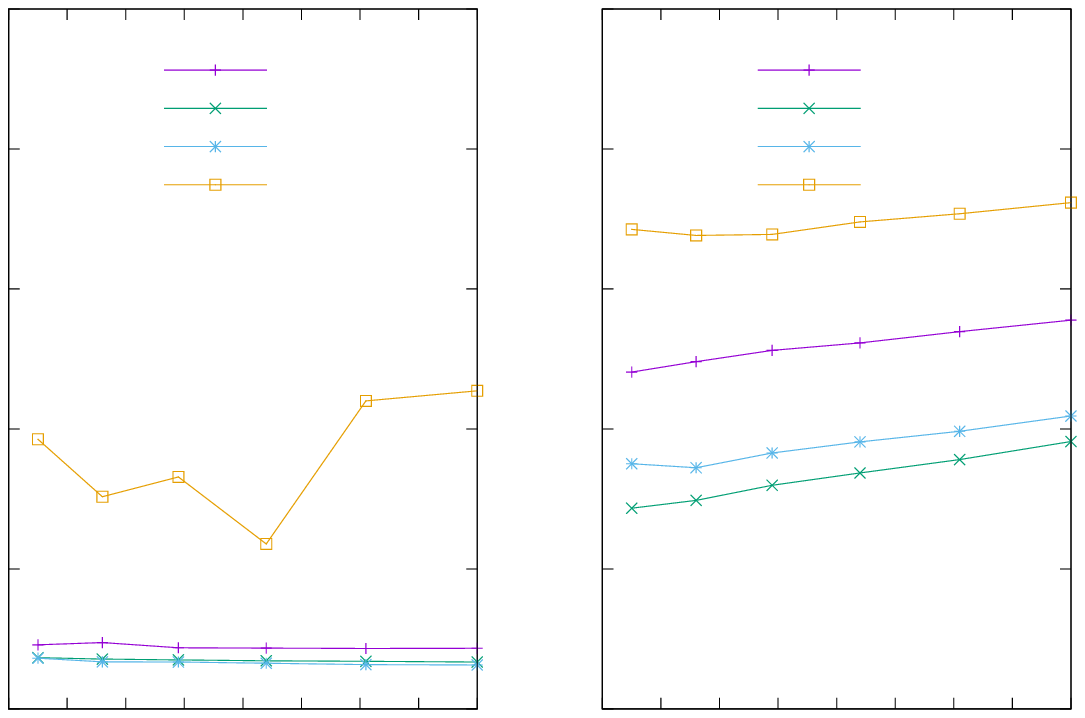}}%
    \gplfronttext
  \end{picture}%
\endgroup